\documentclass[aps, pra, 10pt, superscriptaddress, twocolumn, . longbibliography, floatfix]{revtex4-1}

\usepackage[nointlimits]{amsmath}
\usepackage{amsthm}
\usepackage{graphicx,nicefrac}
\usepackage{subfigure}
\usepackage{color}
\usepackage{txfonts}
\usepackage{mathtools}
\usepackage{hyperref}
\usepackage{tabularx}
\hypersetup{colorlinks=true, citecolor=blue, linkcolor=blue}

\usepackage{amsfonts,amsmath,amssymb}
\usepackage{array,bm,color}
\usepackage{epsfig,graphicx,nomencl,revsymb4-1,upgreek,url}
\usepackage{hyperref}
\usepackage{algpseudocode}
\usepackage{xspace}
\usepackage{tabularx}
\usepackage[normalem]{ulem}
\usepackage{changes}
\usepackage{enumitem}
\usepackage{booktabs}
\usepackage{mathtools}

\def\bra#1{{\left\langle #1 \right|}}
\def\ket#1{{\left| #1 \right\rangle}}

\newcommand{\cphase}{\mathsf{cphase}}
\newcommand{\cz}{\mathsf{cz}}
\newcommand{\cs}{\mathsf{cs}}
\newcommand{\csd}{\mathsf{cs}^{\dagger}}
\newcommand{\gate}[1]{#1} 

\makeatletter
\newcommand{\vast}{\bBigg@{4}}
\newcommand{\Vast}{\bBigg@{5}}
\makeatother

\newcommand{\g}[1]{\mathsf{#1}}
\newcommand{\tcp}[2]{\mathcal{T}(#1, #2)}
\newcommand{\rc}[3]{\mathcal{R}(#2 #1 #3)}
\newcommand{\rand}[1]{\mathsf{#1}}

\newcommand{\cnot}{\mathsf{cnot}}
\newcommand{\x}{\mathsf{x}}
\newcommand{\z}{\mathsf{z}}
\newcommand{\Q}[1]{\texttt{Q#1}}

\DeclareMathOperator*{\E}{\mathbb{E}}

\newcommand{\braa}[1]{\mathinner{\langle\!\langle #1\rvert}}
\newcommand{\kett}[1]{\mathinner{\lvert#1\rangle\!\rangle}}

\newtheorem{lemma}{Lemma}

\begin{document}
\title{Demonstrating scalable randomized benchmarking of universal gate sets}
\author{Jordan Hines}
\thanks{jordanh@berkeley.edu}
\affiliation{Department of Physics, University of California, Berkeley, CA 94720}
\affiliation{Quantum Performance Laboratory, Sandia National Laboratories, Albuquerque, NM 87185 and Livermore, CA 94550}
\author{Marie Lu}
\affiliation{Quantum Nanoelectronics Laboratory, Department of Physics, University of California, Berkeley, CA 94720}
\author{Ravi K. Naik}
\affiliation{Quantum Nanoelectronics Laboratory, Department of Physics, University of California, Berkeley, CA 94720}
\affiliation{Computational Research Division, Lawrence Berkeley National Laboratory, Berkeley, CA 94720}
\author{Akel Hashim}
\affiliation{Quantum Nanoelectronics Laboratory, Department of Physics, University of California, Berkeley, CA 94720}
\affiliation{Computational Research Division, Lawrence Berkeley National Laboratory, Berkeley, CA 94720}
\author{Jean-Loup Ville}
\thanks{Now at Alice \& Bob, 53 Bd du Général Martial Valin, 75015 Paris, FR}
\affiliation{Quantum Nanoelectronics Laboratory, Department of Physics, University of California, Berkeley, CA 94720}
\author{Brad Mitchell}
\affiliation{Quantum Nanoelectronics Laboratory, Department of Physics, University of California, Berkeley, CA 94720}
\affiliation{Computational Research Division, Lawrence Berkeley National Laboratory, Berkeley, CA 94720}
\author{John Mark Kriekebaum}
\thanks{Now at Google Quantum AI, Mountain View, CA, USA 94043.}
\affiliation{Quantum Nanoelectronics Laboratory, Department of Physics, University of California, Berkeley, CA 94720}
\affiliation{Materials Sciences Division, Lawrence Berkeley National Laboratory, Berkeley, CA 94720}
\author{David I. Santiago}
\affiliation{Quantum Nanoelectronics Laboratory, Department of Physics, University of California, Berkeley, CA 94720}
\affiliation{Computational Research Division, Lawrence Berkeley National Laboratory, Berkeley, CA 94720}
\author{Stefan Seritan}
\author{Erik Nielsen}
\author{Robin Blume-Kohout}
\author{Kevin Young}
\affiliation{Quantum Performance Laboratory, Sandia National Laboratories, Albuquerque, NM 87185 and Livermore, CA 94550}
\author{Irfan Siddiqi}
\affiliation{Quantum Nanoelectronics Laboratory, Department of Physics, University of California, Berkeley, CA 94720}
\affiliation{Computational Research Division, Lawrence Berkeley National Laboratory, Berkeley, CA 94720}
\affiliation{Materials Sciences Division, Lawrence Berkeley National Laboratory, Berkeley, CA 94720}
\author{Birgitta Whaley}
\affiliation{Department of Chemistry, University of California, Berkeley, CA 94720}
\author{Timothy Proctor}
\thanks{tjproct@sandia.gov}
\affiliation{Quantum Performance Laboratory, Sandia National Laboratories, Albuquerque, NM 87185 and Livermore, CA 94550}
\begin{abstract} 
Randomized benchmarking (RB) protocols are the most widely used methods for assessing the performance of quantum gates. However, the existing RB methods either do not scale to many qubits or cannot benchmark a universal gate set. Here, we introduce and demonstrate a technique for scalable RB of many universal and continuously parameterized gate sets, using a class of circuits called randomized mirror circuits. Our technique can be applied to a gate set containing an entangling Clifford gate and the set of arbitrary single-qubit gates, as well as gate sets containing controlled rotations about the Pauli axes. We use our technique to benchmark universal gate sets on four qubits of the Advanced Quantum Testbed, including a gate set containing a controlled-S gate and its inverse, and we investigate how the observed error rate is impacted by the inclusion of non-Clifford gates. Finally, we demonstrate that our technique scales to many qubits with experiments on a 27-qubit IBM Q processor. We use our technique to quantify the impact of crosstalk on this 27-qubit device, and we find that it contributes approximately $\nicefrac{2}{3}$ of the total error per gate in random many-qubit circuit layers.
\end{abstract}
\maketitle

\section{Introduction}
Quantum computers suffer from a diverse range of errors that must be quantified if their performance is to be understood and improved. Errors that are localized to single qubits or pairs of qubits can be studied in detail using tomographic techniques \cite{Nielsen2021gate, Rudinger2021-hu}. However, many-qubit circuits are often subject to large additional errors, such as crosstalk \cite{gambetta2012characterization, sarovar2019detecting, proctor2018direct, proctor2021scalable, proctor2020measuring, McKay2020-no}, that are not apparent in isolated one- or two-qubit experiments. There are now techniques for partial tomography on individual many-qubit circuit layers (also called ``cycles''), including cycle benchmarking \cite{erhard2019characterizing} and Pauli noise learning \cite{harper2019efficient, flammia2019efficient, Flammia2021-dn}. But quantum computers can typically implement exponentially many different circuit layers, and it is only feasible to characterize a small subset of them.

\begin{figure}[t!]
\includegraphics{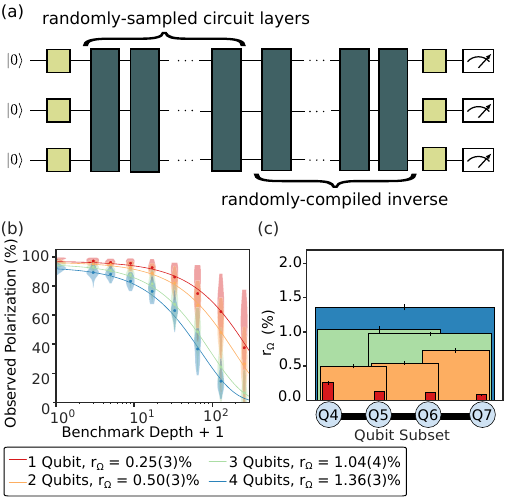} 
\caption{\textbf{Scalable randomized benchmarking of universal gate sets.} (a) Randomized mirror circuits combine a simple reflection structure with randomized compiling to enable scalable and robust RB of universal gate sets. (b) Data and fits to an exponential obtained by using our method---MRB of universal gate sets---to benchmark a universal gate set on $n=1,2,3,4$ qubits of the Advanced Quantum Testbed, and the average error rates of $n$-qubit layers ($r_{\Omega}$, where $\Omega$ is the layer sampling distribution) extracted from these decays. (c) We benchmarked each connected set of $n$ qubits for $n=1,2,3,4$, enabling us to map out the average layer error rate ($r_{\Omega}$) for each subset of qubits.}
\label{fig:1}
\end{figure}

Randomized benchmarks \cite{proctor2021scalable, proctor2020measuring,  McKay2020-no, emerson2005scalable, emerson2007symmetrized, magesan2011scalable, magesan2012characterizing, knill2008randomized, carignan2015characterizing, cross2016scalable, brown2018randomized, hashagen2018real, magesan2011scalable, magesan2012characterizing,  carignan2015characterizing, cross2016scalable, brown2018randomized, hashagen2018real, helsen2018new, Helsen2020-it, Claes2020-cy,  Helsen2020-mb, Morvan2020-ck, proctor2018direct, boixo2018characterizing, arute2019quantum, liu2021benchmarking,  cross2018validating, Mayer-df}  make it possible to quantify the rate of errors in an average $n$-qubit layer, by probing a quantum computer's performance on random $n$-qubit circuits. However, established randomized benchmarks cannot measure the performance of \emph{universal} layer sets in the many-qubit regime, where quantum computational advantage may be possible. Those randomized benchmarks that can be applied to universal layer sets, such as standard randomized benchmarking (RB) \cite{magesan2011scalable, magesan2012characterizing} and cross-entropy benchmarking (XEB) \cite{liu2021benchmarking, boixo2018characterizing, arute2019quantum}, require classical computations that scale exponentially in the number of qubits ($n$). XEB requires classical simulation of random circuits that are famously infeasible to simulate for more than  approximately $50$ qubits \cite{arute2019quantum}. This is because XEB requires estimating the (linear) cross-entropy between each circuit's actual and ideal output distributions. Standard RB of a universal layer set is restricted to even smaller $n$, because it requires compiling and running Haar random $n$-qubit unitaries \cite{magesan2011scalable}. This compilation requires classical computations that are exponentially expensive in $n$, and results in circuits containing $O(2^n)$ two-qubit gates \cite{shende2004minimal}.  Due to the large overhead, even standard RB on Clifford gates---which has lower overheads and non-exponential scaling---has only been implemented on up to 5 qubits \cite{proctor2021scalable, proctor2018direct, McKay2020-no}. Existing RB protocols can be used for heuristic estimates of the performance of a universal gate set---e.g., by synthesizing Clifford gates from a universal gate set \cite{Barends_2014} or by separately benchmarking Clifford gates with standard RB and a non-universal set of non-Clifford gates with dihedral RB \cite{carignan2015characterizing} or interleaved RB \cite{Garion2020-gi, helsen2018new,chasseur2017hybrid, harper2017estimating, dugas2016efficiently}. However, these approaches do not holistically assess a universal gate set, and they typically require strong assumptions on the types of gate errors to be accurate.

In this paper we introduce and demonstrate a simple and scalable technique for RB of a broad class of universal gate sets. Our technique uses a novel kind of \emph{randomized mirror circuits}, shown in Fig.~\ref{fig:1} (a), and advances on a recently introduced method---\emph{mirror} RB (MRB)---that enables scalable RB of Clifford gates \cite{proctor2021scalable}. Mirror circuits \cite{proctor2020measuring, proctor2021scalable, Flammia2021-dn} use a layer-by-layer inversion structure that enables classically efficient circuit construction and prediction of that circuit's error-free output. The idea of layer-by-layer inversion was explored in the earliest work on RB \cite{emerson2005scalable, emerson2007symmetrized}, and recently it was shown that the addition of Pauli frame randomization \cite{knill2005quantum} to Clifford mirror circuits enables reliable error rate estimation \cite{proctor2020measuring, proctor2021scalable, Flammia2021-dn}. The randomized mirror circuits we introduce here combine layer-by-layer inversion with a form of randomized compilation \cite{wallman2015noise} to enable reliable and efficient RB of universal gate sets. MRB on universal gate sets consists of running randomized mirror circuits of varied depths and computing their mean \emph{observed polarization} \cite{proctor2021scalable}, a quantity that is closely related to success probability. The mean observed polarization versus circuit depth is fit to an exponential decay, as shown in Fig.~\ref{fig:1} (b). As in standard RB, the estimated decay rate is then simply rescaled to estimate the average error rate of $n$-qubit layers.  MRB therefore preserves the core strengths and simplicity of standard RB and XEB, while avoiding the classical simulation and compilation roadblocks that have prevented scalable and efficient RB of universal gate sets. 

We use MRB to study errors in two different quantum computing systems, based on superconducting qubits. We demonstrate our method on 4 qubits of the Advanced Quantum Testbed (AQT) \cite{AQT-wp} and on all of the qubits of a 27-qubit IBM Q quantum computer (\texttt{ibmq\_montreal}) \cite{ibmq}. In our experiments on AQT we use MRB to quantify and compare the performance of three different layer sets on each subset of $n$ qubits (for $n=1,2,3,4$), including a layer set containing non-Clifford two-qubit gates [see Fig.~\ref{fig:1} (b-c)]. In our demonstration on  \texttt{ibmq\_montreal} we show that our method scales to many qubits by performing MRB on a universal gate set on up to 27 qubits. 

Multi-qubit MRB enables probing and quantifying crosstalk, which is an important source of error in contemporary many-qubit processors \cite{sarovar2019detecting, proctor2018direct, proctor2020measuring, gambetta2012characterization} that cannot be quantified by only testing one or two qubits in isolation. We quantify the contribution of crosstalk errors to the observed error rates in our experiments on AQT and further divide the error into contributions from individual layers and gates. The techniques we introduce for these analyses complement other established RB-like methods for estimating the error rates of individual gates---such as interleaved RB \cite{magesan2012efficient, Garion2020-gi, harper2017estimating} and cycle benchmarking \cite{erhard2019characterizing}. We use MRB to study how crosstalk errors vary on \texttt{ibmq\_montreal} as $n$ increases, with $n$ ranging from $n=1$ up to $n=27$. We find that crosstalk errors dominate in circuit layers with $n \gg 1$ qubits.

This paper is structured as follows: In Section~\ref{sec:prelim} we introduce our notation and define the error rate that our method measures. In Section~\ref{sec:mrb} we define the MRB protocol. In Section~\ref{sec:theory} we present theory and simulations that show that MRB is reliable. In Sections~\ref{sec:aqt} and~\ref{sec:ibm} we present the results of our experiments on AQT and demonstration on IBM Q's quantum processors, respectively.

\section{Definitions and Preliminaries}\label{sec:prelim}
In this section, we introduce our notation and background information related to our method. In Section~\ref{sec:defs} we introduce the notation and definitions used throughout this paper. In Section~\ref{sec:omega_distributed} we introduce the type of random circuits whose error MRB is designed to measure. In Section~\ref{sec:gate-set} we describe the gate sets that our method can be used to benchmark, i.e., we state the conditions that a gate set must satisfy if it is to be benchmarked with MRB.

\subsection{Definitions}\label{sec:defs}
We begin by introducing our notation and definitions. A $k$-qubit \emph{gate} $\gate{g}$ is an instruction to perform a particular unitary operation $U(\gate{g}) \in \mathbb{SU}(2^k)$ on $k$ qubits. We will only consider $k=1,2$, and we use $\mathbb{G}_1$ and $\mathbb{G}_2$ to denote a set of one- and two-qubit gates, respectively. In this work $\mathbb{G}_2$ will only contain controlled rotations about the $X$, $Y$, or $Z$ axis, denoted $\gate{CP}_{\theta}$ and defined by
\begin{equation}
U(\gate{CP}_{\theta}) = \ket{0}\bra{0} \otimes \mathbb{I} +  \ket{1}\bra{1} \otimes e^{-i \frac{\theta}{2} P},
\end{equation}
where $\theta$ is the angle of rotation and $P$ is the axis of rotation. Our experiments use four of these gates, which we write as $\cs = \gate{CZ}_{\nicefrac{\pi}{2}}$, $\csd = \gate{CZ}_{-\nicefrac{\pi}{2}}$, $\cphase = \gate{CZ}_{\pi}$, and $\cnot = \gate{CX}_{\pi}$.  We denote the single-qubit gate that is a rotation by $\theta$ about  $P$ by $P_{\theta}$.
An $n$-qubit, depth-$d$ \textit{circuit} is a length-$d$ sequence of $n$-qubit layers $\gate{C} = \gate{L}_d\gate{L}_{d-1}\cdots \gate{L}_2 \gate{L}_1$. An $n$-qubit \emph{layer} $\gate{L}$ is an instruction to perform a particular unitary operation $U(\gate{L}) \in \mathbb{SU}(2^n)$ on those $n$ qubits. In this work, we use layers that consist of parallel applications of only  one-qubit gates or only two-qubit gates. We use $\mathbb{L}(\mathbb{G})$ to denote the set of all layers constructed by parallel applications of gates from the gate set $\mathbb{G}$. Often it will be convenient to think of random circuits and layers as random variables, and when we do so we use the $\rand{L}$ font, e.g., we often use $\rand{L}$ to denote a layer-valued random variable, meaning that $\rand{L} = \gate{L}$ with probability $\Omega(\gate{L})$ for some distribution $\Omega$ over $\mathbb{L}(\mathbb{G})$. We use $\gate{L}^{-1}$ to denote an instruction to perform the operation $U(\gate{L})^{-1}$. 

For a layer or circuit $\gate{L}$, we use $\mathcal{U}(\gate{L})$ and $\phi(\gate{L})$ to denote the superoperator for its perfect and imperfect implementations, respectively, so $\mathcal{U}(\gate{L})[\rho] = U(\gate{L}) \rho U^{\dagger}(\gate{L})$. We assume that $\phi(\gate{L})$ is a completely positive trace-preserving (CPTP) map. We often represent superoperators as matrices, acting on states represented as vectors in Hilbert-Schmidt space (denoted by $\kett{\rho}$). A layer $\gate{L}$'s error map is defined by $\mathcal{E}(\gate{L}) = \phi(\gate{L})\mathcal{U}^{\dagger}(\gate{L})$. The \emph{entanglement fidelity} (also called the \emph{process fidelity}) of $\phi(\gate{L})$ to $\mathcal{U}(\gate{L})$ is defined by
\begin{align}
    F\bigl(\phi(\gate{L}), \mathcal{U}(\gate{L})\bigr) = F(\mathcal{E}) & = \langle \varphi | \bigl(\mathbb{I} \otimes \mathcal{E}(\gate{L})\bigr)[|\varphi \rangle \langle \varphi |]|\varphi \rangle \\
    & = \frac{1}{4^n}Tr(\mathcal{U}(L)^{\dag}\phi(L)),
\end{align}
where $\varphi$ is any maximally entangled state of $2n$ qubits \cite{nielsen2002simple}. Throughout, we use the term ``(in)fidelity'' to refer to the entanglement (in)fidelity. 

Our theory will make use of the \emph{polarization} of a channel $\mathcal{E}$, which is a rescaling of $\mathcal{E}$'s fidelity given by 
\begin{equation}
    \gamma(\mathcal{E}) = \frac{4^n}{4^n-1} F(\mathcal{E}) -  \frac{1}{4^n-1},  \label{eq:pol_def}
\end{equation}
 as well as stochastic Pauli channels. An $n$-qubit stochastic Pauli channel $\mathcal{E}_{\text{pauli}, \{\epsilon_{P}\}}$ is parameterized by a probability distribution $ \{\varepsilon_{P}\}$ over the $4^n$ Pauli operators ($\mathbb{P}_n$), and it has the action
\begin{equation}
    \mathcal{E}_{\text{pauli}, \{\varepsilon_{P}\}}[\rho] = \sum\limits_{P \in \mathbb{P}_n} \varepsilon_{P} P\rho P, \label{eq:pauli_channel}
\end{equation}
with $\sum_{P \in \mathbb{P}_n} \varepsilon_{P} = 1$. For a stochastic Pauli channel, the total probability of a fault, i.e., the probability it applies a non-identity Pauli operator, is
$1 - \varepsilon_{\mathbb{I}_n} = 1 - F(\mathcal{E}_{\text{pauli}, \{\varepsilon_{Q}\}})$.

\subsection{$\Omega$-distributed random circuits}
\label{sec:omega_distributed}

In this work, we aim to estimate the average error rate $\epsilon_{\Omega}$ of circuit layers sampled from a distribution $\Omega$. We now introduce a natural family of circuits---which we call $\Omega$-distributed random circuits---that we use in our method in order to estimate $\epsilon_{\Omega}$. $\Omega$-distributed random circuits are similar to the circuits used in XEB and other benchmarking routines. They are defined in terms of a customizable gate set $\mathbb{G}$ and sampling distribution $\Omega$ over that gate set. This gate set consists of one- and two-qubit gate sets $\mathbb{G} = (\mathbb{G}_1, \mathbb{G}_2)$, and $\Omega$ is determined by two probability distributions $\Omega_1$ and $\Omega_2$ over $n$-qubit layer sets $\mathbb{L}(\mathbb{G}_1)$ and $\mathbb{L}(\mathbb{G}_2)$, respectively. An $\Omega$-distributed random circuit with a \emph{benchmark depth} of $d$ is a circuit-valued random variable $\rand{C}_d=\rand{L}_{2d} \cdots \rand{L}_2\rand{L}_1$ where the $d$ odd-indexed layers are $\Omega_1$-distributed and the $d$ even-indexed layers are $\Omega_2$-distributed. These circuits consist of interleaved layers of one and two-qubit gates, so it is useful to define a \emph{composite layer} to be a pair of layers of the form $\rand{L} = \rand{L_2L_1}$ where $\rand{L}_1 \in \mathbb{L}(\mathbb{G}_1)$ is a layer of one-qubit gates and $\rand{L}_2 \in \mathbb{L}(\mathbb{G}_2)$ a layer of one-qubit gates. We denote the set of all composite layers by $\mathbb{L(\mathbb{G})}$. An $\Omega$-distributed random circuit of benchmark depth $d$ then consists of $d$ composite layers that are $\Omega$-distributed over $\mathbb{L}(\mathbb{G})$ with $\Omega(\gate{L}_2\gate{L}_1) = \Omega_1(L_1)\Omega_2(L_2)$.

\subsection{The gate set and sampling distributions}\label{sec:gate-set}
Our technique requires certain conditions of the gate set $\mathbb{G} = (\mathbb{G}_1, \mathbb{G}_2)$ and the sampling distributions $\Omega_1$ and $\Omega_2$. In order to construct the circuits required for our method, the gate set and sampling distributions must satisfy the following properties:
\begin{enumerate}
    \item $\mathbb{G}_2$ is a set of $\gate{CP}_{\theta}$ gates (defined in Section \ref{sec:defs}) and is closed under inverses. Examples of valid $\mathbb{G}_2$ are $\{\g{cnot}\}$ and $ \{\g{cs}, \g{cs}^{\dagger}\}$.  
    \item $\mathbb{G}_1$ is closed under inverses, conjugation by Pauli operators, and multiplication by the single-qubit Pauli axis rotation $P_{\theta}$ for each $\gate{CP}_{\theta} \in \mathbb{G}_2$. This is guaranteed to hold if $\mathbb{G}_1$ is the set of all single-qubit gates $\mathbb{SU}(2)$. 
\end{enumerate}

In addition, we require that our circuits are highly scrambling. To ensure that our circuits are highly scrambling, we require that our gate set and sampling distributions satisfy the following conditions:
\begin{enumerate}
    \item $\mathbb{G}_1$ is a unitary 2-design over $\mathbb{SU}(2)$. Examples of valid $\mathbb{G}_1$ are the set of all single-qubit gates $\mathbb{SU}(2)$ and the set of all 24 single-qubit Clifford gates $\mathbb{C}_1$.
    \item $\mathbb{G}_2$ contains at least one gate with $\theta \neq 0$, i.e., it must contain at least one entangling gate.
    \item $\Omega$-distributed layers quickly randomize and delocalize errors. Informally, this means that any Pauli error is mapped to a distribution over many different errors before another error is likely to have occurred. Formally, we require that for all Pauli operators $P, P' \neq \mathbb{I}_n$, there exists a constant $k \ll \nicefrac{1}{\varepsilon}$ such that
    \begin{equation}
        \frac{1}{4^n}\E\limits_{\rand{L}_1, \cdots, \rand{L}_k} Tr(\mathcal{P}' \mathcal{U}(\rand{L}_k\cdots \rand{L}_1)\mathcal{P} \mathcal{U}(\rand{L}_k \cdots \rand{L}_1)^{-1}) \leq \delta + \frac{1}{4^n},\label{eqn:scrambling}
    \end{equation}
    where $\mathcal{P}[\rho] = P\rho P$ and $\mathcal{P}'[\rho] = P'\rho P'$,$\rand{L_1}, \dots, \rand{L_k}$ are $\Omega$-distributed random layers, $\delta \ll 1$, and $\varepsilon$ is the expected infidelity of an $\Omega$-distributed random layer. While we require that $k \ll \nicefrac{1}{\varepsilon}$ for our theory, this condition on $k$ can be relaxed when $n \gg 1$ because errors that occur on spatially separated qubits cannot cancel even if they occur in sequential circuit layers. Note that Eq.~\eqref{eqn:scrambling} is not equivalent to requiring that a length $k$ sequence of $\Omega$-distributed layers is a good approximation to a unitary 2-design [because we do not require that $\delta=O(\nicefrac{1}{4^n})$].
    \item $\Omega_1$ is the uniform distribution over $\mathbb{G}_1$.
    \item $\Omega_2$ is invariant under exchanging any subset of the gates in a two-qubit gate layer $L$ with their inverses.
\end{enumerate}

The above conditions are sufficient to ensure our circuits are highly scrambling, but not necessary. In particular, our method can be generalized to single-qubit gate sets $\mathbb{G}_1$ that only \emph{generate} a unitary 2-design and to distributions $\Omega_1$ other than the uniform distribution. However, this complicates the analysis, so we do not consider this more general case herein.

\section{Scalable randomized benchmarking of universal gate sets}\label{sec:mrb}
In this section we introduce our method for MRB of universal gate sets. In Section~\ref{sec:rmc_construction} we introduce the family of randomized mirror circuits used in MRB. In Section~\ref{sec:mrb-protocol} we explain the MRB data analysis and define the complete MRB protocol.

\subsection{Randomized mirror circuits for universal gate sets}\label{sec:rmc_construction}

Our protocol uses a novel family of \emph{randomized mirror circuits} \cite{proctor2020measuring, proctor2021scalable, Mayer-df} that we now introduce. The structure of these randomized mirror circuits allows our protocol to measure $\epsilon_{\Omega}$, the average error rate of $n$-qubit layers sampled from $\Omega$ (see Section~\ref{sec:epsilon-def} for the precise definition of $\epsilon_{\Omega}$), without expensive classical computation. One approach to estimating $\epsilon_{\Omega}$ is to run $\Omega$-distributed random circuits of varied depths, and then estimate the decay in the (linear) cross entropy between these circuits' ideal and actual output probability distributions \cite{boixo2018characterizing, liu2021benchmarking}. This is because the decay rate of this cross entropy is known to be approximately equal to $\epsilon_{\Omega}$  \cite{boixo2018characterizing, liu2021benchmarking}. The problem with this method is that the classical computation cost of computing the ideal output probability distribution scales exponentially in the number of qubits ($n$) when the gate set is universal \cite{arute2019quantum}, limiting it to $n \lesssim 50$. To estimate $\epsilon_{\Omega}$ without expensive classical computation our protocol runs $\Omega$-\emph{distributed randomized mirror circuits}, which use an inversion structure to transform an $\Omega$-distributed random circuit into a circuit with an efficiently computable outcome.

\begin{figure}[t!]
\includegraphics{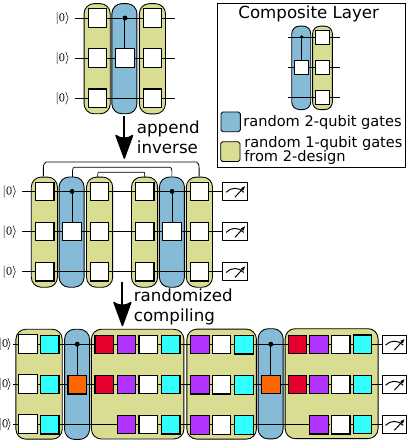} 
\caption{\textbf{Randomized mirror circuits over universal gate sets.} To construct a randomized mirror circuit of benchmark depth $d$ (and total depth $2d+2$) we first sample a random depth $d+1$ circuit. This circuit alternates between layers of randomly sampled one-qubit gates and layers of randomly sampled two-qubit gates. It can be thought of as consisting of a single initial layer of random one-qubit gates followed by $\nicefrac{d}{2}$ composite layers (see inset). We then append to this circuit its inverse, i.e., the circuit in reverse with each layer replaced with its inverse. This creates a depth $2d+2$ circuit that will, if run perfectly, always return the all zeros bit string. This circuit is susceptible to systematic addition or cancellation of errors between the two halves of the circuit. To prevent this unwanted effect we then apply randomized compiling to the circuit. We insert a layer of random single-qubit Pauli gates (cyan) after each one-qubit gate layer. In order to guarantee that this randomly compiled circuit still always, if run perfectly, returns a single bit string $s$, our procedure (1) changes the rotation angles in the two-qubit gates (orange) if these gates are not Clifford gates, (2) adds in single-qubit Pauli axis rotations following the two-qubit gates (red) and, (3) adds in correction Pauli gates (purple) prior to each single-qubit gate layer. The yellow boxes show gates that are compiled together to create the final circuit of depth $2d+2$. This circuit contains $d$ composite layers, which we call its benchmark depth.}
\label{fig:rmcs}
\end{figure}

We construct a specific randomized mirror circuit on $n$ qubits with benchmark depth $d$ via the three-step procedure shown in Fig.~\ref{fig:rmcs}. This procedure consists of first sampling a circuit $\gate{C}_1$ consisting of an $\Omega$-distributed random circuit preceded by an initial layer of random single-qubit gates that randomizes the state input into the circuit (enabling estimation of the circuit's fidelity using the method of Ref.~\cite{proctor2022establishing}). We then append the inverse of $C_1$ to obtain $\gate{C}_2$, a simple form of mirror (or motion-reversal) circuit which, if run perfectly, always outputs a single bit string. Finally, $\gate{C}_2$ is randomly compiled, to prevent systematic coherent addition or cancellation of errors between the $\Omega$-distributed random circuit and its inverse---which is essential for reliable estimation of $\epsilon_{\Omega}$. The exact procedure is as follows:
\newcounter{itemcounter}
\begin{list}
{\arabic{itemcounter}.}
{\usecounter{itemcounter}\leftmargin=1.4em}

	\item\label{forward_ckt_construction} (Sample a random circuit) Construct a circuit $C_1 =\gate{L}_{\nicefrac{d}{2}}\gate{L}_{\theta_{\nicefrac{d}{2}}} \cdots \gate{L_1L_{\theta_1}L_0}$ consisting of:
	
\begin{enumerate}
\item[(a)]\label{first_layer} A layer $\gate{L_0}$ sampled from $\Omega_1$, which consists of a single-qubit gate on each qubit.

\item[(b)]\label{forward_ckt} $\nicefrac{d}{2}$ composite layers $\gate{L_iL_{\theta_i}}$, where $\gate{L_i}$ is sampled from $\Omega_1$, and $\gate{L_{\theta_i}}$ is sampled from $\Omega_2$.
\end{enumerate}

\item\label{simple_mirror_construction} (Construct simple mirror circuit) Add to the circuit $C_1$ the layers in step \ref{forward_ckt} in reverse order, with each layer replaced with its inverse. The result is a circuit 
\begin{equation}
    \gate{C}_2 = \gate{L_{0}^{-1}L_{\theta_1}^{-1}L_1^{-1}}\cdots \gate{L_{\theta_{\nicefrac{d}{2}}}^{-1}L_{\nicefrac{d}{2}}^{-1} L_{\nicefrac{d}{2}}L_{\theta_{\nicefrac{d}{2}}} } \cdots \gate{L_1L_{\theta_1}L_0}, \label{eq:simple-mirror-circ}
\end{equation}
such that $U(\gate{C}_2)=\mathbb{I}$.

\item\label{rc} (Randomized compiling) Construct a new circuit $M$ by starting with $C_2$ and replacing layers using the following randomized compilation procedure, which reduces to standard Pauli frame randomization \cite{wallman2015noise} when the two-qubit gates are all Clifford gates. To specify our procedure, we first write $C_2$ [Eq.~\eqref{eq:simple-mirror-circ}] in the form 
\begin{equation*}
    \gate{C}_2 = \gate{L_{d+1}L_{\theta_{d+1}}L_d}\cdots \gate{L_{\theta_{\nicefrac{d}{2}+2}}L_{\nicefrac{d}{2}+1}L_{\theta_{\nicefrac{d}{2}+1}} L_{\nicefrac{d}{2}}L_{\theta_{\nicefrac{d}{2}}} } \cdots \gate{L_1L_{\theta_1}L_0},
\end{equation*} where $\gate{L}_{\theta_{\nicefrac{d}{2}+1}}$ is a dummy (empty) 2-qubit gate layer, so that $\gate{C}_2$ consists of alternating layers of one- and two-qubit gates. Then:

\begin{enumerate}
\item[(a)] For each single-qubit gate layer $\gate{L}_i$ in $\gate{C}_2$, sample a uniformly random layer of Pauli gates $\gate{P}_i$, that in the following procedure is inserted after and then compiled into $\gate{L}_i$.

\item[(b)] Replace each two-qubit gate layer $\gate{L_{\theta_i}}$ in $\gate{C}_2$ with a new two-qubit gate layer $\tcp{\gate{L}_{\theta_i}}{\gate{P}_{i-1}}$ that is constructed as follows: For each gate $\gate{CP_{\theta}}$ in $\gate{L_{\theta_i}}$ with control qubit $q_j$ and target qubit $q_k$, consider the instructions in $\gate{P_{i-1}}$ acting on $q_j$ and  $q_k$, denoted by $\gate{P_{i-1, j}}$ and $\gate{P_{i-1, k}}$, respectively. If $U(P_{{i-1}, j}) = I$ or $Z$, then add $\gate{CP_{\phi}}$ acting on $(q_j, q_k)$ to $\tcp{\gate{L}_{\theta_i}}{\gate{P}_{i-1}}$ where $\phi = \theta$ if $[U(P), U(\gate{P_{i-1, k}})] = 0$ and $\phi = -\theta$ otherwise. If $U(P_{{i-1}, j}) = X$ or $Y$, then add $\gate{CP_{\phi}}$ acting on $(q_j, q_k)$ to $\tcp{\gate{L}_{\theta_i}}{\gate{P}_{i-1}}$ where $\phi = -\theta$ if $[U(P), U(\gate{P_{i-1, k}})] = 0$ and $\phi = \theta$ otherwise. 

\item[(c)] For each single-qubit gate layer $\gate{L}_i$ in $\gate{C}_2$ with $i>0$, we define a layer of single-qubit gates $\gate{P}_{i-1}^c$ that undoes the effect of adding $\gate{P}_{i-1}$ into the circuit---meaning a layer such that $U(\gate{P}_{i-1}^c\tcp{\gate{L}_{\theta_{i}}}{\gate{P}_{i-1}}\gate{P}_{i-1}) =  U(\gate{L}_{{\theta}_i})$. Because $\mathbb{G}_2$ is restricted to only controlled Pauli-axis rotations, the correction takes the form $U(\gate{P}_{i-1}^c) = U(\gate{P_{i-1}}\gate{P_{\tilde{\theta}_i}})$, where $\gate{P}_{\tilde{\theta}_i}$ consists of single-qubit Pauli axis rotations. If $\gate{L}_i$ is not immediately preceded by a two-qubit gate layer, then $\gate{P}_{{\tilde{\theta}}_i} = \mathbb{I}$. Otherwise,
\begin{equation}\label{eqn:2q_transformation}
    U(\gate{P}_{\tilde{\theta}_i}) = U\bigl(\gate{P}_{i-1}\gate{L}_{{\theta}_i}\gate{P}_{i-1}\tcp{\gate{L}_{\theta_i}}{\gate{P}_{i-1}}^{-1}\bigr).
\end{equation}
 
\item[(d)] Replace each single-qubit gate layer $\gate{L}_i$ in $\gate{C}_2$ with a recompiled layer $\rc{\gate{L}_i}{\gate{P}_i}{ \gate{P}^c_{i-1}}$, defined by
\begin{equation}\label{eqn:1q_transformation}
    U\bigl(\rc{\gate{L}_i}{\gate{P}_i}{ \gate{P}^c_{i-1}}\bigr) = U(\gate{P_i L_i P^c_{i-1}}).
\end{equation}
\end{enumerate}
This randomized compilation step transforms the layer pair $\gate{L_iL_{{\theta}_i}}$ into $\rc{\gate{L}_i}{\gate{P}_i}{ \gate{P}^c_{i-1}}\tcp{\gate{L}_{\theta_i}}{\gate{P}_{i-1}}$, where 
\begin{equation}
    U\bigl(\rc{\gate{L}_i}{\gate{P}_i}{ \gate{P}^c_{i-1}}\tcp{\gate{L}_{\theta_i}}{\gate{P}_{i-1}}\bigr) = U(\gate{P_i L_i L_{{\theta}_i} P_{i-1}}).
\end{equation} 
The final circuit produced by this procedure ($\gate{M}$) has the property that $U(\gate{M})=U(\gate{P}_{d+1})$, i.e., its overall action is an $n$-qubit Pauli operator. So, if run perfectly, $\gate{M}$ returns a single bit string ($s_M$) that is determined during circuit construction with no additional computation needed.
\end{list}
The final depth-$d$ randomized mirror circuit has the form
\begin{equation}
    \gate{M}  = \rc{\gate{L}_0^{-1}}{ \gate{P}_{d+1}}{ \gate{P}_d^c} \, \gate{\tilde{M}} \, \rc{\gate{L}_0}{\gate{P}_{0}}{}, \label{eqn:mc}
\end{equation}
where
\begin{multline}
\gate{\tilde{M}}  = \tcp{\gate{L}_{\theta_1}^{-1}}{\gate{P}_d}\rc{\gate{L}_1^{-1}}{\gate{P}_{d}}{ \gate{P}_{d-1}^c} \cdots \rc{\gate{L}_{\nicefrac{d}{2}}^{-1}}{ \gate{P}_{\nicefrac{d}{2}+1}}{ \gate{P}_{\nicefrac{d}{2}}} \nonumber \\
\rc{\gate{L}_{\nicefrac{d}{2}}}{ \gate{P}_{\nicefrac{d}{2}}}{ \gate{P}^c_{\nicefrac{d}{2}-1}}\cdots \rc{\gate{L}_1}{ \gate{P}_{1}}{ \gate{P}_0^c}\tcp{\gate{L}_{\theta_1}}{\gate{P}_0},
\end{multline}
is the circuit obtained after applying randomized compilation to the $\nicefrac{d}{2}$ composite layers sampled from $\Omega$ and their inverses.

\subsection{RB with non-Clifford randomized mirror circuits}\label{sec:mrb-protocol}
We now introduce our protocol---MRB for universal gate sets.
Our protocol has the same general structure as standard RB \cite{magesan2011scalable} and many of its variants: an exponential decay is fit to data from random circuits. However, our data analysis method is different from standard RB. We use the same analysis technique as MRB of \emph{Clifford} gate sets \cite{proctor2021scalable}. In particular, for each $n$-qubit circuit $\gate{C}$ that we run, we estimate its \emph{observed polarization} \cite{proctor2021scalable} 
\begin{equation}
S =  \frac{4^n}{4^n-1}\left[\sum_{k=0}^{n} \left(-\frac{1}{2}\right)^k h_k\right] - \frac{1}{4^n -1},
\label{eq:S}
\end{equation}
where $h_k$ is the probability that the circuit outputs a bit string with Hamming distance $k$ from its target bit string ($s_C$). As shown in Ref.~\cite{proctor2021scalable} and discussed further below, the simple additional analysis in computing $S$ simulates an $n$-qubit 2-design twirl using only local state preparation and measurement.

A specific MRB experiment is defined by a gate set $\mathbb{G}$, a sampling distribution $\Omega$, and the usual RB sampling parameters (a set of benchmark depths $d$, the number of circuits $K$ sampled per depth, and the number of times $N$ each circuit is run). Our protocol is the following:
\begin{enumerate}
\item For a range of integers $d \geq 0$, sample $K$ randomized mirror circuits that have a benchmark depth of $d$, using the sampling distribution $\Omega$, and run each one $N\geq 1$ times.
\item Estimate each circuit's observed polarization $S$.
\item Fit $\bar{S}_d$, the mean of $S$ at benchmark depth $d$, to 
\begin{equation}
\bar{S}_d = Ap^d,
\label{eq:decay}
\end{equation}
where $A$ and $p$ are fit parameters, and then compute 
\begin{equation}
r_{\Omega} = (4^n - 1)(1 - p)/4^n. \label{eq:r}
\end{equation}
\end{enumerate}

\begin{figure*}
    \centering
    \includegraphics{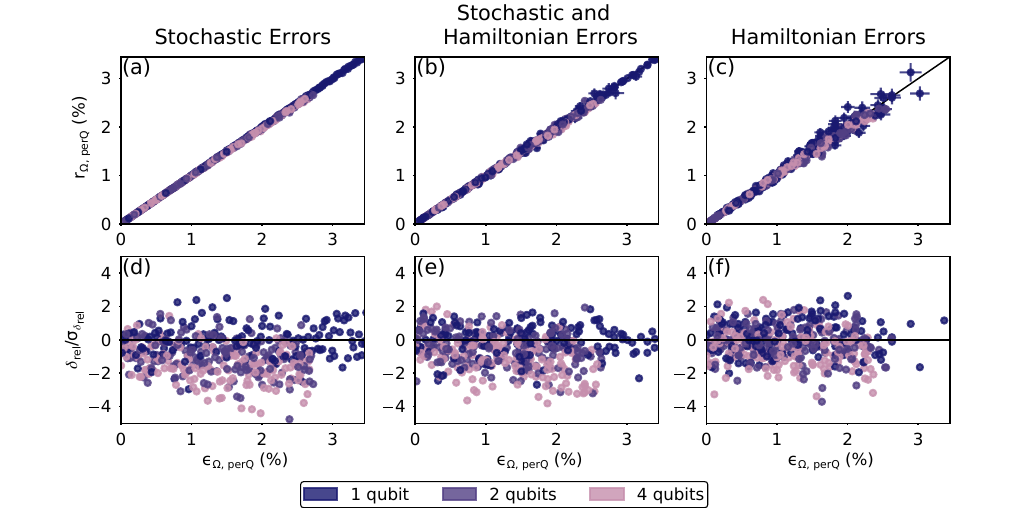}
    \caption{\textbf{Investigating the reliability of MRB using simulations.} We simulated MRB on $n$ all-to-all-connected qubits for $n=1,2,4$ on the gate set $(\mathbb{G}_1, \mathbb{G}_2) = (\mathbb{SU}(2), \{\cs, \csd\})$ with randomly-sampled gate-dependent errors. From left to right, the columns show results from simulations with crosstalk-free error models consisting of only stochastic errors (a,d), a combination of stochastic and Hamiltonian errors (b,e), and only Hamiltonian errors (c,f) (see Section~\ref{sec:theory-nc-2Q-gates} for details). (a-c): The MRB error rate per qubit [$r_{\Omega, \,\textrm{perQ}} = 1-(1-r_{\Omega})^{\nicefrac{1}{n}}$] versus the average composite layer error rate per qubit [$\epsilon_{\Omega, \,\textrm{perQ}} = 1-(1-\epsilon_{\Omega})^{\nicefrac{1}{n}}$] for each randomly sampled error model. The MRB error rate $r_{\Omega} $ closely approximates $\epsilon_{\Omega}$, and the agreement is closest under purely stochastic errors. (d-f): The relative error $\delta_{\textrm{rel}} = \nicefrac{(r_{\Omega, \,\textrm{perQ}} - \epsilon_{\Omega, \,\textrm{perQ}})}{\epsilon_{\Omega, \,\textrm{perQ}}}$ divided by its uncertainty $\sigma_{\delta_{\textrm{rel}}}$ for each randomly sampled error model ($\sigma_{\delta_{\textrm{rel}}}$ is calculated via a standard non-parametric bootstrap). The MRB error rate $r_{\Omega}$ is biased towards very slightly underestimating $\epsilon_{\Omega}$ for $n>2$ qubits, which is expected from our theory (see main text).}
    \label{fig:simulations}
\end{figure*}

\section{Theory and Simulations of MRB on Universal Gate Sets}
\label{sec:theory}
\label{sec:analytics}
In this section we present a theory for MRB of universal gate sets that shows that our method is reliable. We show that the average observed polarization ($\bar{S}_{d}$) decays exponentially, and that the MRB error rate ($r_{\Omega}$) approximately equals the average error rate of $\Omega$-distributed layers ($\epsilon_{\Omega}$). In Section \ref{sec:epsilon-def} we define $\epsilon_{\Omega}$, the error rate that MRB is designed to measure. In Section \ref{sec:r_eps_relation} we show that $r_{\Omega} \approx \epsilon_{\Omega}$ assuming Pauli stochastic error on each circuit layer. In Sections \ref{sec:general_errors_theory} and \ref{sec:theory-nc-2Q-gates} we present theory and simulations of the performance of MRB under general Markovian errors to further validate our method. In particular, we show that the randomized compilation step of our circuit construction guarantees that all errors in the circuit are twirled into Pauli stochastic error (implying that $r_{\Omega} \approx \epsilon_{\Omega}$) under the assumption that all two-qubit gates are Clifford gates. 

\subsection{The error rate of $\Omega$-distributed random circuits}\label{sec:epsilon-def}

Our claim is that $r_{\Omega}$ is a reliable estimate of the average error rate $\epsilon_{\Omega}$ of $\Omega$-distributed $n$-qubit circuit layers. We now make this claim precise by defining $\epsilon_{\Omega}$. Surprisingly, defining the error rate that our method (or any other RB method) should aim to estimate is challenging. RB protocols are often formulated as methods for measuring the mean infidelity of a set of $n$-qubit gates or layers, but this is subtly flawed: the mean infidelity is not an observable property of a set of physical gates---it is not ``gauge-invariant'' \cite{proctor2017randomized}. One solution to this problem, which we adopt herein, was introduced in Ref.~\cite{Carignan_Dugas_2018}: the rate of decay of the mean fidelity of a family of random circuits, as a function of increasing circuit depth, is (approximately) gauge-invariant. This decay rate can therefore be what an RB protocol aims to measure.

Our protocol aims to estimate the rate at which the fidelity of $\Omega$-distributed random circuits decays with depth. The average fidelity of $\Omega$-distributed random circuits with benchmark depth $d$ ($\bar{F}_d$) is given by
\begin{equation}
    \bar{F}_d = \E\limits_{\rand{C}_d}
    F\bigl(\mathcal{U}(\rand{C}_d),\phi(\rand{C}_d)\bigr). \label{eq:f_d_def}
\end{equation}
The requirement that our $\Omega$-distributed circuits are highly scrambling, which is guaranteed by our restrictions on $\mathbb{G}$ and $\Omega$ (see Section \ref{sec:gate-set}), ensures that $\bar{F}_d$ decays exponentially, and therefore has a well-defined rate of decay.  In Section \ref{sec:analytics} we show that $\bar{F}_d$ decays exponentially in depth for our circuits, i.e., $\bar{F}_d \approx Ap_{\textrm{rc}}^d+B$, for constants $A$ and $B$. We then define 
\begin{equation}
\epsilon_{\Omega} = (4^n - 1)(1 - p_{\textrm{rc}})/4^n. \label{eq:epsilon}
\end{equation}
We choose $\epsilon_{\Omega}$ to be this particular rescaling of $p_{\textrm{rc}}$ because $p_{\textrm {rc}}$ corresponds to the effective polarization of a random composite layer in an $\Omega$-distributed random circuit---i.e., the polarization in a depolarizing channel that would give the same fidelity decay---and so $\epsilon_{\Omega}$ is the effective average infidelity of a layer sampled from $\Omega$. When stochastic Pauli errors are the dominant source of error, $\epsilon_{\Omega}$ is approximately equal to the average layer entanglement infidelity (see Appendix \ref{app:exp_decay}). 

\subsection{MRB with stochastic Pauli errors}
\label{sec:r_eps_relation}

We now show that $r_{\Omega} \approx \epsilon_{\Omega}$ under the assumption of stochastic Pauli errors on each circuit layer. A more detailed derivation can be found in Appendix~\ref{app:r_eps_relation}. Throughout this section, we will treat circuits and circuit layers as random variables. We assume each circuit layer has gate-dependent Markovian error, $\phi(\rand{L}) = \mathcal{E}(\rand{L})\mathcal{U}(\rand{L})$. 
We will model the error on state preparation and measurement (SPAM) and the first and last circuit layers of a randomized mirror circuit [$\rc{\rand{L}_0}{\gate{P}_{0}}{}$ and $\rc{\rand{L}_0^{-1}}{\gate{P}_{d}}{\gate{P}_{d-1}^c}$, respectively] as a single global depolarizing error channel $\mathcal{E}_{\textrm{SPAM}}[\rho] = \gamma_{\textrm{SPAM}}\rho + (1-\gamma_{\textrm{SPAM}})\frac{\mathbb{I}}{2^n}$ occurring immediately before the final circuit layer. We assume $\mathcal{E}_{\textrm{SPAM}}$ is independent of $\rand{L}_0$ and the target bit string of the circuit. 

We start by showing that the mean observed polarization [Eq.~\eqref{eq:S}] of randomized mirror circuits, which is measured in the MRB protocol, equals the mean polarization of the overall error map of a randomized mirror circuit. An implementation of the depth-$d$ randomized mirror circuit $\rand{M}_d$ [whose structure is given in Eq.~\eqref{eqn:mc}] can be expressed in terms of its error and its target evolution $\mathcal{U}(\rand{P}_{d+1})$ as
\begin{align}
\phi(\rand{M}_d) = \mathcal{U}(\rand{P}_{d+1})\mathcal{U}(\rand{L}_0^{-1})\mathcal{E}_{\textrm{eff}}(\rand{M}_d)\mathcal{U}(\rand{L}_0), \label{eqn:effective_error}
\end{align}
where 
 \begin{align}
 \mathcal{E}_{\textrm{eff}}(\rand{M}_d) & = \mathcal{E}_{\textrm{SPAM}}\mathcal{E}_{\textrm{eff}}(\rand{\tilde{M}}_d) \label{eqn:error_channel_m}\\ 
 & =  \mathcal{E}_{\textrm{SPAM}}\mathcal{E}'_{\tcp{\rand{L}_{\theta_1}^{-1}}{\rand{P}_{d}}}\cdots \mathcal{E}'_{\rand{L}_{\nicefrac{d}{2}}^{-1}} \mathcal{E}'_{\rand{L}_{\nicefrac{d}{2}}}\cdots\mathcal{E}'_{\tcp{L_{\theta_1}}{\rand{P}_0}} \label{eqn:error_channel}
 \end{align} 
and 
\begin{equation}
    \mathcal{E}'_{\rand{L}_i} = \mathcal{U}(\rand{L}_1)^{-1} \cdots \mathcal{U}(\rand{L}_{i})^{-1}\mathcal{E}(\rand{L}_i)\mathcal{U}(\rand{L}_i)\cdots \mathcal{U}(\rand{L}_1). 
\end{equation}

Eq.~\eqref{eqn:error_channel_m} defines an overall error map for $\rand{M_d}$, which includes the error from the $\nicefrac{d}{2}$ $\Omega$-distributed circuit layers and their inverses (after randomized compilation). To extract the polarization [Eq.~\eqref{eq:pol_def}] of this error map, we average over the initial circuit layer $\rand{L}_0$, making use of a fidelity estimation technique that requires only single-qubit gates: the fidelity of any error channel $\mathcal{E}$ can be found by averaging over a tensor product of single-qubit 2-designs \cite{proctor2022establishing}. In particular, for any bit string $y \in \{0,1\}^n$, 
\begin{eqnarray}
    \gamma(\mathcal{E}) = \frac{4^n}{4^n-1}\smashoperator[lr]{\sum_{x\in \{0,1\}^n}}\left(-\nicefrac{1}{2}\right)^{h(x,y)} \braa{x+y} \bar{\mathcal{E}} \kett{0} - \frac{1}{4^n-1}, \label{eq:2_design_avg}
\end{eqnarray}
where $\bar{\mathcal{E}} = \E_{\rand{L}}[\mathcal{U}(\rand{L})^{\dagger} \mathcal{E} \mathcal{U}(\rand{L})]$ and $\rand{L} = \otimes_{i=1}^n \rand{L}_i$, where each $\rand{L}_i$ is a independent, single-qubit 2-design \cite{proctor2022establishing}. Applying Eq.~\eqref{eq:2_design_avg} to Eq.~\eqref{eqn:effective_error}, we find that
\begin{align}
      \gamma\bigl(\mathcal{E}_{\textrm{eff}}(\rand{M}_d)\bigr) = \E\limits_{\rand{L}_0}S(\rand{M}_d) \label{eq:c_pol}
\end{align}
where $S(\rand{M}_d)$ denotes the observed polarization [Eq.~\eqref{eq:S}] of $\rand{M}_d$. Therefore, the mean observed polarization over all depth-$d$ randomized mirror circuits is
\begin{equation}
    \bar{S}_{d} = \E\limits_{\rand{M}_d} \gamma\bigl(\mathcal{E}_{\textrm{eff}}(\rand{M}_d)\bigr). \label{eqn:s_d_avg_e_eff}
\end{equation}
Equation~\eqref{eqn:s_d_avg_e_eff} says that the average observed polarization $ \bar{S}_{d}$, which is estimated in the MRB protocol, is equal to the expected polarization of the error channel of a depth-$d$ randomized mirror circuit.

We now show how $\bar{S}_d$ depends on the error rate of layers sampled from $\Omega$ ($\epsilon_{\Omega}$). To do so, we use the fact that a depth-$d$ randomized mirror circuit consists of randomized compilation of a circuit consisting of a depth-$\nicefrac{d}{2}$ $\Omega$-distributed random circuit $\rand{C}_{\nicefrac{d}{2}}$ followed by its inverse. These two depth-$\nicefrac{d}{2}$ circuits are both $\Omega$-distributed (even after randomized compilation), but they are correlated. In particular,
\begin{align}
    \bar{S}_{d} & = \gamma(\mathcal{E}_{\textrm{SPAM}})\E_{\rand{C}_{\nicefrac{d}{2}}} \gamma\bigl(\mathcal{U}(\rand{C}_{\nicefrac{d}{2}})\bar{\mathcal{E}}_{\textrm{eff}}(\rand{C}_{\nicefrac{d}{2}}^{-1})\mathcal{U}(\rand{C}_{\nicefrac{d}{2}})^{-1}\mathcal{E}_{\textrm{eff}}(\rand{C}_{\nicefrac{d}{2}})\bigr),  \label{eqn:s_d_avg_2}
\end{align}
where $\mathcal{E}_{\textrm{eff}}(\rand{C}_{\nicefrac{d}{2}})$ is the overall error map for $\rand{C}_{\nicefrac{d}{2}}$ [i.e., $\phi(\rand{C}_{\nicefrac{d}{2}}) = \mathcal{U}(\rand{C}_{\nicefrac{d}{2}})\mathcal{E}_{\textrm{eff}}(\rand{C}_{\nicefrac{d}{2}})$] and $\bar{\mathcal{E}}_{\textrm{eff}}(\rand{C}_{\nicefrac{d}{2}}^{-1})$ denotes the average error map over all possible circuits $\rand{C}'$ resulting from applying randomized compilation to $\rand{C}_{\nicefrac{d}{2}}^{-1}$. Expressing Eq.~\eqref{eqn:s_d_avg_2} in terms of the mean observed polarization of the overall error map on an $\Omega$-distributed random circuit, we have
\begin{equation}
    \bar{S}_{d}  = \gamma(\mathcal{E}_{\textrm{SPAM}})\left( \bar{\Gamma}_{\nicefrac{d}{2}}^2 - \Delta_{\Omega}\right), \label{eqn:s_relation}
\end{equation}
where 
\begin{equation}
    \bar{\Gamma}_{d} = \E_{\rand{C}_d} \gamma\bigl(\mathcal{E}_{\textrm{eff}}(\rand{C}_d)\bigr) \label{eqn:s_d_c}
\end{equation}
and 
\begin{multline}
    \Delta_{\Omega} = \E_{\rand{C}_{\nicefrac{d}{2}}} \gamma\bigl(\mathcal{U}(\rand{C}_{\nicefrac{d}{2}})\bar{\mathcal{E}}_{\textrm{eff}}(\rand{C}_{\nicefrac{d}{2}}^{-1})\mathcal{U}(\rand{C}_{\nicefrac{d}{2}})^{-1}\mathcal{E}_{\textrm{eff}}(\rand{C}_{\nicefrac{d}{2}})\bigr)\\
    - \Bigl(\E_{\rand{C}_{\nicefrac{d}{2}}}\gamma\bigl(\mathcal{E}_{\textrm{eff}}(\rand{C}_{\nicefrac{d}{2}})\bigr)\Bigr)^2. \label{eqn:delta_omega}
\end{multline}

Equation~\eqref{eqn:s_relation} shows that $\bar{S}_{d}  \approx \gamma(\mathcal{E}_{\textrm{SPAM}}) \bar{\Gamma}_{\nicefrac{d}{2}}^2$ if $|\Delta_{\Omega}|$ is small. If $\bar{S}_d$ and $\bar{\Gamma}_{\nicefrac{d}{2}}$ decay exponentially, Eq.~\eqref{eqn:s_relation} relates their decay rates---i.e., $r_{\Omega} = \epsilon_{\Omega}$ if $\Delta_\Omega=0$. $\Delta_{\Omega}$ quantifies the correlation between the overall error map of a depth-$\nicefrac{d}{2}$ $\Omega$-distributed random circuit and the overall error map of its randomly compiled inverse. We conjecture that $|\Delta_{\Omega}|$ is typically small for physically relevant errors, which is supported by our simulations (see Section~\ref{sec:theory-nc-2Q-gates}) and prior work \cite{proctor2021scalable}. 

We now show that the expected polarization of $\Omega$-distributed random circuits ($\bar{\Gamma}_d$) decays exponentially. Together with the assumption that $|\Delta_{\Omega}|$ is small, this implies that $\bar{S}_d$ decays exponentially. To show that $\bar{\Gamma}_d$ decays exponentially, we will assume that the error on each composite layer $\mathcal{E}(\rand{L})$ is a stochastic Pauli channel [Eq.~\eqref{eq:pauli_channel}]. This assumption implies that $\mathcal{E}_{\textrm{eff}}(\rand{C}_d)$ [Eq.~\eqref{eqn:error_channel}] is the composition of a stochastic Pauli channel for each composite layer of $\rand{\tilde{M}}_d$, each rotated by a unitary. This allows us to relate the polarization of $\mathcal{E}_{\textrm{eff}}(\rand{\tilde{M}}_d)$ to the polarizations of the error channels of individual circuit layers using the scrambling condition required for MRB [Eq.~\eqref{eqn:scrambling}].

Due to the scrambling condition on the gate set and sampling distribution for MRB [Eq.~\eqref{eqn:scrambling}], the polarization of the effective error channel of $\Omega$-distributed random circuits is approximately equal to the product of the polarizations of the layers' error channels. Specifically, the expected polarization of the overall error map is
\begin{align}
\bar{\Gamma}_{d} & =  \E\limits_{\rand{L}} \gamma\bigl(\mathcal{E}({\rand{L}})\bigr)^d + \tilde{\delta}, \label{eqn:gamma_to_prc}
\end{align} 
where $\tilde{\delta} = O\left(d\varepsilon(\delta+k\varepsilon)\right)$, and $\varepsilon$ is the average layer infidelity. Because circuits longer than $d = O(\nicefrac{1}{\varepsilon})$ have negligible polarization, we need only consider the case where $d\varepsilon=O(1)$. Because $k\varepsilon$ and $\delta$ are small, $\tilde{\delta}$ is negligible. In the small $n$ limit, Eq.~\eqref{eqn:gamma_to_prc} follows because Eq.~\eqref{eqn:scrambling} implies that depth-$k$ $\Omega$-distributed random circuits rapidly converge to a unitary 2-design (as a function of $k$). In this case, errors in $\Omega$-distributed random circuits are rapidly scrambled into global depolarizing errors, which implies that the polarizations of the circuit layers approximately multiply. For $n \gtrsim 3$, our circuits do not quickly converge to a 2-design, but in  Appendix~\ref{app:exp_decay} we show that Eq.~\eqref{eqn:scrambling} implies that error cancellation is negligible in $\Omega$-distributed random circuits, from which it follows that $\bar{\Gamma}_{d}$ decays exponentially at a rate determined by the expected layer polarization.

We have shown that  the expected polarization of the overall error map of $\Omega$-distributed random circuits decays exponentially, and we now relate its decay rate to the decay rate of the observed polarization of randomized mirror circuits, thereby relating $r_{\Omega}$ and $\epsilon_{\Omega}$. Combining Eq.~\eqref{eqn:gamma_to_prc} with Eq.~\eqref{eqn:s_relation}, we have
\begin{equation}
\bar{S}_{d}  = \gamma(\mathcal{E}_{\textrm{SPAM}})\left(\left(\E\limits_{\rand{L}} \gamma\bigl(\mathcal{E}({\rand{L}})\bigr)\right)^d  - \Delta_{\Omega}\right), \label{eqn:s_d_result}
\end{equation}
Assuming that $\Delta_{\Omega}$ is small, Eq.~\eqref{eqn:s_d_result} implies that $\bar{S}_d$ and $\bar{\Gamma}_d$ have approximately the same decay rate, which implies that $r_{\Omega} \approx \epsilon_{\Omega}$.

\subsection{MRB with general errors}
\label{sec:general_errors_theory}
The theory presented above (Section~\ref{sec:r_eps_relation}) shows that MRB is reliable whenever stochastic Pauli errors dominate over all other possible errors (e.g., coherent errors). In practice, stochastic error is not always dominant, which our protocol addresses with the randomized compilation step [see Fig.~\ref{fig:rmcs}]. The purpose of this step is to, upon averaging, convert all types of errors into stochastic Pauli errors \cite{wallman2015noise}---in which case the theory presented above can be used to infer that $r_{\Omega}\approx \epsilon_{\Omega}$. When MRB is implemented on a gate set in which all of the two-qubit gates are Clifford gates, this noise tailoring follows from standard randomized compilation theory \cite{wallman2015noise}. In Appendix \ref{app:theory-c-2Q-gates}, we show that with a Clifford two-qubit gate set, the error in MRB circuits is twirled into Pauli stochastic noise under the assumption that the error map on the single-qubit gates is independent of the Pauli gates with which they are compiled. In actual devices it is common for the single-qubit gate layers to have errors that are gate-dependent but much smaller than the two-qubit gate errors, in which case this result holds approximately \cite{wallman2015noise}.

Our MRB protocol can be applied to all controlled rotations around Pauli axes, i.e., all  $\gate{CP}_{\theta}$ gates. When the two-qubit gates are not all Clifford gates (i.e., when $\theta \neq 0, \pi$), the randomized compilation method used in our circuits is not equivalent to standard randomized compilation.  In this case, we cannot use standard randomized compilation theory to guarantee that all coherent errors on the two-qubit gates are twirled into stochastic Pauli errors. Ineffective twirling of coherent errors on two-qubit gates could result in coherent cancellation of the errors in a layer of two-qubit gates and its inversion layer in the second half of the mirror circuit (as happens in a simple mirror circuit, or standard Loschmidt echo \cite{proctor2020measuring}). In Appendix~\ref{app:theory-nc-2Q-gates} we prove that our randomized compilation method largely---but not entirely---prevents this error cancellation. We consider the sensitivity of our method to general Hamiltonian errors on each gate $g \in \mathbb{G}_2$. We model these errors by an error map $\mathcal{E}(g) = e^{M_{g}}$, where 
\begin{equation}
M_g = \sum\limits_{P_a,P_b}\varepsilon_{P_a,P_b}^{g}H_{P_a,P_b},
\end{equation}
and $H_{P_a,P_b}$ is the two-qubit Hamiltonian error generator indexed by the Pauli operators $P_a$ and $P_b$, as defined in Ref.~\cite{blume2021taxonomy}. We show that $r_{\Omega}$ depends on all Hamiltonian errors in $\gate{CP_{\theta}}$ gates \emph{except} one particular linear combination of the Hamiltonian errors on $\gate{CP}_{\theta}$ and $\gate{CP}_{-\theta}$ gates, when $\theta \neq 0,\pi$ (i.e., when $\gate{CP}_{\theta}$ is not a Clifford gate). In particular, $r_{\Omega}$ is insensitive (at first order) to $\varepsilon_{P,P}^{\gate{CP}_{\theta}}+\varepsilon_{P,P}^{\gate{CP}_{-\theta}}$ when $\theta \neq 0,\pi$. This is the sum of over- and under-rotation Hamiltonian errors in the $\gate{CP}_{\theta}$ gate and its inverse. In Appendix~\ref{app:adapations} we discuss how our technique could be adapted to remove this limitation. Note that if $\mathbb{G}_2 = \{\cs, \csd\}$, as is the case in our simulations (below) and some of our experiments (Section~\ref{sec:aqt}), then $r_{\Omega}$ is insensitive (at first order) to $\varepsilon_{Z,Z}^{\mathsf{cs}}+\varepsilon_{Z,Z}^{\mathsf{cs}^{\dag}}$. However, it is sensitive to all other linear combinations of the Hamiltonian errors on the $\mathsf{cs}$ and $\mathsf{cs}^{\dag}$ gates.

\subsection{Simulations}
\label{sec:theory-nc-2Q-gates}
We now use numerical simulations to investigate the robustness of MRB, studying whether the MRB error rate ($r_{\Omega}$) closely approximates the error rate of $\Omega$-distributed layers ($\epsilon_{\Omega}$). Our theory for MRB suggests that MRB is particularly robust when the two-qubit gates are Clifford gates and when all errors are stochastic Pauli errors. Therefore we simulated MRB with non-Clifford two-qubit gates and for both stochastic and coherent errors. We simulated MRB for $n$-qubit layer sets constructed from the gate set $\mathbb{G}_1 = \mathbb{SU}(2)$ and $\mathbb{G}_2 = \{\g{cs}, \g{cs}^{\dag}\}$ and $n=1,2,4$, with all-to-all connectivity. We used a sampling distribution $\Omega_2$ for which the two-qubit gate density is $\xi = \nicefrac{1}{2}$ \footnote{Here and throughout this paper we use the ``edge grab'' sampler, which is parameterized by the two-qubit gate density $\xi$, defined in the supplementary material of Ref.~\cite{proctor2020measuring}}. In these simulations (and our experiments) each single-qubit gate is decomposed into the following sequence of $\mathsf{x}_{\nicefrac{\pi}{2}}$ and $\mathsf{z}_{\theta}$ gates:
\begin{equation}\label{eqn:decomp}
\mathsf{u}(\theta,\phi,\lambda) = \mathsf{z}_{-\phi-\nicefrac{\pi}{2}}\,\mathsf{x}_{\nicefrac{\pi}{2}}\,\mathsf{z}_{\pi - 2\theta}\,\mathsf{x}_{\nicefrac{\pi}{2}}\,\mathsf{z}_{-\lambda+\nicefrac{\pi}{2}}.
\end{equation}
Here $\mathsf{x}_{\nicefrac{\pi}{2}}$ is a $\nicefrac{\pi}{2}$ rotation around the $X$ axis and $\g{z}_{\theta}$ is a rotation around the $Z$ axis by $\theta \in [0, 2\pi)$. Note that even when a shorter sequence of gates can implement the required unitary (e.g., $\mathsf{u}(0,0,0)$ implements the identity so it could be implemented with no gates) we always use this sequence of five gates. Therefore, the only difference between any two single-qubit gates is the angles of the $\g{z}_{\theta}$ gates.

We simulated three different families of error model: stochastic Pauli errors, Hamiltonian errors, and stochastic and Hamiltonian errors. These error models are specified using the error generator framework of Ref.~\cite{blume2021taxonomy}, and they consist of gate-dependent errors specified by randomly sampling error rates for each type of error and each gate. We simulated error models that are crosstalk free (note that our theory encompasses crosstalk errors) so each error model is specified by the rates of each type of local error on each gate. In particular, for an $m$-qubit gate we randomly sample $4^m-1$ stochastic error generators, or $4^m-1$ Hamiltonian error generators, or both, depending on the error model family. We sampled the error rates so that the infidelity of each two-qubit gate was approximately $q$, and the infidelity of each one-qubit gate was approximately $0.1q$, where $q$ is a parameter swept over a range of values (See Appendix \ref{app:simulations}). These error models have perfect state preparation and measurements. The effect of SPAM error on the polarization is approximately independent of benchmark depth, and therefore we expect MRB to be robust to SPAM error. In Appendix~\ref{app:simulations} we present simulations compare the MRB error rate in error models with perfect measurements to  error models with bit flip and amplitude damping measurement error. We find that these measurement errors do not significantly impact the resulting MRB error rate.

Figure \ref{fig:simulations} shows the results of our main simulations. It compares the true average layer error rate per qubit 
\begin{equation}
\epsilon_{\Omega, \,\textrm{perQ}} =  1-(1-\epsilon_{\Omega})^{\nicefrac{1}{n}} \approx \nicefrac{\epsilon_{\Omega}}{n}
\end{equation}
to the observed MRB error rate per qubit
\begin{equation}
    r_{\Omega, \,\textrm{perQ}} = 1-(1-r_{\Omega})^{\nicefrac{1}{n}} \approx \nicefrac{r_{\Omega}}{n}
\end{equation}in each simulation, separated into the three families of error model (1$\sigma$ error bars are shown, computed using a standard bootstrap). Figure \ref{fig:simulations}(a)-(c) shows that $r_{\Omega} \approx \epsilon_{\Omega}$ in every simulation, which means that our method closely approximates the error rate of $\Omega$-distributed layers for all of these error models.

For stochastic error models [Fig.~\ref{fig:simulations} (a)], the relative error $\delta_{\textrm{rel}} = (r_{\Omega, \,\textrm{perQ}} - \epsilon_{\Omega, \,\textrm{perQ}})/\epsilon_{\Omega, \,\textrm{perQ}}$ in the MRB estimate of $\epsilon_{\Omega, \textrm{perQ}}$ is small: $|\delta_{\textrm{rel}}| < 0.04$ and the mean $|\delta_{\textrm{rel}}|$ is 0.007 for all sampled error models. This is consistent with, and supports, our theory for MRB with stochastic errors. The relative error is larger for Hamiltonian error models---the mean relative error is 0.04 and $|\delta_{\textrm{rel}}| < 0.21$ for all error models. We expect larger relative error for some Hamiltonian error models, because MRB is insensitive to some Hamiltonian errors (see Section \ref{sec:general_errors_theory})---but note that the uncertainty due to finite sample fluctuations ($\sigma$) are larger in these simulations. For stochastic Pauli errors [Fig.~\ref{fig:simulations} (a)], the uncertainty in $r_{\Omega, \textrm{perQ}}$ is small, because there is little variation in the performance of circuits of the same depth (the mean uncertainty in $r_{\Omega, \textrm{perQ}}$ is $0.5\%$). For Hamiltonian errors [Fig.~\ref{fig:simulations} (c)], the uncertainty in $r_{\Omega, \textrm{perQ}}$ is larger (the mean uncertainty is $3\%$), as individual circuit performance varies widely due to coherent addition or cancellation of error being highly dependent on the circuit structure (as in all RB methods, we expect coherent errors to add or cancel in individual MRB circuits).

Arguably the most relevant simulations for real-world quantum computers are those with both stochastic and coherent errors [Fig.~\ref{fig:simulations} (b)]. In these simulations we sampled random combinations of stochastic and Hamiltonian errors (so the dominant source of error varies across these models). We find that $r_{\Omega} \approx \epsilon_{\Omega}$ holds to a good approximation for typical error models sampled from this ensemble (the mean relative error is $0.017$, and $|\delta_{\textrm{rel}}| < 0.11$ for all models, and the mean uncertainty in $r_{\Omega, \textrm{perQ}}$ is $1.4\%$).

To investigate whether there is evidence for $r_{\Omega}$ systematically under (or over) estimating $\epsilon_{\Omega}$ we plot the relative error divided by its uncertainty $\sigma_{\delta_{\textrm{rel}}}$ [Fig.~\ref{fig:simulations} (d-f)]. For $n=1$ qubit, there is no evidence that MRB is significantly biased towards under or overestimating $\epsilon_{\Omega}$ with these error models. In contrast, we find that MRB slightly but systematically underestimates $\epsilon_{\Omega}$ for $n >1$ qubits. This underestimate can be explained by the correlation between the error in an $\Omega$-distributed circuit and its randomly-compiled inverse, which determines the difference between $r_{\Omega}$ and $\epsilon_{\Omega}$ (see Section~\ref{sec:r_eps_relation}). When the circuits contain two-qubit gates---which in our simulations (and in most real systems) have higher error rates than one-qubit gates---the error in a circuit is typically highly correlated with the number of two-qubit gates in the circuit. As a result, the correlation between a circuit and its randomly-compiled inverse is typically larger when the circuits contain a variable number of two-qubit gates, causing $r_{\Omega}$ to \emph{slightly} underestimate $\epsilon_{\Omega}$.

\section{Experiments on the Advanced Quantum Testbed}\label{sec:aqt}
\begin{figure}
    \centering
    \includegraphics[width=6.5cm]{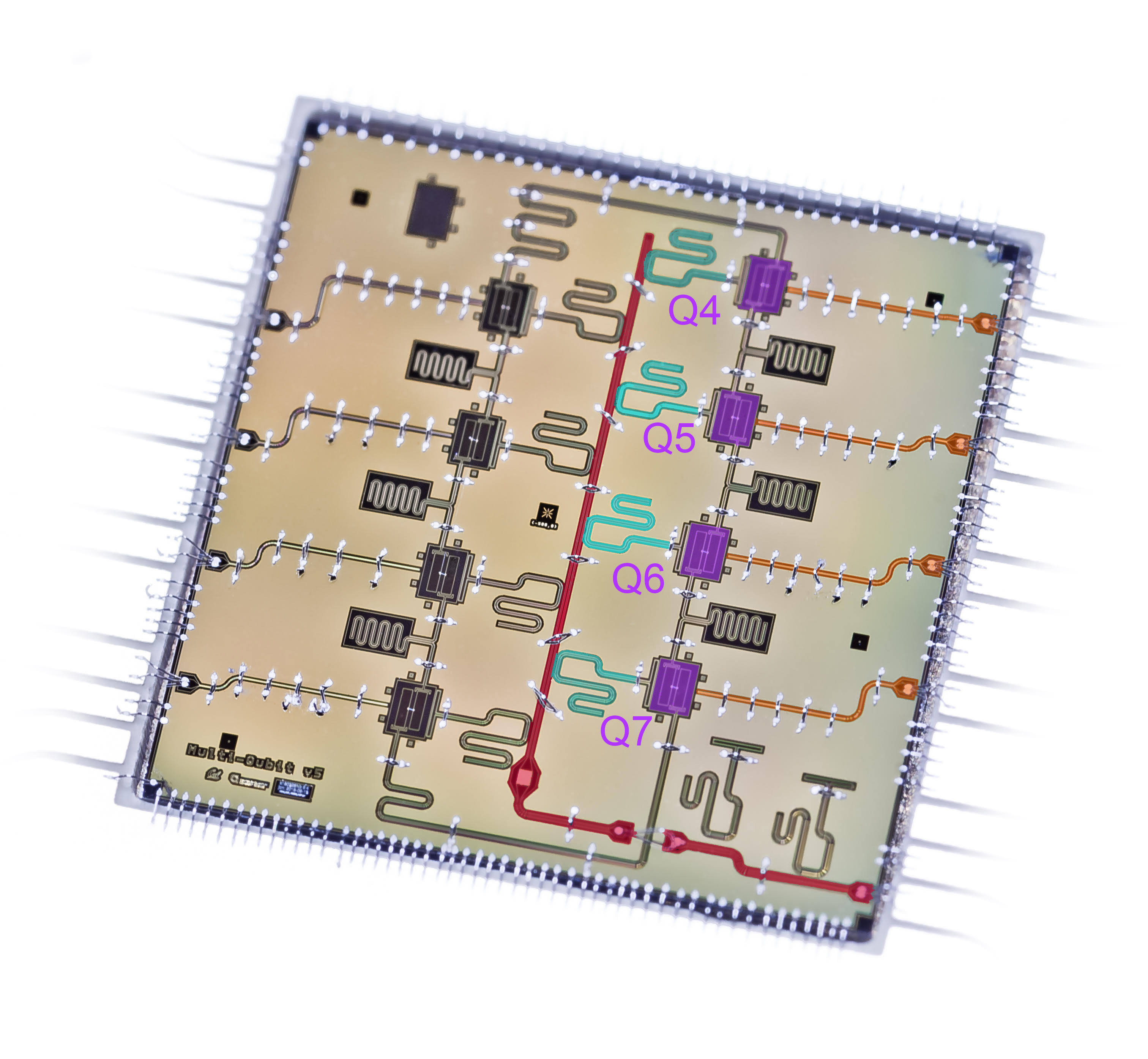}
    \caption{\textbf{The Advanced Quantum Testbed.} We performed MRB experiments on four qubits (\texttt{Q4}-\texttt{Q7}) of AQT's eight-qubit superconducting transmon processor (\texttt{AQT@LBNL Trailblazer8-v5.c2}). The processor includes 8 fixed frequency transmons coupled in a ring geometry. Each qubit (purple) has its own control line (orange) and readout resonator (cyan) coupled to a shared readout bus (red) for multiplexed readout.}
    \label{fig:aqt_chip}
\end{figure}

We used MRB to benchmark universal gate sets on the Advanced Quantum Testbed (AQT) \cite{AQT-wp}, a quantum computing testbed platform based on superconducting qubits. We performed our experiments on four qubits (\texttt{Q4}-\texttt{Q7}) of an eight-qubit superconducting transmon processor (\texttt{AQT@LBNL Trailblazer8-v5.c2}). These four qubits are coupled to their nearest neighbors in a linear geometry (see Fig.~\ref{fig:aqt_chip}). Below and throughout this paper, estimated quantities include error bars where possible \footnote{All uncertainties are calculated using a standard bootstrap.}. All error bars are $1\sigma$ and are written using standard concise notation, i.e., $r=1.2(3)\%$ means $r=1.2\%$ with a standard error of $0.3\%$. 

\begin{figure*}[t!]
\centering
\includegraphics{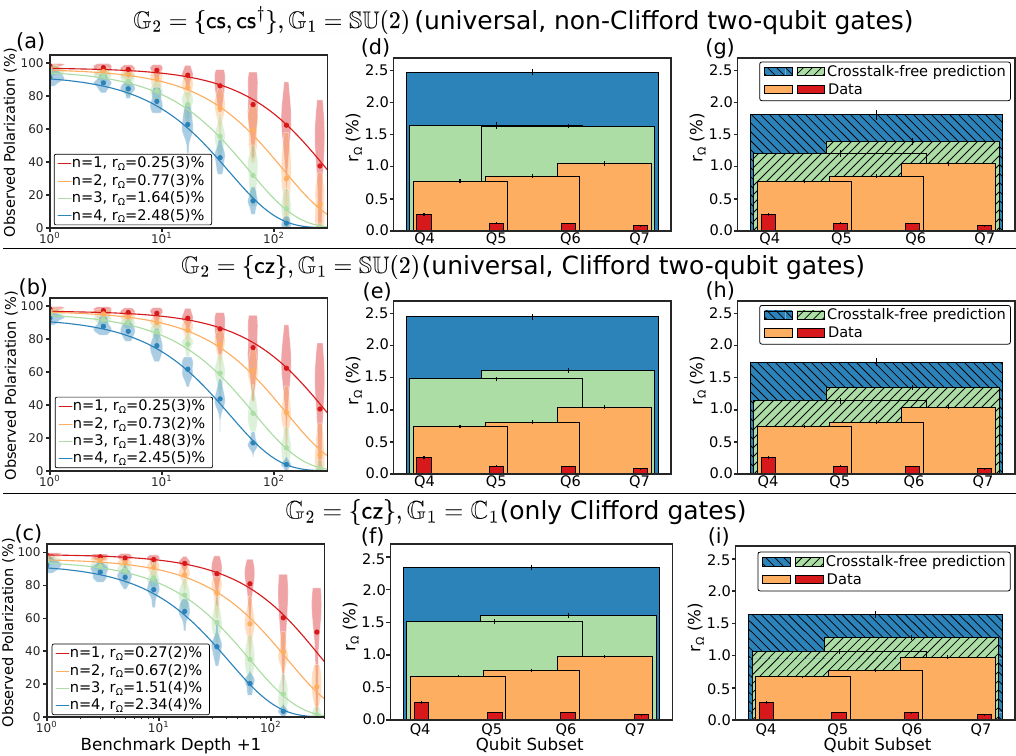} 
\caption{\textbf{Randomized benchmarking of universal gate sets on four qubits of the Advanced Quantum Testbed.} We used MRB to benchmark $n$-qubit layers constructed from three different gate sets, on each connected $n$-qubit subset of a linearly-connected set of four qubits $\{\Q{4},\Q{5},\Q{6},\Q{7}\}$ in an eight-qubit superconducting transmon processor (\texttt{AQT@LBNL Trailblazer8-v5.c2}). The rows correspond to results from three different choices of gate set, each consisting of a two-qubit gate set $\mathbb{G}_2$ and a single-qubit gate set $\mathbb{G}_1$. From top to bottom, the rows correspond to: a universal gate set containing two non-Clifford entangling gates and the set of all single-qubit gates [$\mathbb{G}_2=\{\cs,\csd\}$, $\mathbb{G}_1=\mathbb{SU}(2)$]; a universal gate set containing a Clifford entangling gate and the set of all single-qubit gates [$\mathbb{G}_2=\{\cz\}$, $\mathbb{G}_1=\mathbb{SU}(2)$]; and a non-universal, Clifford gate set [$\mathbb{G}_2=\{\cz\}$, $\mathbb{G}_1=\mathbb{C}_1$ where $\mathbb{C}_1$ is the one-qubit Clifford group]. (a-c): MRB decays for the qubit subsets $\{\Q4\}$, $\{\Q4,\Q5\}$, $\{\Q4,\Q5, \Q6\}$, and $\{\Q4,\Q5,\Q6,\Q7\}$. Violin plots and points show the distribution and mean, respectively, of the MRB circuit's observed polarization ($S_d$) versus benchmark depth ($d$). The curve is a fit of the mean of $S_d$ ($\bar{S}_d$) to $\bar{S}_d = Ap^d$. The average error rate of an $n$-qubit layer ($r_{\Omega}$) is given by $r_{\Omega}=(4^n - 1)(1 - p)/4^n$. The observed $\bar{S}_d$ decays exponentially, as predicted by our theory for MRB. (d-f): The estimated error rate $r_{\Omega}$ for each qubit subset that we benchmarked. (g-i): Predictions for the average layer error rate of 3- and 4-qubit subsets (hatched) based on the experimental 1- and 2-qubit error rates (un-hatched) and the assumption of no crosstalk errors. The difference between (d-f) and (g-i) quantifies the contribution of crosstalk errors to the average error rate of an $n$-qubit layer, for $n=3,4$. For all three gate sets and $n=4$, we see that crosstalk errors are contributing approximately $0.7\%$ error to $r_{\Omega}$, which is approximately $\nicefrac{1}{3}$ of $r_{\Omega}$.}
\label{fig:aqt_main}
\end{figure*}

\subsection{Experiment design}
One of the advantages of MRB is that it can benchmark a wide variety of $n$-qubit layer sets, and we used this flexibility to explore the performance of three distinct layer sets on AQT. Each layer set is defined by a set of single-qubit gates $\mathbb{G}_1$, a set of two-qubit gates $\mathbb{G}_2$, a two-qubit gate density $\xi$, and the connectivity of the qubit subset (see Section~\ref{sec:prelim}). In our experiments we investigated three different choices for $(\mathbb{G}_1,\mathbb{G}_2)$: $(\mathbb{SU}(2), \{\cs, \csd \})$, $(\mathbb{SU}(2), \{\cz \})$, and $(\mathbb{C}_1, \{\cz \})$, where $\mathbb{C}_1$ is the set of all 24 single-qubit Clifford gates. These circuits contain strict barriers between all layers, including between the single- and two-qubit gate layers that make up each composite layer.

MRB enables benchmarking each layer set on any connected set of qubits, and the error rates on subsets of a device can be used to learn about the location and type of errors. We benchmarked $n$-qubit layer sets for every possible connected set $\mathbb{Q} \subseteq \{\Q4, \Q5, \Q6, \Q7 \}$ of $n$ qubits with $n=1,2,3,4$, resulting in 10 different qubit subsets.  Independently benchmarking every connected subset of qubits allows us to study the spatial variation in gate performance in detail and determine the size of crosstalk error in circuits with $3$ and $4$ qubits (see Section~\ref{sec:aqt-crosstalk}). For each RB experiment, we sampled $K=30$ circuits at a set of exponentially-spaced benchmarking depths ($d = 0,2,4,8\dots$).

For each of the three gate sets $(\mathbb{G}_1,\mathbb{G}_2)$, and each qubit subset $\mathbb{Q}$, we ran experiments with a two-qubit gate density of $\xi=\nicefrac{1}{2}$. To investigate the effect of varying $\xi$, we also ran experiments with $\xi=\nicefrac{1}{8}$ for one of the gate sets---$( \mathbb{SU}(2), \{\cs, \csd \})$---and every $\mathbb{Q}$. For each qubit subset we therefore ran 4 MRB experiments, defined by \footnote{For the four one-qubit subsets, three of the cases coincide---as they differ only by the two-qubit gate set or the two-qubit gate density, which are unused parameters in one-qubit circuits. In that case we only sample and run only one of the three identical MRB designs.}:
\begin{enumerate}
    \item $\mathbb{G}_1 = \mathbb{SU}(2)$, $\mathbb{G}_2 = \{\cs, \csd \}$, and $\xi=\nicefrac{1}{8}$.
    \item $\mathbb{G}_1 = \mathbb{SU}(2)$, $\mathbb{G}_2 = \{\cs, \csd \}$, and $\xi=\nicefrac{1}{2}$.
    \item $\mathbb{G}_1 = \mathbb{SU}(2)$, $\mathbb{G}_2 = \{\cz \}$, and $\xi=\nicefrac{1}{2}$.
    \item $\mathbb{G}_1 = \mathbb{C}_1$, $\mathbb{G}_2 = \{\cz \}$, and $\xi=\nicefrac{1}{2}$.
\end{enumerate}
Further experiment details are provided in Appendix~\ref{app:aqt}.

\subsection{Estimating average error rates of universal layer sets}\label{sec:aqt_results}
Figure \ref{fig:aqt_main} summarizes the results of the $3 \times 10$ MRB experiments in which we vary the gate set $(\mathbb{G}_1, \mathbb{G}_2)$---corresponding to each row of Fig.~\ref{fig:aqt_main}---and the subset of qubits benchmarked $\mathbb{Q}$, but we keep the expected two-qubit gate density constant ($\xi = \nicefrac{1}{2}$). The main output of an MRB experiment is an average layer error rate ($r_{\Omega}$), obtained by fitting the mean observed polarization [$\bar{S}_d$, defined in Eq.~\eqref{eq:S}] to an exponential decay. This error rate is a function of $(\mathbb{G}_1, \mathbb{G}_2, \mathbb{Q}, \xi)$, so we denote our estimated error rates by $r(\mathbb{G}_1, \mathbb{G}_2, \mathbb{Q}, \xi)$ whenever we need to refer to a particular error rate. These error rates quantify the performance of random circuits on this device and enable us to compare the average performance of the gate sets we tested.

Figure~\ref{fig:aqt_main} (a-c) shows MRB data and fits to an exponential, for each of the three gate sets and $\xi=\nicefrac{1}{2}$. For each MRB experiment, we show the mean observed polarization ($\bar{S}_d$) versus benchmark depth, the distribution of the observed polarization versus benchmark depth, and the fit of $\bar{S}_d$ to $\bar{S}_d = Ap^d$. Data for a single representative subset of qubits of each size ($n=1,2,3,4$) are shown. In all cases, we observe that $\bar{S}_d$ is consistent with an exponential decay in $d$, providing experimental evidence for our claim that $\bar{S}_d$ will decay exponentially under a broad range of conditions. 

Figure~\ref{fig:aqt_main} (d-f) shows the estimated error rates ($r_{\Omega}$) for each qubit subset that we benchmarked, for each of the three different gate sets. Each $r_{\Omega}$ is a rescaling of the decay rate of the fitted exponential [see Eq.~\eqref{eq:decay}]. By comparing Fig.~\ref{fig:aqt_main} (d), (e) and (f) we can compare the average error rates of $n$-qubit layers constructed from three different gate sets, two of which are universal and one of which contains only Clifford gates and therefore is not. By comparing (e) and (f), we find that the average error rate of a layer set is approximately independent of whether single-qubit gates are sampled from $\mathbb{SU}(2)$ or from $\mathbb{C}_1$ (the single-qubit Clifford group)---that is, $r(\mathbb{SU}(2), \{\cz \},\mathbb{Q}, \nicefrac{1}{2}) \approx r(\mathbb{C}_1, \{\cz \}, \mathbb{Q}, \nicefrac{1}{2})$ for all ten subsets of qubits $\mathbb{Q}$. All single-qubit gates in our experiments are implemented using a composite $\g{u}(\theta,\phi,\lambda)$ gate [see Eq.~\eqref{eqn:decomp}] that contains two $\x_{\nicefrac{\pi}{2}}$ gates and three $\z_{\theta}$ gates. This is the case even for unitaries that do not require two $\x_{\nicefrac{\pi}{2}}$ pulses, such as the identity. The difference between any two single-qubit gates is therefore only in the angles of the three $\z_{\theta}$ gates within $\g{u}(\theta,\phi,\lambda)$. These gates are implemented by in-software phase updates on later pulses \cite{McKay2017-eq}, so it is expected that these ``virtual gates'' cause negligible errors. The observed similarity between the average performance of these two gate sets is consistent with this expectation (numerical values for all estimated $r_{\Omega}$ are included in Table~\ref{tab:error_rates}). Note, however, that the observed 
similarity between the average success rates of circuits in which the single qubit-gate gates $\g{u}(\theta, \phi, \lambda)$ are sampled from two different distributions does \emph{not} imply that the success rate of an individual circuit is independent of the values of $\theta$, $\phi$ and $\lambda$ in its $\g{u}(\theta, \phi, \lambda)$ gates --- see Appendix~\ref{app:per_c} for further discussions.

Our experiments included MRB on $n$-qubit layers containing two non-Clifford two-qubit gates---$\cs$ and $\csd$---and we now turn to these results. Comparing Figs.~\ref{fig:aqt_main} (d) and (f), we observe that the error rates for layers containing $\cs$ and $\csd$ gates are all almost equal to, but slightly larger than, the error rates for layers containing $\cz$ gates. The largest relative difference is in the experiments on the 3-qubit set $\{\Q{4},\Q5,\Q6\}$: $r(\mathbb{SU}(2), \{\mathsf{cs},\mathsf{cs}^{\dagger}\}, \{ \Q{4}, \Q{5}, \Q{6} \}, \nicefrac{1}{2})=1.64(5)\%$ and $r(\mathbb{SU}(2), \{\cz\}, \{ \Q{4}, \Q{5}, \Q{6} \}, \nicefrac{1}{2}) = 1.48(4)\%$. The three different two-qubit gates ($\cs$, $\csd$, and $\cz$) on each qubit pair were \emph{a priori} expected to have similar error rates, due to their similar calibration procedures. The slightly larger error rates for $\cs$ and $\csd$ were cross-validated using cycle benchmarking \cite{erhard2019characterizing} (see Section~\ref{sec:gate_errors} for a quantitative comparison). Therefore, these results are experimental evidence for the robustness of MRB with non-Clifford two-qubit gates (see Sections~\ref{sec:general_errors_theory} and \ref{sec:theory-nc-2Q-gates} for discussion of and theory for MRB of non-Clifford two-qubit gates).

\subsection{Estimating crosstalk errors}\label{sec:aqt-crosstalk}

\begin{figure}
    \centering
    \includegraphics{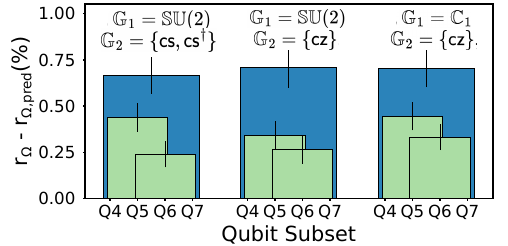}
    \caption{\textbf{Estimating crosstalk errors on AQT.} We estimate the contribution of crosstalk errors to the layer error rate $r_{\Omega}$ for $n=3,4$ qubits by taking the difference between each experimental error rate ($r_{\Omega}$) and a corresponding prediction ($r_{\Omega,\textrm{pred}}$) obtained from the experimental one- and two-qubit error rates and the assumption of no crosstalk. We find that crosstalk contributes approximately $0.2\%-0.4\%$ to $r_{\Omega}$ for $n=3$ (which is $\nicefrac{1}{8}-\nicefrac{1}{4}$ of $r_{\Omega}$), and approximately $0.7\%$ to $r_{\Omega}$ for $n=4$ (which is $\nicefrac{1}{3}$ of $r_{\Omega}$).}
    \label{fig:aqt_crosstalk}
\end{figure}

Crosstalk is an important type of error in current quantum processors, but it is challenging to quantify \cite{sarovar2019detecting}. Multi-qubit MRB captures crosstalk errors, and it enables us to quantify the contribution of crosstalk errors to the average error rate of $n$-qubit layers. To do so, we compare the observed increase in $r_{\Omega}$ with $n$ [Fig.~\ref{fig:aqt_main} (d-f)] to predictions for $r_{\Omega}$ that assume no crosstalk errors. The excess observed error above these predictions is then attributed to crosstalk.

We predict $r_{\Omega}$ for sets of three or more qubits from the observed $r_{\Omega}$ values for each one- and two-qubit subset (note, however, that this is not the only possible way to predict $r_{\Omega}$). This prediction is built on a simple theory for MRB. We model $r_{\Omega}$ by
\begin{equation}\label{eqn:r_omega}
    r_{\Omega} = \sum_{\gate{L} \in \mathbb{L}} \Omega(\gate{L}) \epsilon_\gate{L},
\end{equation}
where $\epsilon_{\gate{L}}$ is the infidelity of a \emph{$\mathbb{G}_1$-dressed layer} $\gate{L}$, which consists of a specific layer of two-qubit gates---i.e., $\gate{L}$ is labelled by the two-qubit gate layer---followed by a layer of random single-qubit gates (either from $\mathbb{SU}(2)$ or $\mathbb{C}_1$). Equation~\eqref{eqn:r_omega} is justified by our theory for MRB (see Section~\ref{sec:theory}), but note that it only holds approximately, unless each layer's error channel is an $n$-qubit depolarizing channel. The fidelity $F = 1 - \epsilon$ of a tensor product of channels is the product of those channels' fidelities. So, under the assumption that there are no crosstalk errors, the infidelity of $\gate{L}$ is given by $\epsilon_{\gate{L}} = \prod_{\gate{g} \in \gate{L}}F_{\gate{g}}$, where $\gate{g}$ are the $\mathbb{G}_1$-dressed gates in the $\mathbb{G}_1$-dressed layer $\gate{L}$, and $F_{\gate{g}}$ is the fidelity of $\gate{g}$. Therefore,
\begin{equation}\label{eqn:layer_error_est}
\epsilon_{\gate{L}} = 1 - \prod\limits_{\gate{g} \in \gate{L}}(1-\epsilon_{\gate{g}}),
\end{equation}
where $\epsilon_{\gate{g}} = 1 - F_{\gate{g}}$. 

To predict $\epsilon_{\gate{L}}$ using Eq.~\eqref{eqn:layer_error_est} [and then $r_{\Omega}$ using Eq.~\eqref{eqn:r_omega}] we need estimates for $\epsilon_{\gate{g}}$ for every possible $\mathbb{G}_1$-dressed gate $\gate{g}$. That is, we need estimates for (1) $\epsilon_{\g{idle}(\Q{i})}$ for each qubit $\Q{i} \in \{\Q{4},\Q{5},\Q{6},\Q{7}\}$ where  $\g{idle}(\Q{i})$ is the $\mathbb{G}_1$-dressed idle gate on $\Q{i}$, and (2) $\epsilon_{\gate{g}(\Q{i},\Q{j})}$ for each connected pair of qubits $\{\Q{i},\Q{j}\}$ where $\gate{g}(\Q{i},\Q{j})$ is a two-qubit gate on $\{\Q{i},\Q{j}\}$ uniformly sampled from $\mathbb{G}_2$. Each of these quantities can be estimated from the observed one- and two-qubit MRB error rates. Using Eq.~\eqref{eqn:r_omega} we have
\begin{equation}\label{eq:r_omega_1Q}
r(\mathbb{G}_1,\{\Q{i}\}) = \epsilon_{\g{idle}(\Q{i})},
\end{equation}
because each single-qubit MRB circuit simply consists of repeating the $\mathbb{G}_1$-dressed idle gate. Similarly, using Eq.~\eqref{eqn:r_omega} we have
\begin{multline}
r(\mathbb{G}_1,\mathbb{G}_2,\{\Q{i},\Q{j}\}, \xi) = \xi\epsilon_{\gate{g}(\Q{i},\Q{j})} + \\ (1-\xi)(1-\epsilon_{\g{idle}(\Q{i})})(1-\epsilon_{\g{idle}(\Q{j})}) , \label{eq:r_omega_2Q}
\end{multline}
because each $\mathbb{G}_1$-dressed layer in a two-qubit MRB circuit is either (with probability $\xi$) a $\mathbb{G}_1$-dressed two-qubit gate sampled uniformly at random from $\mathbb{G}_2$, or a $\mathbb{G}_1$-dressed idle on each qubit (with probability $1-\xi$).

Using Eqs.~\eqref{eqn:layer_error_est}--\eqref{eq:r_omega_2Q} and explicit expressions for $\Omega(\gate{L})$, we obtain analytic expressions for our crosstalk-free predictions of $r_{\Omega}$ for the 3- and 4-qubit layers. These predictions are shown in Fig.~\ref{fig:aqt_main} (g-i). The crosstalk-free predictions are significantly smaller than the observed experimental values, shown in Fig.~\ref{fig:aqt_main} (d-f). For each gate set, the predicted 4-qubit $r_{\Omega}$ is approximately $25\%$ smaller than the observed value. The crosstalk-free predictions for $\{\Q4,\Q5,\Q6\}$ are 13\%--19\% smaller than their observed values, and the crosstalk-free predictions for $\{\Q5,\Q6,\Q7\}$ are 20\%--27\% smaller than their observed values. The difference between the experimental error rates and the crosstalk free predictions, shown in Fig.~\ref{fig:aqt_crosstalk}, is a quantification of the contribution of crosstalk errors to the average rate of errors in 3- and 4-qubit random circuits in this system. We note that one contribution to the difference between the observed $r_{\Omega}$ and the crosstalk-free prediction is the difference between idle gates that occur in parallel with a two-qubit gate and idle gates that occur in single-qubit circuits. The idle that occurs in parallel with a two-qubit gate is a 200 ns idle (the duration of a two-qubit gate on this device), whereas the idle gate that occurs in a one-qubit circuit is a 60 ns idle. Our prediction methodology implicitly assumes that these two idle gates have the same error rate. However, we conjecture that the contribution from this difference is small, because idle gates in this system are relatively low error.

\begin{figure}
    \centering
    \includegraphics{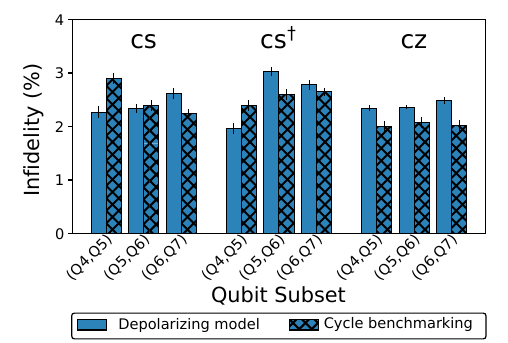}
    \caption{\textbf{Estimating the infidelity of dressed 4-qubit layers.} Estimates of the error rates of individual $\mathbb{G}_1$-dressed layers containing a single 2-qubit gate ($\cs$, $\csd$, or $\cz$), obtained by fitting an $n$-qubit depolarizing model to the 4-qubit MRB data. This scalable analysis technique enables extraction of additional information about each layer's error from MRB data. To validate our results against an established technique, we compare to infidelities independently estimated using cycle benchmarking \cite{erhard2019characterizing}. We observe qualitative agreement. The cycle benchmarking experiments measure the infidelities of layers dressed with one-qubit gates sampled from a different gate set (the Pauli group) to those used in our MRB experiments, so exact agreement is not expected.}
    \label{fig:cb_comparison}
\end{figure}

\begin{figure*}[ht!]
    \centering
    \includegraphics{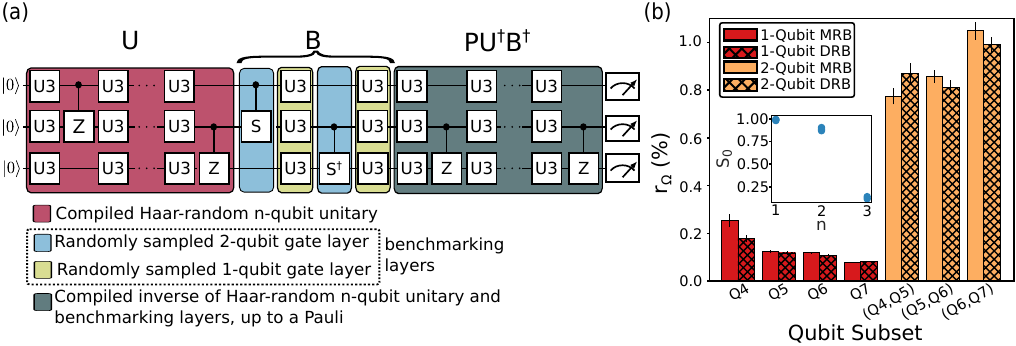}
    \caption{\textbf{Validating MRB by comparison to DRB.} (a) The structure of DRB circuits, which is a method for benchmarking an $n$-qubit layer set, when applied to a universal gate set. DRB is known to be reliable, but it is exponentially expensive in $n$ for universal gate sets---because, for a universal gate set, its circuits start by implementing a Haar random unitary from $\mathbb{SU}(2^n)$. (b) The error rates obtained when running equivalent DRB and MRB experiments, on every one- and two-qubit subset of the 4 qubits we benchmarked on AQT. The close agreement between the DRB and MRB error rates is experimental evidence that MRB is reliable. The inset shows the polarization at benchmark depth $d=0$ ($S_0$) for $n$-qubit DRB with $n=1,2,3$. The rapid decay in $S_0$ is due to the overhead in implementing a Haar-random unitary, and it makes DRB of universal gate sets infeasible on more than around 2--3 qubits. }\label{fig:drb}
\end{figure*}

\subsection{Estimating the error rates of individual gates}
\label{sec:gate_errors}
An MRB experiment is primarily designed to estimate a single error rate ($r_{\Omega}$) that quantifies the average error rate of an $n$-qubit layer. However, it is also often useful to quantify the error in specific layers, e.g., to identify high-error gates. Information about the error rates of individual layers is contained within the MRB data (e.g., RB data can even be used for full tomography \cite{kimmel2014robust, Nielsen2021-cb}), and we extract it using a scalable model fitting method. Specifically, we fit a 4-qubit depolarizing error model to the 4-qubit MRB data to estimate the error rates of individual $\mathbb{G}_1$-dressed layers [Fig.~\ref{fig:cb_comparison}]. To validate our results, we compare the infidelities we estimate to independent estimates obtained from an established technique: cycle benchmarking \cite{erhard2019characterizing}, which is a method for estimating the infidelity of individual many-qubit gate layers. Fig.~\ref{fig:cb_comparison} shows that our estimates are broadly similar to the those obtained from cycle benchmarking, differing by at most $23\%$, and note that we would not expect exact agreement \footnote{The cycle benchmarking experiments measure the infidelities layers dressed with a different single-qubit gate set (the Pauli group) to those used in our MRB experiments, and these experiments were implemented on a different day than the MRB experiments, so exact agreement is not expected.}.  This demonstrates the potential of MRB to go beyond average error rate estimation, and provides an alternative to, e.g., interleaved RB. In Appendix~\ref{app:gate_errors} we discuss the depolarizing model fit as well as two additional methods for estimating the error rate of individual layers from MRB data, and compare their predictions.

\subsection{Comparison to direct RB}
\label{sec:exp_drb}

One of the purposes of our experiments is to test the reliability of MRB.
To investigate whether $r_{\Omega}\approx \epsilon_{\Omega}$ in experiment (as claimed by our theory), we compare the results of MRB to an alternative, established RB technique: direct RB (DRB) \cite{proctor2018direct}. DRB is a streamlined variant of standard RB. Both DRB and standard RB are inefficient when applied to universal gate sets---as they have costs that scale exponentially with the number of qubits---but they are feasible in the very few qubit regime. We chose to compare MRB to DRB because these two methods have the same flexible circuit sampling and they are designed to measure the same error rate: $\epsilon_{\Omega}$. In contrast, standard RB benchmarks a gate set that forms a group, e.g., $\mathbb{SU}(2^n)$, and it measures an error rate for a uniformly random element of that group---so this error rate cannot be directly compared to $r_{\Omega}$.

An $n$-qubit, benchmark depth $d$ DRB circuit for a universal layer set is constructed by first sampling a depth-$d$ circuit $\gate{C}$ with layers sampled from some distribution $\Omega$---exactly as with MRB. As shown in Fig.~\ref{fig:drb} (a), this circuit $\gate{C}$ is then embedded between (1) a circuit that implements an $n$-qubit Haar random unitary, and (2) a circuit that returns the qubits to the computational basis. Note that both (1) and (2) require circuits of one- and two-qubit gates whose size grows exponentially in $n$ (we compile a $\mathbb{SU}(2^n)$ unitary into a circuit of $\x_{\nicefrac{\pi}{2}}$, $\z_{\theta}$ and $\cz$ gates using the \texttt{Qsearch} package \cite{osti_1769276, 9259942}).
We therefore ran DRB on all $n$-qubit subsets only up to $n = 3$.

In our DRB experiments we used the same layer sampling distribution as in our $\mathbb{G}_1=\mathbb{SU}(2)$, $\mathbb{G}_2=\{\cs,\csd\}$, and $\xi=\nicefrac{1}{2}$ MRB experiments. So the DRB error rates we are measuring---which we denote by $r_{\textrm{DRB}}(\mathbb{SU}(2),\{\cs,\csd\},\mathbb{Q},\nicefrac{1}{2})$ for qubit subset $\mathbb{Q}$---will be equal to the equivalent MRB error rates $r(\mathbb{SU}(2),\{\cs,\csd\},\mathbb{Q},\nicefrac{1}{2})$ if both DRB and MRB are working correctly. Figure \ref{fig:drb} compares these DRB and MRB error rates for each one- and two-qubit subset. For each of these qubit subsets, the two error rates differ by no more than $2\sigma$. Due to the overhead in implementing a Haar-random unitary from $\mathbb{SU}(2^n)$, the 3-qubit DRB circuits were so large that the polarization of all $n=3$ DRB circuits was $S_d \approx 0$, even for the $d=0$ circuits, so we were not able to obtain reliable estimates of $r_{\textrm{DRB}}$ for either 3-qubit subset. The rapid decrease in the $d=0$ polarization ($S_0$) with increasing $n$ is shown in the inset of Fig. \ref{fig:drb} (b). This demonstrates that DRB cannot be used to benchmark universal gate sets on more than around 2-3 qubits (and note that standard RB requires running even larger circuits than those used in DRB).

\begin{figure*}
    \centering
    \includegraphics{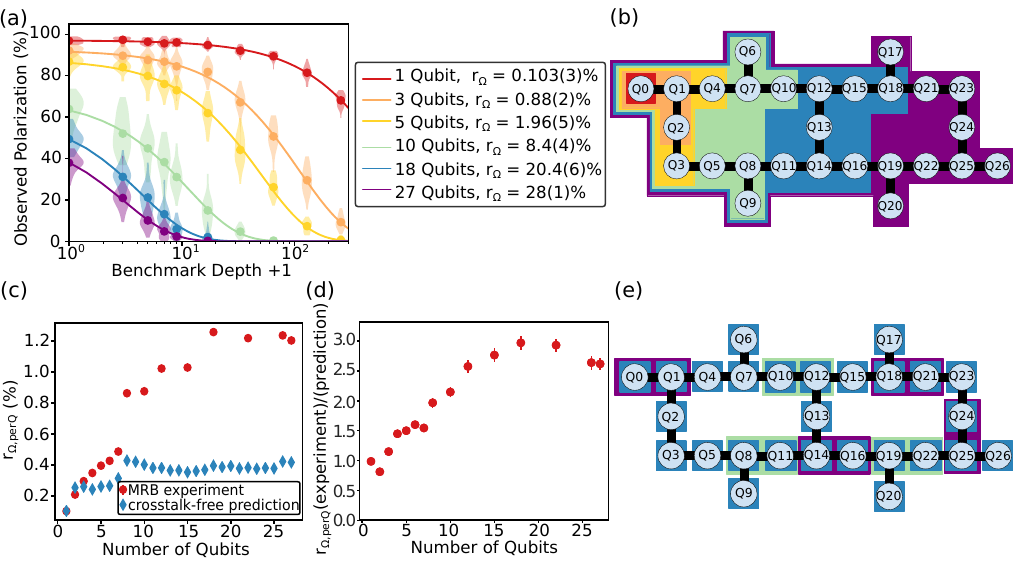}
    \caption{\textbf{Randomized benchmarking of a universal gate set on a 27-qubit IBM Q processor.} We ran MRB on $n$-qubit subsets of the \texttt{ibmq\_montreal} processor, for 15 exponentially spaced $n$ from $n=1$ to $n=27$. (a) The MRB decays and fits to an exponential for the six subsets of qubits illustrated in (b). The observed polarization decays exponentially in all cases. Due to the minimal overhead in MRB circuits, we obtain an exponential decay even for 27 qubits and can extract a low-uncertainty estimate of the average error rate of 27-qubit layers [$r_{\Omega}=28(1)\%$]. (c) The observed error rate per qubit $r_{\Omega, \textrm{perQ}} = 1 - (1 -r_{\Omega})^{\nicefrac{1}{n}} \approx \nicefrac{r_{\Omega}}{n}$ (red circles) versus $n$ increases rapidly with $n$, even though the circuits have a constant expected two-qubit gate density $\xi = \nicefrac{1}{4}$. This increase in $r_{\Omega, \textrm{perQ}}$ is due to two-qubit gate crosstalk, not spatial variations in gate error rates. This is confirmed by comparison to predictions for $r_{\Omega, \textrm{perQ}}$ (blue diamonds) obtained from one- and two-qubit error rates, for each one-qubit and connected two-qubit subset, and the assumption of no crosstalk. (d) The ratio  of the observed ($r_{\Omega, \textrm{perQ}}$) to predicted ($r_{\Omega, \textrm{perQ},\textrm{pred}}$) per-qubit error rate shows that crosstalk errors cause the per-qubit error rate $r_{\Omega, \textrm{perQ}}$ to increase by approximately 250\%--300\% when $n\geq 15$. (e) The one- and two-qubit error rates were obtained using simultaneous one-qubit MRB on all 27 qubits (blue boxes), and two-qubit MRB, on all pairs of connected qubits, run simultaneously on the qubit pairs in eight distinct groupings (the purple and green boxes show two such groups).} \label{fig:ibm}
\end{figure*}

\section{27-qubit IBM Q demonstration}\label{sec:ibm}
To investigate what many-qubit MRB can reveal about errors in current many-qubit hardware, we ran MRB on a 27-qubit IBM Q device ($\texttt{ibmq\_montreal}$, a $\texttt{Falcon r4}$ processor). We used the universal gate set $\mathbb{G}_1=\mathbb{SU}(2)$ and $\mathbb{G}_2=\{\cnot\}$, and we sampled layers with a two-qubit gate density of $\xi=\nicefrac{1}{4}$. Our circuits contain barriers between each layer of gates, as in our experiments on AQT (Section~\ref{sec:aqt}). choose a single qubit subset $\mathbb{Q}$ containing $n$ qubits for 15 exponentially spaced $n$ up to $n=27$.
This is illustrated in Fig.~\ref{fig:ibm} (b), for 6 of the 15 qubit subsets. For each qubit subset, we sampled and ran 25 circuits at each of a set of exponentially spaced depths.

Figure \ref{fig:ibm} (a) shows the observed polarization versus benchmark depth for six representative values of $n$. Even for $n=27$, where we observe an average layer error rate of $r_{\Omega}=28(1)\%$, we obtain a $d=0$ average observed polarization of $S_0 \approx 40\%$. This demonstrates that MRB is practical on many qubits, even when the error rate per layer is $O(10\%)$. For all $n$, we observe that the mean observed polarization is consistent with an exponential decay, as expected. Fig.~\ref{fig:ibm} (b) shows the error rate per qubit ($r_{\Omega, \,\textrm{perQ}} = 1-(1-r_{\Omega})^{\nicefrac{1}{n}} \approx \nicefrac{r_{\Omega}}{n}$) versus $n$. Our circuits have a fixed expected two-qubit gate density (of $\xi=\nicefrac{1}{4}$). Therefore, $r_{\Omega, \,\textrm{perQ}}$ will be independent of $n$ for $n\geq 2$ if (1) the error rate of one-qubit gates and the error rate of two-qubit gates is invariant across the device, and (2) there are no crosstalk errors. Instead, we observe that $r_{\Omega, \,\textrm{perQ}}$ rapidly increases from $r_{\Omega, \,\textrm{perQ}} \approx 0.2\%$ for $n=2$ up to $r_{\Omega, \,\textrm{perQ}} \approx 1.2\%$---an increase of approximately $500\%$.

To quantify the contribution of crosstalk errors to the observed increase in the per-qubit error rate with $n$, we first need to quantify the spatial variations in the one- and two-qubit gate error rates (meaning the error rates of those gates when all other qubits are idle). We used one- and two-qubit MRB to measure the error rates of each one-qubit subset and each connected two-qubit subset of the 27-qubits. Because of the large number of qubits, it would require running more circuits than was feasible to implement independent one-qubit MRB experiments on each qubit (27 MRB experiments) and independent two-qubit MRB experiments on each connected pair of qubits (30 MRB experiments). Instead, we implemented all 27 one-qubit MRB experiments simultaneously \cite{gambetta2012characterization}. The resultant one-qubit MRB error rates therefore include contributions from single-qubit gate crosstalk errors. We ran the 30 two-qubit MRB experiments in eight groups, selected to minimize the closeness in frequency space of the qubits in each group. These two-qubit MRB error rates will therefore include some contributions from two-qubit gate crosstalk, but the experiments have been designed with the aim of minimizing this contribution. We also ran five isolated two-qubit MRB experiments and observed that the simultaneous two-qubit MRB error rates were a factor of between 1.5 and 2.5 times larger than the corresponding isolated MRB error rates (see Table \ref{tab:1_2_q_ibm}).

We use the set of measured one- and two-qubit MRB error rates to predict the $n$-qubit $r_{\Omega}$ that would be observed if there are no two-qubit gate crosstalk errors, using Eqs.~\eqref{eqn:r_omega}--\eqref{eq:r_omega_2Q} \footnote{In this case we do not have a closed form expression for $\Omega(\gate{L})$. Instead, we sample 10000 layers from $\Omega$, compute $\epsilon_{\gate{L}}$ for each layer from the one- and two-qubit MRB error rates, and then average all 10000 $\epsilon_{\gate{L}}$ to estimate $r_{\Omega}$.}. 
Figure \ref{fig:ibm} (c) shows the predictions for the per-qubit error rate $r_{\Omega, \,\textrm{perQ}}$. For $n\gg 1$ these predictions (blue diamonds) are much smaller than the observations (red circles). This prediction accounts for spatial variations in the one- and two-qubit error rates, and includes contributions from one-qubit gate crosstalk errors (and some contributions from two-qubit gate crosstalk). Therefore, we can conclude that the additional observed error is due to crosstalk caused by the two-qubit gates, and it lower bounds the total contribution of crosstalk errors to $r_{\Omega}$. Figure \ref{fig:ibm} (d) shows the ratio $R$ of the observed to the predicted error rate per qubit $r_{\Omega, \,\textrm{perQ}}$, versus $n$. $R$ grows approximately linearly from $R \approx 0.2$ at $n=2$ up to $R \approx 2.5$ at $n \approx 13$ and then saturates at between $R \approx 2.5$ and $R \approx 3.0$. One possible explanation for this is two-qubit gate crosstalk errors with finite spatial radius, i.e., two-qubit gates cause increased errors on other qubits within some distance of the target qubits.

\section{Discussion}\label{sec:discussion}
Scalable benchmarking methods are needed to quantify the integrated performance of medium- and large-scale quantum processors. In this paper, we introduced a scalable method for RB of universal gate sets that uses a novel and customizable family of randomized mirror circuits. We presented a theory for our method, showing that it reliably measures the error rate of a random $n$-qubit circuit layer sampled from a user-specified distribution $\Omega$. We demonstrated MRB on multiple gate sets in both simulations and experiments, demonstrating that it is reliable and that it is a powerful tool for understanding errors in many-qubit circuits. Our method can be viewed as both an adaptation of standard RB and its variants, to enable efficient and scalable benchmarking of universal gate sets, and as an adaptation of XEB that removes XEB's inefficient circuit simulation step. It therefore provides a link between two widely used benchmarking methodologies, and so we anticipate that the ideas introduced here will lead to further advances in randomized benchmarking.

Using two quantum processors, we demonstrated MRB of a gate set consisting of $\g{cnot}$ and arbitrary single-qubit gates on up to 27 qubits and MRB of a gate set with non-Clifford two-qubit gates ($\g{cs}$ and $\g{cs}^{\dag}$) on up to 4 qubits. Our results provide evidence that MRB with non-Clifford gates is a robust method for determining a processor's error rate per gate layer, and that these error rates can be used to understand the magnitude of various types of errors. Additionally, our results show that MRB on many qubits reveals and quantifies errors not present in one- and two-qubit circuits, highlighting the importance of scalable benchmarks. Comparisons of RB error rates predicted from crosstalk-free models and our experimental results show evidence of large crosstalk errors in both of the devices we benchmarked and, importantly, our methods make it possible to quantify the size of these crosstalk errors.

We anticipate that a variety of interesting benchmarking methods can be constructed using MRB and extensions or adaptations of this method. For example, we anticipate that MRB can form the foundation of methods for estimating the error rates of individual gates and layers, within the context of many-qubit circuits. In this work we demonstrated a simple example of such a technique---fitting MRB data to a depolarizing model---and we expect that a variety of robust methods could be developed, that would complement or advance on existing methods for this task  \cite{magesan2012efficient, erhard2019characterizing, Flammia2021-dn}, such as interleaved RB. For example, MRB can potentially be adapted to extend the averaged circuit eigenvalue sampling protocol \cite{Flammia2021-dn} to universal gate sets. Furthermore, we anticipate that MRB can be adapted to construct scalable ``full-stack'' benchmarks based on random circuits, such as a scalable variant of the widely used quantum volume benchmark \cite{cross2018validating}.

\section*{Acknowledgements} 

This material is based upon work supported by the Laboratory Directed Research and Development program at Sandia National Laboratories and the U.S. Department of Energy, Office of Science, National Quantum Information Science Research Centers, Quantum Systems Accelerator. This work was also supported by the U.S. Department of Energy, Office of Science, Office of Advanced Scientific Computing Research Quantum Testbed Program under Contract No.~DE-AC02-05CH11231. Sandia National Laboratories is a multi-mission laboratory managed and operated by National Technology \& Engineering Solutions of Sandia, LLC (NTESS), a wholly owned subsidiary of Honeywell International Inc., for the U.S. Department of Energy’s National Nuclear Security Administration (DOE/NNSA) under contract DE-NA0003525. This written work is authored by an employee of NTESS. The employee, not NTESS, owns the right, title and interest in and to the written work and is responsible for its contents. Any subjective views or opinions that might be expressed in the written work do not necessarily represent the views of the U.S. Government. The publisher acknowledges that the U.S. Government retains a non-exclusive, paid-up, irrevocable, world-wide license to publish or reproduce the published form of this written work or allow others to do so, for U.S. Government purposes. The DOE will provide public access to results of federally sponsored research in accordance with the DOE Public Access Plan.

We acknowledge the use of IBM Quantum services for this work. The views expressed are those of the authors, and do not reflect the official policy or position of IBM or the IBM Quantum team.

\section*{Code and data availability}\label{sec:code}
Data and code for our simulations and experiments will be provided upon reasonable request. All circuit sampling and simulations were performed using \texttt{pyGSTi} \cite{nielsen2020probing}.

\bibliography{Bibliography}

\appendix
\onecolumngrid
\section{MRB with stochastic Pauli errors}
\label{app:r_eps_relation}
In this appendix, we provide further details on the theory presented in Section~\ref{sec:r_eps_relation}, showing that the MRB error rate ($r_{\Omega}$) approximately equals $\epsilon_{\Omega}$ [Eq.~\eqref{eq:epsilon}] under the assumption of stochastic Pauli errors. 
\subsection{Determining the observed polarization of MRB circuits}
\label{app:s_mrb}

We start by proving Eq.~\eqref{eqn:s_d_avg_e_eff}, which says that the mean observed polarization [Eq.~\eqref{eq:S}] of randomized mirror circuits equals the mean polarization of the overall error map of a randomized mirror circuit. We start by defining an overall error map for our mirror circuit, which captures all error in the circuit. We can define an overall error map for a circuit $\rand{C}$ by rewriting $\phi(\rand{C})$ with all error moved to the beginning of the circuit. For a general depth-$l$ circuit with gate-dependent errors,
\begin{align}
    \phi(\rand{C}) & = \mathcal{E}(\rand{L}_l)\mathcal{U}(\rand{L}_l) \cdots \mathcal{E}(\rand{L}_2)\mathcal{U}(\rand{L}_2)\mathcal{E}(\rand{L}_1)\mathcal{U}(\rand{L}_1) \nonumber \\
    & = \mathcal{U}(\rand{L}_l) \cdots \mathcal{U}(\rand{L}_2)\mathcal{U}(\rand{L}_1)\mathcal{E}'_{\rand{L}_l} \cdots \mathcal{E}'_{\rand{L}_2}\mathcal{E}'_{\rand{L}_1}, \label{eqn:app_rearranged_error}
\end{align}
where 
\begin{equation}
    \mathcal{E}'_{\rand{L}_i} = \mathcal{U}(\rand{L}_1)^{-1} \cdots \mathcal{U}(\rand{L}_{i})^{-1}\mathcal{E}(\rand{L}_i)\mathcal{U}(\rand{L}_i)\cdots \mathcal{U}(\rand{L}_1). 
\end{equation}
Applying Eq.~\eqref{eqn:app_rearranged_error} to our randomized mirror circuit allows us to express the error in  $\tilde{\rand{M}}_d$ (which is the mirror circuit without the initial and final layers), as a single error channel following the initial randomized state preparation layer $\rand{L}_0$. We find that $\phi(\rand{M}_d)$ can be expressed as
\begin{align}
\phi(\rand{M}_d) & = \phi\bigl(\rc{\rand{L}_0^{-1}}{ \rand{P}_{d+1}}{ \rand{P}_d^c}\bigr) \phi(\rand{\tilde{M}}_d) \phi\bigl(\rc{\rand{L}_0}{\rand{P}_{0}}\bigr) \nonumber \\
& = \mathcal{U}\bigl(\rc{\rand{L}_0^{-1}}{ \rand{P}_{d+1}}{ \rand{P}_d^c}\rand{\tilde{M}}_d\bigr)\mathcal{U}(\rand{P}_0)\mathcal{E}_{\textrm{SPAM}}\mathcal{E}_{\textrm{eff}}(\rand{\tilde{M}}_d)\mathcal{U}(\rand{L}_0) \nonumber \\
& = \mathcal{U}(\rand{P}_{d+1})\mathcal{U}(\rand{L}_0^{-1})\mathcal{E}_{\textrm{eff}}(\rand{M}_d)\mathcal{U}(\rand{L}_0), \label{eqn:app_effective_error}
\end{align}
where 
 \begin{align}
 \mathcal{E}_{\textrm{eff}}(\rand{M}_d) & = \mathcal{E}_{\textrm{SPAM}}\mathcal{E}_{\textrm{eff}}(\rand{\tilde{M}}_d)\\
 & =  \mathcal{E}_{\textrm{SPAM}}\mathcal{E}'_{\tcp{\rand{L}_{\theta_1}^{-1}}{\rand{P}_{d}}}\cdots \mathcal{E}'_{\rand{L}_{\nicefrac{d}{2}}^{-1}} \mathcal{E}'_{\rand{L}_{\nicefrac{d}{2}}}\cdots\mathcal{E}'_{\tcp{L_{\theta_1}}{\rand{P}_0}}. \label{eqn:app_error_channel}
 \end{align} 
To obtain Eq.~\eqref{eqn:app_effective_error}, we use the reflection structure of randomized mirror circuits---in particular, $\mathcal{U}(\rc{\rand{L}_0^{-1}}{ \rand{P}_{d+1}}{ \rand{P}_d^c}\rand{\tilde{M}}_d) = \mathcal{U}(\rand{P}_{d+1}\rand{L}_0^{-1}\rand{P}_0)$, where $\rand{P}_0$ and $\rand{P}_{d+1}$ are the Pauli gates that are recompiled into $\rand{L}_0$ and $\rand{L}_0^{-1}$, respectively, in the randomized compilation step. The Pauli gate $\rand{P}_{d+1}$ determines the target bit string of $\rand{M}_d$---i.e., $\mathcal{U}(\rand{M})\kett{0}=\mathcal{U}(\rand{P}_{d+1})\kett{0}=\kett{b}$.
The overall error map $\mathcal{E}_{\textrm{eff}}(\rand{\tilde{M}_d})$ [Eq.~\eqref{eqn:app_error_channel}] contains the error from the $\nicefrac{d}{2}$ $\Omega$-random circuit layers and their inverses (after randomized compilation), and it is composed of unitary rotations of the error channels associated with each circuit layer.

In the MRB protocol, we compute each circuit's observed polarization $S$ [Eq.~\eqref{eq:S}]. We now show that the observed polarization $S(\rand{M}_d)$ is related to the polarization [Eq.~\eqref{eq:pol_def}] of $\rand{M}_d$'s overall error map (introduced above). Using the expression for $\phi(\rand{M}_d)$ in Eq.~\eqref{eqn:app_effective_error}, the probability of measuring bit string $x$ on circuit $\rand{M}_d$ is given by
\begin{align}
    \mathcal{P}_x = \braa{x}\mathcal{U}(\rand{P}_{d+1})\mathcal{U}(\rand{L}_0^{-1})\mathcal{E}_{\textrm{eff}}(\rand{M}_d)\mathcal{U}(\rand{L}_0)\kett{0} \\
    = \braa{x+b}\mathcal{U}(\rand{L}_0^{-1})\mathcal{E}_{\textrm{eff}}(\rand{M}_d)\mathcal{U}(\rand{L}_0)\kett{0}. \label{eq:px}
\end{align}
The layer $\rand{L}_0$ consists of single-qubit gates independently sampled from single-qubit unitary 2-designs. We now average over the initial circuit layer $\rand{L}_0$, making use of a fidelity estimation technique based on single-qubit gates: the fidelity of any error channel $\mathcal{E}$ can be found by averaging over a tensor product of single-qubit 2-designs \cite{proctor2022establishing}. In particular, for any bit string $y \in \{0,1\}^n$, 
\begin{eqnarray}
    \gamma(\mathcal{E}) = \frac{4^n}{4^n-1}\smashoperator[lr]{\sum_{x\in \{0,1\}^n}}\left(-\nicefrac{1}{2}\right)^{h(x,y)} \braa{x+y} \bar{\mathcal{E}} \kett{0} - \frac{1}{4^n-1}, \label{eq:app_2_design_avg}
\end{eqnarray}
where $\bar{\mathcal{E}} = \E_{\rand{L}}[\mathcal{U}(\rand{L})^{\dagger} \mathcal{E} \mathcal{U}(\rand{L})]$ \cite{proctor2022establishing} and $\rand{L} = \otimes_{i=1}^n \rand{L}_i$, where each $\rand{L}_i$ is a independent, single-qubit 2-design. This implies that the expected observed polarization of $\rand{M}_d$ over $\rand{L}_0$ is
\begin{align}
     \E\limits_{\rand{L}_0}S(\rand{M}_d) & =  \frac{4^n}{4^n-1}\left[\sum_{k=0}^{n}\sum\limits_{\langle x,b \rangle = k} \left(-\frac{1}{2}\right)^k \E\limits_{\rand{L_0}}\mathcal{P}_x\right] - \frac{1}{4^n -1} \nonumber \\
    & =  \frac{4^n}{4^n-1}\left[\sum\limits_{x\in\{0,1\}^n} \left(-\frac{1}{2}\right)^{h(x,b)} \E\limits_{\rand{L_0}} \mathcal{P}_x\right] - \frac{1}{4^n -1} \nonumber \\
     & = \gamma(\mathcal{E}_{\textrm{eff}}(\rand{M}_d)) \label{eq:app_c_pol}
\end{align}
where $\gamma(\mathcal{E})$ denotes the \emph{polarization} of $\mathcal{E}$ [Eq.~\eqref{eq:pol_def}]. Eq.~\eqref{eq:app_c_pol} follows from Eq.~\eqref{eq:app_2_design_avg}.  Averaging over all depth-$d$ randomized mirror circuits, the mean observed polarization is
\begin{equation}
    \bar{S}_{d} = \E\limits_{\rand{M}_d} \gamma\bigl(\mathcal{E}_{\textrm{eff}}(\rand{M}_d)\bigr). \label{eq:s_d_avg_e_eff}
\end{equation}
Equation~\eqref{eq:s_d_avg_e_eff} says that the average observed polarization $ \bar{S}_{d}$, which is estimated by the MRB protocol, is equal to the expected polarization of the error channel of a depth-$d$ mirror circuit.
\subsection{Relating the observed polarization of MRB circuits and $\Omega$-distributed random circuits}
\label{app:r_and_epsilon}
Above, we related the mean observed polarization ($\bar{S}_{d}$), which determines the MRB error rate, to the expected polarization of the overall error map of a depth-$d$ randomized mirror circuit. We now use this result to derive Eq.~\eqref{eqn:s_relation}, which relates the mean observed polarization of depth-$d$ randomized mirror circuits to the expected polarization of the overall error map of a depth-$\nicefrac{d}{2}$ $\Omega$-distributed circuit. In combination with the theory in Section~\ref{app:exp_decay}---which shows that $\bar{S}_d$ and the mean polarization of the overall error map of $\Omega$-distributed random circuits decay exponentially---the relationship we derive here implies that $r_{\Omega} \approx \epsilon_{\Omega}$.

Our goal is to relate the rate of decay of $\bar{S}_d$ to the rate of decay of the fidelity of $\Omega$-distributed circuits ($\bar{F}_d$) [Eq.~\eqref{eq:f_d_def}]. We start by expressing $\bar{F}_d$ in terms of the expected polarization of the overall error map of a depth-$d$ $\Omega$-distributed circuit. Applying Eq.~\eqref{eqn:app_rearranged_error} to a depth-$d$, $\Omega$-distributed random circuit $\rand{C}_d = \rand{L}_{d} \rand{L_{\theta_{d}}}\cdots \rand{L_0}\rand{L_{\theta_1}}$, we obtain an overall error map for $\rand{C}_d$, $\mathcal{E}_{\textrm{eff}}(\rand{C}_d)$, which is defined by $\phi(\rand{C}_d) = \mathcal{U}(\rand{C}_d)\mathcal{E}_{\textrm{eff}}(\rand{C}_d)$. We define $
\Gamma_{d}$ to be the average polarization of the error map of a depth-$d$ mirror circuit:
\begin{equation}
    \bar{\Gamma}_{d} = \E_{\rand{C}_d} \gamma\bigl(\mathcal{E}_{\textrm{eff}}(\rand{C}_d)\bigr). \label{eq:s_d_c}
\end{equation}

To relate $\bar{S}_d$ to $\bar{\Gamma}_d$, we use the fact that a depth-$d$ randomized mirror circuit consists of randomized compilation of a depth-$\nicefrac{d}{2}$ $\Omega$-distributed random circuit followed by its inverse. These two depth-$\nicefrac{d}{2}$ circuits are both $\Omega$-distributed (even after randomized compilation), but they are correlated. Below, we show that the polarization of the mirror circuit's overall error map depends on the covariance between the error in a depth-$\nicefrac{d}{2}$ $\Omega$-distributed circuit and its randomly compiled inverse. We can write the overall error map in Eq.~\eqref{eqn:app_error_channel} as a composition of two error maps---an overall error map for a random circuit and an overall error map for its randomly compiled inverse:
\begin{equation}
\mathcal{E}_{\textrm{eff}}(\rand{M}_d) = \mathcal{E}_{\textrm{SPAM}} \mathcal{E}_{\textrm{eff}, 2}(\rand{M}_d)\mathcal{E}_{\textrm{eff}, 1}(\rand{M}_d),
\label{eq:eff_M_d}
\end{equation}
where
\begin{align}
\mathcal{E}_{\textrm{eff}, 1}(\rand{M}_d) & = \mathcal{E}'_{ \rc{\rand{L}_{\nicefrac{d}{2}}}{ \rand{P}_{\nicefrac{d}{2}}}{ \rand{P}^c_{\nicefrac{d}{2}-1}}}\cdots\mathcal{E}'_{\tcp{\rand{L}_{\theta_1}}{\rand{P}_0}} \nonumber \\ 
\mathcal{E}_{\textrm{eff}, 2}(\rand{M}_d) & = \mathcal{E}'_{\tcp{\rand{L}_{\theta_1}^{-1}}{\rand{P}_d}}\cdots \mathcal{E}'_{ \rc{\rand{L}_{\nicefrac{d}{2}}^{-1}}{ \rand{P}_{\nicefrac{d}{2}+1}}{ \rand{P}_{\nicefrac{d}{2}}}} \nonumber \\
& = \mathcal{U}(\rand{C}) \mathcal{E}_{\textrm{eff}}\bigl(\rand{\tcp{\rand{L}_{\theta_1}^{-1}}{\rand{P}_d} \cdots \rc{\rand{L}_{\nicefrac{d}{2}}^{-1}}{ \rand{P}_{\nicefrac{d}{2}+1}}{ \rand{P}_{\nicefrac{d}{2}}}}\bigr) \mathcal{U}(\rand{C})^{-1}, \nonumber
\end{align}
and \[\rand{C} = \rc{\gate{L}_{\nicefrac{d}{2}}^{-1}}{ \gate{P}_{\nicefrac{d}{2}+1}}{ \gate{P}_{\nicefrac{d}{2}}} \nonumber \\
\rc{\gate{L}_{\nicefrac{d}{2}}}{ \gate{P}_{\nicefrac{d}{2}}}{ \gate{P}^c_{\nicefrac{d}{2}-1}}\cdots \rc{\gate{L}_1}{ \gate{P}_{1}}{ \gate{P}_0^c}\tcp{\gate{L}_{\theta_1}}{\gate{P}_0}.\]
$\rand{C}$ is the first half of $\tilde{\rand{M}}_d$, and it is a depth-$\nicefrac{d}{2}$ $\Omega$-distributed random circuit that has had randomized compilation applied to it. By substituting Eq.~\eqref{eq:eff_M_d} into Eq.~\eqref{eq:s_d_avg_e_eff}, we obtain
\begin{align}
    \bar{S}_{d} & = \E_{\rand{M}_d} \gamma\bigl(\mathcal{E}_{\textrm{SPAM}}\mathcal{E}_{\textrm{eff},2}(\rand{M}_d)\mathcal{E}_{\textrm{eff},1}(\rand{M}_d)\bigr) \label{eq:s_d_avg0} \\
    & = \gamma(\mathcal{E}_{\textrm{SPAM}})\E_{\rand{M}_d} \gamma\bigl(\mathcal{E}_{\textrm{eff},2}(\rand{M}_d)\mathcal{E}_{\textrm{eff},1}(\rand{M}_d)\bigr), \label{eq:s_d_avg_1} 
\end{align}
where, to go from Eq.~\eqref{eq:s_d_avg0} to Eq.~\eqref{eq:s_d_avg_1}, we have used the assumption that $\mathcal{E}_{\textrm{SPAM}}$ is a global depolarizing channel.

Applying randomized compilation to an $\Omega$-distributed random circuit creates a new random circuit that is also $\Omega$-distributed. This is due to the conditions we require of $\Omega_1$ and $\Omega_2$ ($\Omega_1$ is the uniform distribution, and $\Omega_2$ is invariant under replacing a subset of a layer's gates with their inverses). Therefore, we can replace the average over all depth-$d$ randomized mirror circuits in Eq.~\eqref{eq:s_d_avg_1} with an average over all depth-$\nicefrac{d}{2}$ $\Omega$-distributed random circuits:
\begin{align}
    \bar{S}_{d} & = \gamma(\mathcal{E}_{\textrm{SPAM}})\E_{\rand{C}_{\nicefrac{d}{2}}} \gamma\bigl(\mathcal{U}(\rand{C}_{\nicefrac{d}{2}})\bar{\mathcal{E}}_{\textrm{eff}}(\rand{C}_{\nicefrac{d}{2}}^{-1})\mathcal{U}(\rand{C}_{\nicefrac{d}{2}})^{-1}\mathcal{E}_{\textrm{eff}}(\rand{C}_{\nicefrac{d}{2}})\bigr)  \label{eq:s_d_avg_2},
\end{align}
where $\bar{\mathcal{E}}_{\textrm{eff}}(\rand{C}_{\nicefrac{d}{2}}^{-1})$ denotes the average over all possible circuits $\rand{C}'$ resulting from applying randomized compilation to $\rand{C}_{\nicefrac{d}{2}}^{-1}$. Expressing Eq.~\eqref{eq:s_d_avg_2} in terms of $\bar{\Gamma}_{\nicefrac{d}{2}}$ [Eq.~\eqref{eq:s_d_c}], we have

\begin{equation}
    \bar{S}_{d}  = \gamma(\mathcal{E}_{\textrm{SPAM}})\left( \bar{\Gamma}_{\nicefrac{d}{2}}^2 - \Delta_{\Omega}\right), \label{eq:s_relation}
\end{equation}
where 
\begin{equation}
    \Delta_{\Omega} = \E_{\rand{C}_{\nicefrac{d}{2}}} \gamma\bigl(\mathcal{U}(\rand{C}_{\nicefrac{d}{2}})\bar{\mathcal{E}}_{\textrm{eff}}(\rand{C}_{\nicefrac{d}{2}}^{-1})\mathcal{U}(\rand{C}_{\nicefrac{d}{2}})^{-1}\mathcal{E}_{\textrm{eff}}(\rand{C}_{\nicefrac{d}{2}})\bigr) - \left(\E_{\rand{C}_{\nicefrac{d}{2}}}\gamma\bigl(\mathcal{E}_{\textrm{eff}}(\rand{C}_{\nicefrac{d}{2}})\bigr)\right)^2. \label{eq:delta_omega}
\end{equation}

\subsection{Fidelity decay of $\Omega$-distributed random circuits}
\label{app:exp_decay}
In this section, we show that the fidelity of $\Omega$-distributed random circuits decays approximately exponentially in depth, assuming stochastic Pauli errors, when $n$ is sufficiently large that $\nicefrac{1}{4^n}$ is negligible (in the small $n$ case, $\Omega$-distributed random circuits rapidly converge to a 2-design, from which it follows that the fidelity decays approximately exponentially). In this section, we use the notation $\rand{L}_{a;b}$ to denote the sequence of composite layer-valued random variables $\rand{L}_a\rand{L}_{a+1}\cdots \rand{L}_b$. We will assume that each composite layer has a stochasic Pauli error channel, i.e. $\phi(\rand{L}) = \mathcal{E}_{\rand{L}}\mathcal{U}(\rand{L})$, where $\mathcal{\rand{L}}$ is a stochastic Pauli channel. We will used the stacked representation of superoperators, $\mathcal{U} = U \otimes U^{\ast}$. We use $\mathbb{P}_n^{\ast}$ to denote the $n$-qubit Paulis, excluding the identity. We use $\mathcal{P}$ to denote the superoperator representation of a Pauli $P$.

We first prove a useful lemma that follows from the properties of the layer sampling distribution required for MRB (see Section~\ref{sec:gate-set}). MRB requires that the layer sampling distribution is invariant under the randomized compilation procedure defined in Section~\ref{sec:rmc_construction}, which implies that the distribution of unitaries induced by our layer sampling distribution is invariant under left and right multiplication by Paulis, i.e., for all $P, P' \in \mathbb{P}_n$,
\begin{equation}
    \E\limits_{\rand{L}} U(\rand{L}) = \E\limits_{\rand{L}} P'U(\rand{L})P,
\end{equation}
where $\rand{L}$ is sampled from a layer sampling distribution $\Omega$ that satisfies the conditions in Section~\ref{sec:gate-set}. Using this fact, we obtain the following lemma:

\begin{lemma} Let $\rand{L}$ be a circuit layer-valued random variable sampled from an MRB layer sampling distribution $\Omega$. Let $P_1, P_2, P_3, P_4 \in \mathbb{P}_n$, be Pauli operators. If either (i) $P_1 \neq P_2$ or (ii) $P_3 \neq P_4$, then
\begin{equation}
    \E\limits_{\rand{L}} Tr\left(\mathcal{U}(\rand{L})^{-1}(P_1 \otimes P_2^{\ast})\mathcal{U}(\rand{L})(P_3 \otimes P_4^{\ast})\right) = 0.
\end{equation}
\end{lemma}

\begin{proof}
We first consider the case where $P_3 \neq P_4$. Becuase $\Omega$ is invariant under right multiplication by Paulis, we can insert a right-multiplying Pauli $\mathcal{Q}$:
\begin{equation}
    \E\limits_{\rand{L}} Tr\left(\mathcal{U}(\rand{L})^{-1}(P_1 \otimes P_2^{\ast})\mathcal{U}(\rand{L})(P_3 \otimes P_4^{\ast})\right) = \E\limits_{\rand{L}} Tr\left(\mathcal{Q}\mathcal{U}(\rand{L})^{-1}(P_1 \otimes P_2^{\ast})\mathcal{U}(\rand{L})\mathcal{Q}(P_3 \otimes P_4^{\ast})\right), \label{eqn:lemma_eq1}
\end{equation}
where $\mathcal{Q}$ is a Pauli superoperator. We can rewrite Eq.~\eqref{eqn:lemma_eq1} as
\begin{align}
    \E\limits_{\rand{L}} Tr\left(\mathcal{U}(\rand{L})^{-1}(P_1 \otimes P_2^{\ast})\mathcal{U}(\rand{L})(P_3 \otimes P_4^{\ast})\right) & = \E\limits_{\rand{L}} Tr\left(\mathcal{U}(\rand{L})^{-1}(P_1 \otimes P_2^{\ast})\mathcal{U}(\rand{L})(Q \otimes Q^{\ast})(P_3 \otimes P_4^{\ast})(Q \otimes Q^{\ast})\right) \\
    & = \E\limits_{\rand{L}} \eta(Q, P_3)\eta(Q, P_4)Tr\left(\mathcal{U}(\rand{L})^{-1}(P_1 \otimes P_2^{\ast})\mathcal{U}(\rand{L})(P_3 \otimes P_4^{\ast})\right) ,
\end{align}
where $\eta(P,Q)=-1$ if $P$ and $Q$ anticommute and $\eta(P,Q)=1$ if $P$ and $Q$ commute. If $P_3 \neq P_4$, then there exists some Pauli $Q$ such that $Q$ anticommutes with $P_3$ and commutes with $P_4$ (otherwise, we would have $[PQ,P']=0$ for all $P' \in \mathbb{P}_n$, which implies $PQ=\mathbb{I}$ and hence $P=Q$). Taking $Q$ to be such a Pauli, we have
\begin{align}
    \E\limits_{\rand{L}} Tr\left(\mathcal{U}(\rand{L})^{-1}(P_1 \otimes P_2^{\ast})\mathcal{U}(\rand{L})(P_3 \otimes P_4^{\ast})\right) & = -\E\limits_{\rand{L}} Tr\left(\mathcal{U}(\rand{L})^{-1}(P_1 \otimes P_2^{\ast})\mathcal{U}(\rand{L})(P_3 \otimes P_4^{\ast})\right). 
\end{align}
Therefore, $\E\limits_{\rand{L}} Tr\left(\mathcal{U}(\rand{L})^{-1}(P_1 \otimes P_2^{\ast})\mathcal{U}(\rand{L})(P_3 \otimes P_4^{\ast})\right) = 0$.

Similarly, to address the case where $P_1 \neq P_2$, we use the invariance of the sampling distribution under left multiplication by Paulis to obtain
\begin{align}
    \E\limits_L Tr\left(\mathcal{U}(L)^{-1}(P_1 \otimes P_2^{\ast})\mathcal{U}(L)(P_3 \otimes P_4^{\ast})\right) &= \E\limits_L Tr\left(\mathcal{U}(L)^{-1}\mathcal{Q}(P_1 \otimes P_2^{\ast})\mathcal{Q}\mathcal{U}(L)(P_3 \otimes P_4^{\ast})\right) \\
    & = \E\limits_L \left(\eta(Q, P_1)\eta(Q, P_2))Tr(\mathcal{U}(L)^{-1}(P_1 \otimes P_2^{\ast})\mathcal{U}(L)(P_3 \otimes P_4^{\ast})\right).
\end{align}
Using an argument analogous to the previous case, we conclude that if $P_1 \neq P_2$, then $\E\limits_{\rand{L}} Tr\left(\mathcal{U}(\rand{L})^{-1}(P_1 \otimes P_2^{\ast})\mathcal{U}(\rand{L})(P_3 \otimes P_4^{\ast})\right) = 0$.
\end{proof}

We now show that the fidelity of $\Omega$-distributed random circuits decays approximately exponentially. Our theory shows that the expected polarization of $\Omega$-distributed random circuits is given by Eq.~\eqref{eqn:s_d_c}:
\begin{equation}
    \bar{\Gamma}_{d} = \E_{\rand{C}_d} \gamma\bigl(\mathcal{E}_{\textrm{eff}}(\rand{C}_d)\bigr) = \E_{\rand{L}_1, \cdots \rand{L}_d} \gamma\left(\mathcal{U}(\rand{L}_1)^{-1}\cdots \mathcal{U}(\rand{L}_d)^{-1}\mathcal{E}_{\rand{L}_d}\mathcal{U}(\rand{L}_d)\cdots\mathcal{E}_{\rand{L}_1}\mathcal{U}(\rand{L}_1)\right). \label{eqn:l_op_channel} 
\end{equation}
where $\mathcal{E}_{\textrm{eff}}$ is the overall error channel of a depth-$d$ $\Omega$-distributed random circuit. Analogously, the expected fidelity of these circuits is given by
\begin{equation}
    \E_{\rand{C}_d} F\bigl(\mathcal{E}_{\textrm{eff}}(\rand{C}_d)\bigr) = \E_{\rand{L}_1, \cdots \rand{L}_d} F\left(\mathcal{U}(\rand{L}_1)^{-1}\cdots \mathcal{U}(\rand{L}_d)^{-1}\mathcal{E}_{\rand{L}_d}\mathcal{U}(\rand{L}_d)\cdots\mathcal{E}_{\rand{L}_1}\mathcal{U}(\rand{L}_1)\right). \label{eqn:fidelity_c_d} 
\end{equation}
We will now show that the expected process fidelity decays exponentially in benchmark depth ($d$), from which it follows that the polarization $\gamma(\mathcal{E}_{\textrm{eff}})$ decays exponentially (as $n \gg 1$). We will use the \emph{Pauli unravelling} of the circuit: we expand each error channel as $\mathcal{E}_{L_i} = \sum\limits_{P \in \mathbb{P}_n} \gamma_P(L_i) \mathcal{P}$. We will then expand the fidelity (Eq.~\eqref{eqn:fidelity_c_d}) as a sum of terms corresponding to sequences of Paulis in the Pauli unravelling. Eq.~\eqref{eqn:fidelity_c_d} becomes
\begin{align}
    F(\mathcal{E}_{\textrm{eff}}) & =  \frac{1}{4^n}\E_{\rand{L}_1, \dots, \rand{L}_d} \left[ \sum\limits_{P_1, \dots, P_d \in \mathbb{P}_n}  \gamma_{P_d}(\rand{L}_d) \cdots \gamma_{P_1}(\rand{L}_1) Tr\left(\mathcal{U}(\rand{L}_{1;d})^{-1}\mathcal{P}_{d}\mathcal{U}(\rand{L}_{d})\cdots \mathcal{P}_{2}\mathcal{U}(\rand{L}_{2})\mathcal{P}_{1}\mathcal{U}(\rand{L}_{1})\right)\right].  \label{eqn:expansion1}\\
    & =  \frac{1}{4^n} \sum\limits_{P_1, \dots, P_d \in \mathbb{P}_n} \left[\E_{\rand{L}_1, \dots, \rand{L}_d}  \gamma_{P_d}(\rand{L}_d) \cdots \gamma_{P_1}(\rand{L}_1) Tr\left(\mathcal{U}(\rand{L}_{1;d})^{-1}\mathcal{P}_{d}\mathcal{U}(\rand{L}_{d})\cdots \mathcal{P}_{2}\mathcal{U}(\rand{L}_{2})\mathcal{P}_{1}\mathcal{U}(\rand{L}_{1})\right)\right].  \label{eqn:expansion}
\end{align}
The sum in Eq.~\eqref{eqn:expansion} has $4^{nd}$ terms, each with a different sequence of $d$ Pauli superoperators $\mathcal{P}_1, \mathcal{P}_2, \dots, \mathcal{P}_d$, which represents a possible sequence of Pauli errors in a depth-$d$ circuit. We will separate these terms by the number of errors in the Pauli sequence, i.e. the number of indices $i$ such that $\mathcal{P}_i \neq \mathbb{I}_n$. Throughout this section, we will assume that indices for circuit layers and Paulis satisfy $1 \leq i \leq d$. We use the term \emph{error pattern} to refer to a description of the locations in a Pauli sequence where errors occur, which we describe by the set of indices $S$ such that $\mathbb{P}_i \neq \mathbb{I}_n$ if and only if $i \in S$. We call an error \emph{$k$-separated} if there are no errors within $k$ layers of the error, i.e., there is a $k$-separated error on layer $i$ if $P_i \neq \mathbb{I}$ and $j \notin S$ for all $j \neq i$ such that $|i-j|<k$. 

We expand Eq.~\eqref{eqn:expansion} by dividing the terms in the sum over Paulis $P_1, \dots, P_d$ up by their error patterns to get
\begin{align}
    F(\mathcal{E}_{\textrm{eff}}) & = \E_{\rand{L}_1, \dots, \rand{L}_d} \left[\prod_{i=1}^d \gamma_{\mathbb{I}_n}(\rand{L}_i)\right] +  \sum\limits_{j=1}^{d} c_j +  \sum\limits_{j=2}^{d} h_j \\ 
    & = \left(\E_{\rand{L}} \gamma_{\mathbb{I}_n}(\rand{L})\right)^d +  \sum\limits_{j=1}^{d} c_j +  \sum\limits_{j=2}^{d} h_j \label{eqn:c_expansion}
\end{align}
where $c_j$ is the sum of all terms in Eq.~\eqref{eqn:expansion} in which the error pattern has $|S| = j$  errors and contains a $k$-separated error, and $h_j$ is the sum of all terms in Eq.~\eqref{eqn:expansion} with an error pattern with $|S| = j$ that does not contain a $k$-separated error. By the cyclic property of the trace, all terms with exactly one error have no contribution to the fidelity:
\begin{align}
    c_1 & = \left(\frac{1}{4^n}\right)\sum\limits_{i_1=1}^d\sum\limits_{P_{i_1} \neq \mathbb{I}_n}\E_{\rand{L}_1, \dots, \rand{L}_d}  \left[ \left(\prod_{i \neq i_1}\gamma_{\mathbb{I}_n}(\rand{L}_i)\right)\gamma_{P_{i_1}}(\rand{L}_{i_1}) Tr\left(\mathcal{U}(\rand{L}_{1;d})^{-1}\mathcal{U}( \rand{L}_{i_1+1;d})\mathcal{P}_{i_1}\mathcal{U}(\rand{L}_{1;i_1})\right) \right] = 0.
\end{align}    
It remains to show that $c_j$ and $h_j$ are negligible for $j \geq 2$. First, we will use the scrambling condition on our gate set and sampling distribution to show that the $c_j$ are small.

We start by considering the terms that make up $c_2$. We break $c_2$ up into terms for each error pattern $\{i_1, i_2\}$ contributing to $c_2$, and we will bound the contribution of each of these terms:
\begin{equation}
    c_2 = \sum_{\substack{i_2 - i_1 > k}} c_{\{i_1, i_2\}}, \label{eqn:c2_expansion}
\end{equation}
where 
\begin{equation}
    c_{\{i_1, i_2\}} = \left(\frac{1}{4^n}\right)\sum_{P_{i_1}, P_{i_2} \in \mathbb{P}_n^{\ast}} \E_{\rand{L}_1, \dots, \rand{L}_d}  \left[ \left(\prod_{i \neq i_1, i_2}\gamma_{\mathbb{I}_n}(\rand{L}_i)\right)\gamma_{P_{i_1}}(\rand{L}_{i_1})\gamma_{P_{i_2}}(\rand{L}_{i_2}) Tr\left(\mathcal{U}(\rand{L}_d\cdots \rand{L}_1)^{-1}\mathcal{U}(\rand{L}_{i_2+1;d})\mathcal{P}_{i_2}\mathcal{U}( \rand{L}_{i_1+1;i_2})\mathcal{P}_{i_1}\mathcal{U}(\rand{L}_{1;i_1})\right) \right]. \label{eqn:c2_term}
\end{equation}
We will now use the scrambling condition [Eq.~\ref{eqn:scrambling}], to derive a bound on $c_{\{i_1, i_2\}}$. We can simplify Eq.~\eqref{eqn:c2_term} to
\begin{align}
     c_{\{i_1, i_2\}} & = \left(\frac{1}{4^n}\right)  \sum_{P_{i_1}, P_{i_2} \in \mathbb{P}_n^{\ast}}\E_{\rand{L}_1, \dots, \rand{L}_d} \left[ \left(\prod_{i \neq i_1, i_2}\gamma_{\mathbb{I}_n}(\rand{L}_i)\right)\gamma_{P_{i_1}}(\rand{L}_{i_1})\gamma_{P_{i_2}}(\rand{L}_{i_2}) Tr\left(\mathcal{U}(\rand{L}_{i_1+1;i_2})^{-1}\mathcal{P}_{i_2}\mathcal{U}(\rand{L}_{i_1+1;i_2})\mathcal{P}_{i_1}\right) \right]. \label{eqn:c2_1}
\end{align}
Now, we define the unitarily-rotated error channel
\begin{align}
\mathcal{U}(L_{i_2})^{-1} \mathcal{P}_{i_2} \mathcal{U}(L_{i_2}) =  \sum_{P,P'} d_{P, P'} (P \otimes P')\label{eqn:pauli_expansion_2error_case},
\end{align}
where $d_{P,P'}$ depends on $\mathcal{U}(L_{i_2})$ and $\mathcal{P}_{i_2}$, but we have suppressed this dependence. Substituting Eq.~\eqref{eqn:pauli_expansion_2error_case} into Eq.~\eqref{eqn:c2_1} and applying Lemma 1, we get
\begin{align}
     c_{\{i_1, i_2\}} & = \left(\frac{1}{4^n}\right)\sum_{P_{i_1}, P_{i_2}  \in \mathbb{P}_n^{\ast}}\left(\E_{\substack{\rand{L}_j\\ j < i_1, j>i_2}} \left[ \prod_{j}  \gamma_{\mathbb{I}_n}(\rand{L}_j)\right] \E_{\substack{\rand{L}_i\\ i_2 \geq i \geq i_1}} \prod_{i_2 > i > i_1}\left[ \gamma_{\mathbb{I}_n}(\rand{L}_i)\right]\gamma_{P_{i_1}}(\rand{L}_{i_1})\gamma_{P_{i_2}}(\rand{L}_{i_2}) Tr\left(\mathcal{U}(\rand{L}_{i_1+1;i_2})^{-1}\mathcal{P}_{i_2}\mathcal{U}(\rand{L}_{i_1+1;i_2})\mathcal{P}_{i_1}\right)  \right) \label{eqn:c2_2}\\
    & = \left(\frac{1}{4^n}\right)\sum_{P_{i_1}, P_{i_2}  \in \mathbb{P}_n^{\ast}}\left(\E_{\substack{\rand{L}_j\\ j < i_1, j>i_2}} \left[ \prod_{j}  \gamma_{\mathbb{I}_n}(\rand{L}_j)\right] \E_{\substack{\rand{L}_i\\ i_2 \geq i \geq i_1}} \prod_{i_2 > i > i_1}\left[ \gamma_{\mathbb{I}_n}(\rand{L}_i)\right]\gamma_{P_{i_1}}(\rand{L}_{i_1})\gamma_{P_{i_2}}(\rand{L}_{i_2}) \sum_{P, P' \in \mathbb{P}_n} d_{P, P'}Tr\left(\mathcal{U}(\rand{L}_{i_1+1;i_2-1})^{-1}(P \otimes P'^{\ast})\mathcal{U}(\rand{L}_{i_1+1;i_2-1})\mathcal{P}_{i_1}\right)  \right) \label{eqn:c2_3} \\
    & = \left(\frac{1}{4^n}\right)\sum_{P_{i_1}, P_{i_2}  \in \mathbb{P}_n^{\ast}}\left(\E_{\substack{\rand{L}_j\\ j < i_1, j>i_2}} \left[ \prod_{j}  \gamma_{\mathbb{I}_n}(\rand{L}_j)\right] \E_{\substack{\rand{L}_i\\ i_2 \geq i \geq i_1}} \prod_{i_2 > i > i_1}\left[ \gamma_{\mathbb{I}_n}(\rand{L}_i)\right]\gamma_{P_{i_1}}(\rand{L}_{i_1})\gamma_{P_{i_2}}(\rand{L}_{i_2}) \sum_{P \in \mathbb{P}_n} d_{P, P}Tr\left(\mathcal{U}(\rand{L}_{i_1+1;i_1-k})^{-1}\mathcal{P}\mathcal{U}(\rand{L}_{i_1+1;i_1-k})\mathcal{P}_{i_1}\right)  \right). \label{eqn:c2_5} 
\end{align}
We now apply the scrambling condition [Eq.~\eqref{eqn:scrambling}] and use the definition of the layer infidelity to obtain an upper bound on $c_{\{i_1, i_2\}}$:
\begin{align}
c_{\{i_1, i_2\}}    & \leq \sum_{P_{i_1}, P_{i_2}  \in \mathbb{P}_n^{\ast}}\left((1-\varepsilon)^{d-(i_2-i_1+1)} \E_{\substack{\rand{L}_i\\ i_2 > i \geq i_1+1}} \left[ \prod_{i}\gamma_{\mathbb{I}_n}(\rand{L}_i)\right]\E_{\substack{\rand{L}_{i_1}, \rand{L}_{i_2}}} \gamma_{P_{i_1}}(\rand{L}_{i_1})\gamma_{P_{i_2}}(\rand{L}_{i_2}) \sum_{P \in \mathbb{P}_n^{\ast}} d_{P, P}\delta  \right) \label{eqn:c2_6}\\
    & \leq \sum_{P_{i_1}, P_{i_2}  \in \mathbb{P}_n^{\ast}}\left((1-\varepsilon)^{d-(i_2-i_1+1)} \E_{\substack{\rand{L}_i\\ i_2 > i \geq i_1+1}} \left[ \prod_{i}\gamma_{\mathbb{I}_n}(\rand{L}_i)\right]\E_{\substack{\rand{L}_{i_1}, \rand{L}_{i_2}}} \gamma_{P_{i_1}}(\rand{L}_{i_1})\gamma_{P_{i_2}}(\rand{L}_{i_2}) \delta  \right) \label{eqn:c2_7}
\end{align}
Letting $\varepsilon = 1- \E_L [\gamma_{\mathbb{I}_n}(L)]$ be the expected layer infidelity, Eq.~\eqref{eqn:c2_7} becomes
\begin{equation}
 c_{\{i_1, i_2\}}   \leq (1-\varepsilon)^{d-k-2}\varepsilon^2 \delta. \label{eqn:c2_9}
\end{equation}
Eq.~\eqref{eqn:c2_9} bounds the value of each term in the expansion of $c_2$ given in Eq.~\eqref{eqn:c2_expansion}, and the number of terms in $c_{\{i_1, i_2\}}$ is bounded by $\nicefrac{d(d-k)}{2}$.

We now bound $c_j$ for $j > 2$. Again, we will start by bounding the contribution of each individual error pattern $S$ that contributes to $c_j$, i.e., each $S$ such that $|S|=j$ and $S$ contains a $k$-separated error. Let $S = \{i_1, i_2, \dots, i_j\}$ be an error pattern with $j > 2$ and $i_i < i_2 < \dots < i_j$, and assume that $i_{q+1} - i_q > k$ and $i_q - i_{q-1} > k$ for some $q < j$ (we will address cases in which the first or last error is $k$-separated later). We expand the errors before and after the $k$-separated error in layer $i_q$ in terms of tensor products of Paulis, then apply the properties of our layer distribution to bound the value of $c_S$. First, using the cyclic property of the trace, $c_S$ becomes
\begin{align}
c_S & = \left(\frac{1}{4^n}\right)\sum_{P_{i_1}, \dots, P_{i_j} \in \mathbb{P}_n^{\ast}}\E_{\rand{L}_1,\dots, \rand{L}_d} \left[ \left(\prod_{m \notin S}\gamma_{\mathbb{I}_n}(\rand{L}_m)\right)\left(\prod_{m \in S}\gamma_{P_m}(\rand{L}_m)\right) Tr\bigl(\mathcal{U}(\rand{L}_{i_1+1;i_j})^{-1} \mathcal{P}_{i_j} \cdots \mathcal{U}(\rand{L}_{i_2+1;i_3})\mathcal{P}_{i_2}\mathcal{U}(\rand{L}_{i_1+1;i_2})\mathcal{P}_{i_1}\bigr) \right] \label{eqn:cj_1} \\ 
 & = \left(\frac{1}{4^n}\right)\sum_{P_{i_1}, \dots, P_{i_j} \in \mathbb{P}_n^{\ast}}\E_{\rand{L}_1,\dots, \rand{L}_d}  \vast[ \left(\prod_{m \notin S}\gamma_{\mathbb{I}_n}(\rand{L}_m)\right)\left(\prod_{m \in S}\gamma_{P_m}(\rand{L}_m)\right) Tr\bigl(\mathcal{P}_{i_{q}}\mathcal{U}(\rand{L}_{i_q;i_{q-1}+1}) \cdots \mathcal{U}(\rand{L}_{i_2+1;i_3})\mathcal{P}_{i_2}\mathcal{U}(\rand{L}_{i_1+1;i_2})\mathcal{P}_{i_1} \nonumber \\
 & \hspace{6cm} \mathcal{U}(\rand{L}_{i_{1};i_j})^{-1} \mathcal{P}_{i_j} \mathcal{U}(\rand{L}_{i_{j-1}+1;i_{j}}) \cdots \mathcal{P}_{i_{q+1}} \mathcal{U}(\rand{L}_{i_{q}+1;i_{q+1}})\bigr)\vast]. \label{eqn:cj_2} 
\end{align}
We expand two unitaries, corresponding to the layers of the circuit before and after $k$ error-free layers, in terms of tensor products of Paulis:
\begin{align}
\mathcal{P}_{i_{q}}\mathcal{U}(\rand{L}_{i_{q-1}+1;i_q}) \cdots \mathcal{U}(\rand{L}_{i_2+1;i_3})\mathcal{P}_{i_2}\mathcal{U}(\rand{L}_{i_1+1;i_2})\mathcal{P}_{i_1} \mathcal{U}(\rand{L}_{i_{1}+1;i_q})^{-1} = \sum_{P,P' \in \mathbb{P}_n} b_{P, P'} (P \otimes P'^{\ast}) \label{eqn:pauli_expansion1} \\
\mathcal{U}(\rand{L}_{i_{q}+k+1;i_j})^{-1} \mathcal{P}_{i_j} \mathcal{U}(\rand{L}_{i_{j-1}+1;i_j}) \cdots \mathcal{U}(\rand{L}_{i_{q+1}+1;i_{q+2}}) \mathcal{P}_{i_{q+1}}\mathcal{U}(\rand{L}_{i_{q}+1;i_q+k+1}) = \sum_{P,P' \in \mathbb{P}_n} d_{P, P'} (P \otimes P'^{\ast})  \label{eqn:pauli_expansion2}
\end{align}
Using Eq.~\eqref{eqn:pauli_expansion1} and Eq.~\eqref{eqn:pauli_expansion2}, Eq.~\eqref{eqn:cj_2} becomes 
\begin{multline}
    c_S  =  \left(\frac{1}{4^n}\right)\sum_{P_{i_1}, \cdots P_{i_j} \in \mathbb{P}^{\ast}_n}\E_{\rand{L}_1,\dots, \rand{L}_d} \left[\left(\prod_{\substack{m \notin S \\ i_q + 1 \leq m \leq i_q+k}}\gamma_{\mathbb{I}_n}(\rand{L}_m)\right)\left(\prod_{\substack{m \notin S \\ m > i_q+k , m < i_q}}\gamma_{\mathbb{I}_n}(\rand{L}_m)\right)\left(\prod_{m \in S}\gamma_{P_m}(\rand{L}_m)\right)\right] \\
    \left[\sum_{P_1, P_2, P_3, P_4 \in \mathbb{P}_n} \left( b_{P_1, P_2} d_{P_3, P_4} Tr((P_1 \otimes P_2^{\ast}) \mathcal{U}(\rand{L}_{i_q+1;i_{q}+k})^{-1}(P_3 \otimes P_4^{\ast}) \mathcal{U}(\rand{L}_{i_q+1;i_{q}+k}))\right)\right] \label{eqn:cj_6} 
\end{multline}
Using Lemma 1, Eq.~\eqref{eqn:cj_6} becomes
\begin{multline}
c_S =  \left(\frac{1}{4^n}\right)\sum_{P_{i_1}, \dots, P_{i_j} \in \mathbb{P}_n^{\ast}}\E_{\rand{L}_1,\dots, \rand{L}_d} \left[\left(\prod_{\substack{m \notin S \\ i_q + 1 \leq m \leq i_q+k}}\gamma_{\mathbb{I}_n}(\rand{L}_m)\right)\left(\prod_{\substack{m \notin S \\ m > i_q+k , m < i_q}}\gamma_{\mathbb{I}_n}(\rand{L}_m)\right)\left(\prod_{m \in S}\gamma_{P_m}(\rand{L}_m)\right)\right] \\
  \left[ \sum_{P, P' \in \mathbb{P}_n^{\ast}}  \left(b_{P, P} d_{P', P'} Tr\left( \mathcal{P} \mathcal{U}(\rand{L}_{i_q+1;i_{q+1}})^{-1}\mathcal{U}(\rand{L}_{i_q+1;i_{q+1}})\mathcal{P}'\right) + Tr\left(\mathcal{U}(\rand{L}_{i_q+1;i_{q+1}})^{-1}\mathcal{U}(\rand{L}_{i_q+1;i_{q+1}})\right)b_{I,I}d_{I,I}\right)\right] \label{eqn:cj_7} 
\end{multline}
Because $\gamma_{\mathbb{I}_n}(L) \leq 1$ for all $L$,

\begin{equation}
c_S  \leq  \left(\frac{1}{4^n}\right)\sum_{P_{i_1}, \dots, P_{i_j} \in \mathbb{P}_n^{\ast}}\E_{\rand{L}_1,\dots, \rand{L}_d} \left[\left(\prod_{\substack{m \notin S \\ m > i_q+k , m < i_q}}\gamma_{\mathbb{I}_n}(\rand{L}_m)\right)\left(\prod_{m \in S}\gamma_{P_m}(\rand{L}_m)\right)\right] \left( \sum_{P, P' \in \mathbb{P}_n^{\ast}}  b_{P, P} d_{P', P'} Tr\left( \mathcal{U}(\rand{L}_{i_q+1;i_{q+1}})^{-1} \mathcal{P}\mathcal{U}(\rand{L}_{i_q+1;i_{q+1}})\mathcal{P}'\right) + 4^nb_{I,I}d_{I,I}\right). \label{eqn:cj_8} \\
\end{equation}
Applying Eq.~\eqref{eqn:scrambling}, we have
\begin{equation}
c_S \leq  \sum_{P_{i_1}, \dots, P_{i_j} \in \mathbb{P}_n^{\ast}} \E_{\rand{L}_1, \cdots \rand{L}_{i_q}}\E_{\rand{L}_{i_q+k+1}, \dots, \rand{L}_d}\left[\left(\prod_{\substack{m \notin S \\ m > i_q+k , m < i_q}}\gamma_{\mathbb{I}_n}(\rand{L}_m)\right)\left(\prod_{m \in S}\gamma_{P_m}(\rand{L}_m)\right)\right] \left(\sum_{P, P' \in \mathbb{P}_n^{\ast}}  b_{P, P} d_{P', P'} \delta + b_{I,I}d_{I,I}\right). \label{eqn:cj_9} \\ 
\end{equation}
We then bound sums of $d_{P',P'}$  and $b_{P, P}$ coefficients and use the fact that Eq.~\eqref{eqn:scrambling} implies $b_{I,I} \leq \delta$ to bound $c_s$ in terms of the average layer infidelity and $\delta$:
\begin{align}
c_S & \leq  \sum_{P_{i_1}, \dots, P_{i_j} \in \mathbb{P}_n^{\ast}}\E_{\rand{L}_1, \cdots \rand{L}_{i_{q}}}\E_{\rand{L}_{i_q+k+1}, \dots, \rand{L}_d} \left[\left(\prod_{\substack{m \notin S \\ m > i_q+k , m < i_q-k}}\gamma_{\mathbb{I}_n}(\rand{L}_m)\right)\left(\prod_{m \in S}\gamma_{P_m}(\rand{L}_m)\right)\right]  2\delta  \label{eqn:cj_10} \\ 
& \leq 2(1-\varepsilon)^{d-2k-j} \varepsilon^j \delta. \label{eqn:cj_11} 
\end{align}

Where the final inequality follows from the definition of $\varepsilon$. Eq.~\eqref{eqn:cj_11} bounds the contribution of the term with error pattern $S$ to the polarization of depth-$d$ $\Omega$-distributed random circuits. The number of possible error patterns $S$ with $|S| = j$ of the type considered in our argument above (i.e., error patterns with $|S|=j$ and contain a $k$-separated error that is not the first or last error) is bounded by $  k{d-2k \choose j}$.

There are two additional types of error pattern $S$ that contribute to $c_j$: (1) $i_1-i_2 > k$, and (2) $i_{q}-i_{q-1} > k$.
These two cases essentially reduce to the two error case. In case (1), all errors after the first error can be expanded in terms of the Pauli basis:
\begin{align}
c_S & = \left(\frac{1}{4^n}\right)\sum_{P_{i_1}, \dots, P_{i_j} \in \mathbb{P}_n^{\ast}}\E_{\rand{L}_1\cdots \rand{L}_d} \left[ \left(\prod_{m \notin S}\gamma_{I_m}\right)\left(\prod_{m \in S}\gamma_{P_m}(\rand{L}_m)\right) Tr(\mathcal{U}(\rand{L}_{i_1+1;i_j})^{-1} \mathcal{P}_{i_j} \cdots \mathcal{U}(\rand{L}_{i_2+1;i_3})\mathcal{P}_{i_2}\mathcal{U}(\rand{L}_{i_1+1;i_2})\mathcal{P}_{i_1}) \right] \label{eqn:edge_cj_1} \\ 
& = \left(\frac{1}{4^n}\right)\sum_{P_{i_1}, \dots, P_{i_j} \in \mathbb{P}_n^{\ast}}\E_{\rand{L}_1,\dots, \rand{L}_d}\sum_{P, P' \in \mathbb{P}_n}  \left[ \left(\prod_{m \notin S}\gamma_{I_m}\right)\left(\prod_{m \in S}\gamma_{P_m}(\rand{L}_m)\right) Tr(\mathcal{U}(\rand{L}_{i_1+1;i_2})^{-1}d_{P,P'} (P \otimes P') \mathcal{U}(\rand{L}_{i_1+1;i_2})\mathcal{P}_{i_1}) \right] \label{eqn:edge_cj_2} \\ 
& = \left(\frac{1}{4^n}\right)\sum_{P_{i_1}, \cdots P_{i_j}  \in \mathbb{P}_n^{\ast}}\E_{\rand{L}_1,\dots, \rand{L}_d}\sum_{P \in \mathbb{P}_n}\left[ \left(\prod_{m \notin S}\gamma_{I_m}\right)\left(\prod_{m \in S}\gamma_{P_m}(L_m)\right) Tr(\mathcal{U}(L_{i_1+1;i_2})^{-1}d_{P,P} \mathcal{P} \mathcal{U}(L_{i_1+1;i_2})\mathcal{P}_{i_1}) \right] \label{eqn:edge_cj_3} \\ 
& \leq (1-\varepsilon)^{d-k-j}\varepsilon^j \delta
\end{align}
The argument for case (2) is analogous, doing an expansion of all errors except the last error. For each case (1) and case (2), the number of possible error patterns $S$ with $|S| = j$ is bounded by $k  {d-k-1 \choose j-1}$.

As in the previous section, our arguments bound $c_S$ for each valid error pattern $S$ with $|S| \geq 3$. Therefore, 
\begin{align}
    \sum\limits_{j=1}^{d} c_j & < 2k\sum\limits_{j=2}^{d-k} {d-k-1 \choose j-1} (1-\varepsilon)^{d-k-j}\varepsilon^j\delta + (d-2k-2)\sum\limits_{j=2}^{d-2k} {d-2k-1 \choose j-1} (1-\varepsilon)^{d-2k-j}\varepsilon^j\delta \\
    & < 2k\varepsilon\sum\limits_{j=2}^{d-k} {d-k-1 \choose j-1} (1-\varepsilon)^{d-k-j}\varepsilon^j\delta + (d-2k-2)\varepsilon\sum\limits_{j=3}^{d-2k} {d-2k-1 \choose j-1} (1-\varepsilon)^{d-2k-j}\varepsilon^{j-1}\delta \\
    & < d\delta \varepsilon
    \label{eqn:delta_bound}
\end{align}
We need not consider circuits with depth larger than $O(\nicefrac{1}{\varepsilon})$, because the circuit depth at which the polarization becomes negligible is $O(\nicefrac{1}{\varepsilon})$---because when $d \varepsilon \gtrapprox 1$ at least one error is almost certain to occur. Therefore, Eq.~\eqref{eqn:delta_bound} implies that the $c_j$ terms have a negligible contribution to the fidelity $F(\mathcal{E}_{\textrm{eff}})$. 

We have bounded the contributions of the $c_j$ terms to $F(\mathcal{E}_{\textrm{eff}})$. We now argue that that contribution of the $h_j$ terms is also negligible. These terms represent error patterns in which every error is within $k$ layers of another error, so that our scrambling condition cannot guarantee that the probability of error cancellation is negligible. Instead, we will argue that the total probability of these error patterns is negligible. Because the contribution of an error pattern to the fidelity is bounded by its probability, this will bound the contribution of the $h_j$ terms to the fidelity.

\begin{align}
\sum\limits_{j=2}^d h_j &\leq \sum\limits_{l=1}^d Pr(\text{error in layer }l)Pr(\text{an error in layers }l-k, l-k+1, \dots, l-1, l+1, l+2, \dots, l+k\mid\text{error in layer }l) \\
\sum\limits_{j=2}^d h_j &\leq \sum\limits_{l=1}^d  \varepsilon Pr(\text{an error in layers }l-k, l-k+1, \dots, l-1, l+1, l+2, \dots, l+k) \\
\sum\limits_{j=2}^d h_j &\leq \sum\limits_{l=1}^d  k\varepsilon^2 \\
\sum\limits_{j=2}^d h_j &\leq dk\varepsilon^2 , 
\end{align}
Because we only consider $d = O(\nicefrac{1}{\varepsilon})$ (see argument above), we have $d\varepsilon = O(1)$. Since we have $k\varepsilon \ll 1$ from our scrambling condition, it follows that $dk\varepsilon^2$ is small, and hence $h_j$ is small.

We also note that as $d$ gets large, is highly likely that an error pattern contains a $k$-separated error, and we can show that the probability of no $k$-separated layers is exponentially suppressed. We can break a depth $d$ circuit into $\frac{d}{2k+1}$ pieces of $2k+1$ layers to bound the probability of there being no $k$-separated errors---if one of these blocks of $2k+1$ layers consists of an error with $k$ error-free layers before and after it, then there is a $k$-separated error. When $\nicefrac{d\varepsilon}{2k+1} > 1$, we can bound the probability of there being no block of this form (which we call a \emph{k-separated block}) using a Chernoff bound:
\begin{equation}
    Pr(\text{No k-separated block})  \leq \exp\left(-1-\frac{1}{2}\left[\frac{d}{2k+1}\varepsilon(1-\varepsilon)^{2k}+ \frac{2k+1}{d}\frac{1}{\varepsilon(1-\varepsilon)^{2k}}\right]\right) \label{eqn:chernoff}
\end{equation}
this bound implies that $h_j$ is bounded by an $O(e^{-\nicefrac{d\varepsilon}{k}})$ term, which is small in the regime where $dk\varepsilon \gg 1$.

Using the bound given by Eq.~\eqref{eqn:delta_bound} in Eq.~\eqref{eqn:c_expansion}, we obtain an expression for the fidelity of $\Omega$-distributed random circuits:
\begin{align}
    F(\mathcal{E}_{\textrm{eff}})& = \left(\E_{\rand{L}} \gamma_{\mathbb{I}_n}(\rand{L})\right)^d +  \sum\limits_{j=1}^{d} c_j +  \sum\limits_{j=2}^{d} h_j \\
    & = \left(\E_{\rand{L}} \gamma_{\mathbb{I}_n}(\rand{L})\right)^d + \tilde{\delta}_c + \tilde{\delta}_k \\
    & = (1-\varepsilon)^d + \tilde{\delta}
\end{align}
where $\varepsilon$ is the expected infidelity of a layer of an $\Omega$-distributed random circuit, and 
\begin{equation}
    \tilde{\delta} = \tilde{\delta}_c + \tilde{\delta}_k = O\left(d\varepsilon(\delta + k\varepsilon)\right),
\end{equation}
and $\tilde{\delta}$ is negligible for $d = O(\nicefrac{1}{\varepsilon})$. 
Because the average layer polarization is $\gamma = \frac{4^n}{4^n-1}(1-\varepsilon) - \frac{1}{4^n-1},$ and $n \gg 1$ (by assumption), $1-\varepsilon \approx \gamma$, up to a negligible $O(\nicefrac{1}{4^n})$ constant. Therefore, the polarization decays as
\begin{equation}
    \gamma(\mathcal{E}_{\textrm{eff}}) = p_{
\textrm{rc}}^d + \tilde{\delta}, 
\end{equation}
where
$p_{\textrm{rc}} = \mathbb{E}_{\rand{L}}\left(\gamma\left(\mathcal{E}(\rand{L})\right)\right)$ is the expected polarization of an $\Omega$-random layer, and $\tilde{\delta}$ is a negligible constant.

\section{MRB with Clifford Two-Qubit Gates}
\label{app:theory-c-2Q-gates}

The theory presented in Section \ref{sec:r_eps_relation} (which is presented in detail in Appendix~\ref{app:r_eps_relation}) assumes stochastic Pauli noise on each circuit layer to derive the exponential decay of the observed polarization of mirror circuits. However, stochastic error is not always the dominant error in a processor. Our method uses a randomized compilation procedure to convert error into stochastic Pauli error. In this appendix, we prove that when all two-qubit gates in an MRB experiment are Clifford, the error in randomized mirror circuits is twirled into Pauli stochastic error, under the assumption that the error map on the one-qubit gates is independent of the Paulis with which they are compiled.

We consider a depth-$d$ randomized mirror circuit (treated as a random variable), which we write as 
\begin{equation}
    \rand{M}_d = \rc{\rand{L}_0^{-1}}{ \rand{P}_{d+1}}{ \rand{P}_d^c}\tcp{\rand{L}_{\theta_1}^{-1}}{\rand{P}_d}\rc{\rand{L}_1^{-1}}{\rand{P}_{d}}{ \rand{P}_{d-1}^c} \cdots \rc{\rand{L}_{\nicefrac{d}{2}}^{-1}}{ \rand{P}_{\nicefrac{d}{2}+1}}{ \rand{P}_{\nicefrac{d}{2}}}
\rc{\rand{L}_{\nicefrac{d}{2}}}{ \rand{P}_{\nicefrac{d}{2}}}{ \rand{P}^c_{\nicefrac{d}{2}-1}}\cdots \rc{\rand{L}_1}{ \rand{P}_{2}}{ \rand{P}_1^c}\tcp{\rand{L}_{\theta_1}}{\rand{P}_0} \rc{\rand{L}_0}{\rand{P}_{0}}{}.
\end{equation}
When the two-qubit gate layers consist of two-qubit Cliffords of the form $\gate{CP}_{\theta}$,  they are not changed by the randomized compilation step of our circuit construction. Therefore, 
\begin{equation}
    \rand{M}_d  = \rc{\rand{L}_0^{-1}}{ \rand{P}_{d+1}}{ \rand{P}_d^c}\rand{L}_{\theta_1}^{-1}\rc{\rand{L}_1^{-1}}{\rand{P}_{d}}{ \rand{P}_{d-1}^c} \cdots \rc{\rand{L}_{\nicefrac{d}{2}}^{-1}}{ \rand{P}_{\nicefrac{d}{2}+1}}{ \rand{P}_{\nicefrac{d}{2}}}
\rc{\rand{L}_{\nicefrac{d}{2}}}{ \rand{P}_{\nicefrac{d}{2}}}{ \rand{P}^c_{\nicefrac{d}{2}-1}}\cdots \rc{\rand{L}_1}{ \rand{P}_{2}}{ \rand{P}_1^c}\rand{L}_{\theta_1} \rc{\rand{L}_0}{\rand{P}_{0}}{}, \label{eq:mc1}
\end{equation}
We will assume the error on the single-qubit gates is independent of the Paulis they are recompiled with---i.e., $\phi\bigl(\rc{\rand{L}_i}{P'}{P}\bigr)=\mathcal{E}(L_i)\mathcal{U}\bigl(\rc{\rand{L}_i}{P'}{P}\bigr)$. Using this assumption, an implementation of the circuit $\rand{M}_d$ can be written as
\begin{multline}
    \phi(\rand{M}_d) = \mathcal{E}(\rand{L_0^{-1}}) \mathcal{U}\bigl(\rc{\rand{L}_0^{-1}}{ \rand{P}_{d+1}}{ \rand{P}_d^c}\bigr)\mathcal{E}(\rand{L_{\theta_1}}^{-1}) \mathcal{U}(\rand{L}_{\theta_1}^{-1})\mathcal{E}(\rand{L}_1^{-1})\mathcal{U}\bigl(\rc{\rand{L}_1^{-1}}{\rand{P}_{d}}{ \rand{P}_{d-1}^c}\bigr)
    \cdots \mathcal{E}(\rand{L}_{\theta_{\nicefrac{d}{2}}}^{-1})\mathcal{U}(\rand{L_{\theta_{\nicefrac{d}{2}}}^{-1}})\mathcal{E}(\rand{L}_{\nicefrac{d}{2}}^{-1}) \mathcal{U}\bigl(\rc{\rand{L}_{\nicefrac{d}{2}}^{-1}}{ \rand{P}_{\nicefrac{d}{2}+1}}{ \rand{P}_{\nicefrac{d}{2}}}\bigr)\mathcal{E}(\rand{L}_{{\nicefrac{d}{2}}}) \\
    \mathcal{U}\bigl(\rc{\rand{L}_{\nicefrac{d}{2}}}{ \rand{P}_{\nicefrac{d}{2}}}{ \rand{P}^c_{\nicefrac{d}{2}-1}}\bigr)\mathcal{E}(\rand{L}_{\theta_{\nicefrac{d}{2}}})\mathcal{U}(\rand{L_{\theta_{\nicefrac{d}{2}}}})\mathcal{E}(\rand{L}_{{\nicefrac{d}{2}-1}})\mathcal{U}\bigl(\rc{\rand{L}_{\nicefrac{d}{2}-1}}{ \rand{P}_{\nicefrac{d}{2}-1}}{ \rand{P}^c_{\nicefrac{d}{2}-2}}\bigr) 
    \cdots \mathcal{E}(\rand{L}_{\theta_1}) \mathcal{U}(\rand{L}_{\theta_1}) \mathcal{E}(\rand{L}_0) \mathcal{U}\bigl(\rc{\rand{L}_0}{\rand{P}_{0}}{}\bigr). \label{eq:supop1}
\end{multline}
We now push the error on the single-qubit gate layers through the two-qubit gate layers, defining new error channels that represent the error on a composite layer. Eq.~\eqref{eq:supop1} becomes
\begin{multline}
    \phi(\rand{M}_d) = \mathcal{E}(\rand{L_0^{-1}}) \mathcal{U}\bigl(\rc{\rand{L}_0^{-1}}{ \rand{P}_{d+1}}{ \rand{P}_d^c}\bigr)\mathcal{E}'_{d}\mathcal{U}(\rand{L}_{\theta_1}^{-1})\mathcal{U}\bigl(\rc{\rand{L}_1^{-1}}{\rand{P}_{d}}{ \rand{P}_{d-1}^c}\bigr)
    \cdots \mathcal{E}'_{{\nicefrac{d}{2}+1}}\mathcal{U}\bigl(\rand{L_{\theta_{\nicefrac{d}{2}}}^{-1}}) \mathcal{U}(\rc{\rand{L}_{\nicefrac{d}{2}}^{-1}}{ \rand{P}_{\nicefrac{d}{2}+1}}{ \rand{P}_{\nicefrac{d}{2}}}\bigr)\mathcal{E}(\rand{L}_{{\nicefrac{d}{2}}}) \\
    \mathcal{U}\bigl(\rc{\rand{L}_{\nicefrac{d}{2}}}{ \rand{P}_{\nicefrac{d}{2}}}{ \rand{P}^c_{\nicefrac{d}{2}-1}}\bigr)\mathcal{E}'_{{\nicefrac{d}{2}}}\mathcal{U}(\rand{L_{\theta_{\nicefrac{d}{2}}}})\mathcal{U}\bigl(\rc{\rand{L}_{\nicefrac{d}{2}-1}}{ \rand{P}_{\nicefrac{d}{2}-1}}{ \rand{P}^c_{\nicefrac{d}{2}-2}}\bigr) 
    \cdots  \mathcal{E}'_1 \mathcal{U}(\rand{L}_{\theta_1})\mathcal{U}\bigl(\rc{\rand{L}_0}{\rand{P}_{0}}{}\bigr),  \label{eq:supop2}
\end{multline}
where 
\begin{align}
\mathcal{E}'_i & = \mathcal{E}(\rand{L}_{\theta_i})\mathcal{U}(\rand{L}_{\theta_i})\mathcal{E}(\rand{L_{i-1}})\mathcal{U}(\rand{L}_{\theta_i})^{-1} & 1 \leq i \leq \frac{d}{2} \\
\mathcal{E}'_i & = \mathcal{E}(\rand{L}_{\theta_{d-i+1}})\mathcal{U}(\rand{L}_{\theta_{d-i+1}})^{-1}\mathcal{E}(\rand{L}_{i+1})\mathcal{U}(\rand{L}_{\theta_{d-i+1}}) & \frac{d}{2} <i \leq d.
\end{align}

We have grouped the error channels into error channels $\mathcal{E}_i'$ that represent the error in a composite layer. Now, we use the structure of the randomized compilation procedure to twirl the error. The dressed layers can be expanded in terms of the original sampled layer and the Paulis inserted in randomized compilation as 
\begin{align}
    \mathcal{U}(\rand{L}_{\theta_i})\mathcal{U}\bigl(\rc{\rand{L}_{i-1}}{\rand{P}_{i-1}}{\rand{P}_{i-2}^c}\bigr) & = \mathcal{U}\bigl((\rand{P}_{i}^c)^{-1}\rand{L}_{\theta_i}\rand{L}_{i-1}\rand{P}_{i-1}^c\bigr) & 1 \leq i \leq \frac{d}{2} \label{eqn:dressed_layer_forward}\\
     \mathcal{U}(\rand{L}_{\theta_{d-i+1}}^{-1})\mathcal{U}\bigl(\rc{\rand{L}_{d-i+1}^{-1}}{\rand{P}_{i+1}}{\rand{P}_{i}^c}\bigr)& = \mathcal{U}\bigl((\rand{P}^c_{i+1})^{-1}\rand{L}_{\theta_{d-i+1}}^{-1}\rand{L}_{d-i+1}^{-1}\rand{P}^c_{i}\bigr) & \frac{d}{2} <i \leq d. \label{eqn:dressed_layer_inv}
\end{align}
Rewriting Eq.~\eqref{eq:supop2} using these expansions, we have
\begin{multline}
    \phi(\rand{M}_d) = \mathcal{E}(\rand{L_0^{-1}}) \mathcal{U}(\rand{P}_{d+1}\rand{L}_0^{-1} \rand{P}_d^c)\mathcal{E}'_{d}\mathcal{U}\bigl((\rand{P}^c_d)^{-1}\rand{L}_{\theta_1}^{-1}\rand{L}_1^{-1}\rand{P}^c_{d-1}\bigr)
    \cdots \mathcal{E}'_{{\nicefrac{d}{2}+1}}\mathcal{U}\bigl((\rand{P}^c_{\nicefrac{d}{2}+1})^{-1}\rand{L}_{\theta_{\nicefrac{d}{2}}}^{-1}\rand{L}_{\nicefrac{d}{2}}^{-1}\rand{P}^c_{\nicefrac{d}{2}}\bigr)\mathcal{E}(\rand{L}_{{\nicefrac{d}{2}}}) \\
    \mathcal{U}\bigl((\rand{P}_{\nicefrac{d}{2}}^c)^{-1}\rand{L}_{\theta_{\nicefrac{d}{2}}}\rand{L}_{\nicefrac{d}{2}-1}\rand{P}_{\nicefrac{d}{2}-1}^c\bigr)\mathcal{E}'_{{\nicefrac{d}{2}}}
    \mathcal{U}\bigl((\rand{P}_{\nicefrac{d}{2}-1}^c)^{-1}\rand{L}_{\theta_{\nicefrac{d}{2}-1}}\rand{L}_{\nicefrac{d}{2}-2}\rand{P}_{\nicefrac{d}{2}-1}^c\bigr)  \cdots  \mathcal{E}'_1 \mathcal{U}\bigl((\rand{P}_{1}^c)^{-1}\rand{L}_{\theta_1}\rand{L}_0\bigr),  \label{eq:supop3}
\end{multline}
where each correction layer $\rand{P}_i^c$ is a uniform random Pauli, because the two-qubit gates are Clifford. Averaging over the uniform random $n$-qubit Paulis $\rand{P}_0, \rand{P}_1, \dots, \rand{P}_d$, which equivalently averages over the correction Paulis $\rand{P}^c_1, \dots, \rand{P}^c_{d+1}$, performs a Pauli twirl, converting the error channels into Pauli stochastic error channels. Performing this average, Eq.~\eqref{eq:supop3} becomes
\begin{equation}
    \phi(\rand{M}_d) = \mathcal{E}(\rand{L_0^{-1}}) \mathcal{U}(\rand{P}_{d+1}\rand{L}_0^{-1})\mathcal{S}_{d}\mathcal{U}(\rand{L}_{\theta_1}^{-1}\rand{L}_1^{-1})
    \cdots \mathcal{S}_{{\nicefrac{d}{2}+1}}\mathcal{U}(\rand{L}_{\theta_{\nicefrac{d}{2}}}^{-1}\rand{L}_{\nicefrac{d}{2}}^{-1})\mathcal{S}(\rand{L}_{{\nicefrac{d}{2}}})\mathcal{U}(\rand{L}_{\theta_{\nicefrac{d}{2}}}\rand{L}_{\nicefrac{d}{2}-1})\mathcal{S}_{{\nicefrac{d}{2}}}
    \mathcal{U}(\rand{L}_{\theta_{\nicefrac{d}{2}-1}}\rand{L}_{\nicefrac{d}{2}-2})  \cdots  \mathcal{S}_1 \mathcal{U}(\rand{L}_{\theta_1}\rand{L}_0),  \label{eqn:supop4}
\end{equation}
where $\rand{S}_i = \E_{\rand{P}}\rand{P}\mathcal{E}'_i\rand{P}^{-1}$ and $\mathcal{S}(\rand{L}_{\nicefrac{d}{2}}) = \E_{\rand{P}}\rand{P}\mathcal{E}(\rand{L}_{\nicefrac{d}{2}})\rand{P}^{-1}$ are stochastic Pauli channels, each of which captures the error from one composite layer.
All error, except the error on the final circuit layer, is twirled into stochastic Pauli noise by the random Paulis inserted in randomized compilation. Therefore, we expect our method to be sensitive to all errors when the two-qubit gates are chosen to be Clifford gates. 

\section{MRB with non-Clifford two-qubit gates}

In this appendix we show that, when applied to a gate set containing non-Clifford two-qubit gates, MRB is sensitive to all Hamiltonian errors on those two-qubit gates except one linear combination of errors on a non-Clifford two-qubit gate and its inverse. We then discuss possible adaptations to our protocol that would guarantee sensitivity to all Hamiltonian errors on non-Clifford two-qubit gates. 

\subsection{Sensitivity of errors in non-Clifford two-qubit gates} \label{app:theory-nc-2Q-gates}
In Appendix \ref{app:theory-c-2Q-gates}, we showed that when the two-qubit gate set used in MRB contains only Clifford gates, the error in the two-qubit gates is twirled, upon averaging, into stochastic Pauli noise. This guarantees sensitivity to general errors on the two-qubit gates. We now consider circuits with non-Clifford two-qubit gates and show that randomized mirror circuits are sensitive to most Hamiltonian errors on the two-qubit gates, to first order. We will assume there is no crosstalk error, and all two-qubit layers are sampled independently, so that we expect the only systematic coherent cancellation of errors to come from a layer and its inverse. We will also assume there is no error on the single-qubit gates. To see the effect of error in a two-qubit gate on a randomized mirror circuit to first order, it is sufficient to consider mirror circuits resulting from applying our circuit construction procedure to a single two-qubit composite layer $\gate{L} = \gate{L_{1}L_{\theta}}$, where $\gate{L_{\theta}}=\gate{CP}_{\theta}$ is a two-qubit gate and $\gate{L_1}$ is a one-qubit gate layer. After mirroring and randomized compilation on $\gate{L}$, we have the circuit $M = \tcp{L_{\theta}^{-1}}{P_2}\rc{L_1^{-1}}{P_2}{P_1^c}\rc{L_1}{P_1}{P_0^c}\tcp{L_{\theta}}{P_0}$, where $P_0, P_1,$ and $P_2$ are random two-qubit Pauli layers. The ideal operation $\gate{M}$ implements is $\mathcal{U}(M) = (\gate{P}^c_2)^{-1}  P_{0}$. An imperfect implementation of $\gate{M}$ can be expressed as
\begin{align}
   \phi(\gate{M}) & = \mathcal{E}\bigl(\tcp{L_{\theta}^{-1}}{\gate{P}_{2}}\bigr)\mathcal{U}\bigl(\tcp{L_{\theta}^{-1}}{P_2}\rc{L_1^{-1}}{P_2}{P_1^c}\rc{L_1}{P_1}{P_0^c}\bigr) \mathcal{E}\bigl(\tcp{\gate{L_{\theta}}}{\gate{P}_{0}}\bigr)\mathcal{U}\bigl(\tcp{L_{\theta}}{P_0}\bigr) \\ 
    & = \mathcal{E}\bigl(\tcp{L_{\theta}^{-1}}{\gate{P}_{2}}\bigr)\mathcal{U}\bigl(\tcp{L_{\theta}^{-1}}{P_2} P_{2} P^c_{0} \bigr) \mathcal{E}\bigl(\tcp{\gate{L_{\theta}}}{\gate{P}_{0}}\bigr)\mathcal{U}\bigl(\tcp{\gate{L_{\theta}}}{\gate{P}_{0}}\bigr) \\
    & = \mathcal{E}\bigl(\tcp{L_{\theta}^{-1}}{\gate{P}_{2}}\bigr) \mathcal{U}\bigl((P^c_2)^{-1} L_{\theta}^{-1} P^c_{0}\bigr) \mathcal{E}\bigl(\tcp{\gate{L_{\theta}}}{\gate{P}_{0}}\bigr)\mathcal{U}\bigl(\tcp{\gate{L_{\theta}}}{\gate{P}_{0}}\bigr)\\
    & = \mathcal{E}\bigl(\tcp{L_{\theta}^{-1}}{\gate{P}_{2}}\bigr) \mathcal{U}\bigl((\gate{P}^c_2)^{-1}   P_{0} \tcp{\gate{L_{\theta_i}}}{\gate{P}_{0}}^{-1} (P^c_{0})^{-1} P^c_{0}\bigr) \mathcal{E}\bigl(\tcp{\gate{L_{\theta}}}{\gate{P}_{0}}\bigr)\mathcal{U}\bigl(\tcp{\gate{L_{\theta}}}{\gate{P}_{0}}\bigr)\\
    & = \mathcal{E}\bigl(\tcp{L_{\theta}^{-1}}{\gate{P}_{2}}\bigr) \mathcal{U}\bigl((\gate{P}^c_2)^{-1}  P_{0} \tcp{\gate{L_{\theta}}}{\gate{P}_{0}}^{-1}\bigr) \mathcal{E}\bigl(\tcp{\gate{L_{\theta}}}{\gate{P}_{0}}\bigr)\mathcal{U}\bigl(\tcp{\gate{L_{\theta}}}{\gate{P}_{0}}\bigr), \label{eqn:2q_op}
\end{align}
where we have used the definitions of the two-qubit gate layer $\tcp{L_{\theta}}{P}$ and the correction layers $P_i^c$ to rewrite the unitary evolution. 

We now consider the effect of general gate-dependent Hamiltonian errors on the two-qubit gates on $\phi(M)$. We will write the error in terms of elementary error generators, as defined in the error generator formalism of Ref. \cite{blume2021taxonomy}. We model the error on each two-qubit gate $\gate{g}$ as $\mathcal{E}(\gate{g}) = e^{M_{g}}$, where 
\begin{equation}
M_g = \sum\limits_{P_a,P_b}\varepsilon_{P_a,P_b}^{g}H_{P_a,P_b},
\end{equation}
and where $H_{P_a,P_b}$ is the two-qubit Hamiltonian error generator indexed by the Pauli operators $P_a$ and $P_b$. Using this expression for the error and expanding Eq.~\eqref{eqn:2q_op} to first order in the error rates $\varepsilon^g_{P_a, P_b}$, we have
\begin{equation}
   \phi(\gate{M}) \approx \mathcal{U}\bigl((P_{2}^c)^{-1}P_{0}\bigr)\left(\mathcal{U}(\gate{I}) + \left(\sum\limits_{P_a,P_b}\varepsilon_{P_a,P_b}^{\tcp{\gate{L}_{-\theta}}{\gate{P}_{2}}}H_{P_a,P_b}\right) + \mathcal{U}\bigl( \tcp{\gate{L_{\theta}}}{\gate{P}_{0}}^{-1}\bigr) \left( \sum\limits_{P_a,P_b}\varepsilon_{P_a,P_b}^{\tcp{\gate{L_{\theta}}}{\gate{P}_{0}}}H_{P_a,P_b}\right) \mathcal{U}\bigl(\tcp{\gate{L_{\theta}}}{\gate{P}_{0}}\bigr)\right). \label{eq:target_plus_error}
\end{equation}
Eq.~\eqref{eq:target_plus_error} expresses the implementation of $\gate{M}$ in terms of its target evolution and a first order correction. The circuit is insensitive to an error to first order when the correction term vanishes, which occurs when
\begin{equation}
    \left(\sum\limits_{P_a,P_b}\varepsilon_{P_a,P_b}^{\tcp{\gate{L}_{-\theta}}{\gate{P}_{2}}}H_{P_a,P_b}\right)  + \mathcal{U}\bigl(\tcp{\gate{L_{\theta}}}{\gate{P}_{0}}^{-1}\bigr) \left( \sum\limits_{k}\varepsilon_{P_a,P_b}^{\tcp{\gate{L_{\theta_i}}}{\gate{P}_{i-1}}}H_{P_a,P_b}\right) \mathcal{U}\bigl(\tcp{\gate{L_{\theta}}}{\gate{P}_{0}}\bigr) = 0. \label{eqn:first_order_approx}
\end{equation}

Satisfying Eq.~\eqref{eqn:first_order_approx} requires that the coefficient of each elementary error generator $H_{P_a,P_b}$ is 0, which results in a system of 15 linear equations for each of $16^2$ choices of two-qubit random Paulis $\gate{P_{2}}, \gate{P_{0}}$ used in randomized compilation. The randomized mirror circuits are sensitive to an error if for some choice of $\gate{P_0}$ and $\gate{P_2}$, the system cannot be satisfied when the error is nonzero. The two-qubit gate set $\mathbb{G}_2$ is closed under inverses, so in addition to mirroring $\gate{L} = \gate{L_1L_{\theta}}$ as we have done above, we can mirror $\gate{L_1L_{-\theta}}$ to get an analogous set of linear equations. Considering all of the equations from mirroring $\gate{L_1L_{\theta}}$ and $\gate{L_1L_{-\theta}}$, we have a system of $2 \times 16^2 \times 15$ linear equations. The solutions to this system are $\varepsilon_{P_a,P_b}^{\theta} = \varepsilon_{P_a,P_b}^{-\theta} = 0$ $\forall (P_a, P_b) \neq (P , P)$ and $\varepsilon^G_{P,P} = \varepsilon^{G^{\dag}}_{P,P}.$ This means that to first order, the mirror circuits are not sensitive to the sum of $H_{P,P}$ errors on $\gate{CP_{\theta}}$ and $\gate{CP_{-\theta}}$, as we can change $\varepsilon_{P,P}^{G} + \varepsilon_{P,P}^{G^{\dag}}$ without changing the error in any of the mirror circuits. This is a result of the structure we use for our mirror circuits. Below, we discuss how our method can be adapted to address this insensitivity.

\subsection{Adaptations of MRB}
\label{app:adapations}
While our simulations and experiments suggest non-Clifford MRB is a robust method, when our randomized mirror circuits contain non-Clifford two-qubit gates they are not sensitive to some coherent errors on these gates. In Appendix \ref{app:theory-nc-2Q-gates} we showed that MRB circuits containing non-Clifford two-qubit gates are not sensitive to one linear combination of the Hamiltonian errors in these gates because of the correlations between the randomized compilation and the two-qubit gate that is applied, which prevent error from being perfectly twirled into stochastic noise. This shortcoming in our method is due to our choice of structure for our randomized mirror circuits. However, circuit mirroring is a flexible technique that can be applied to a variety of circuit structures, and here we discuss several adaptations of our method utilizing this flexibility that would address the error insensitivity in MRB.

Our method involves sampling random circuits with layers sampled from a user-specified distribution $\Omega$ over circuit layers. Different choices of circuit structure can address the shortcomings of our method and make other scalable benchmarks. We could guarantee sensitivity to all errors with more complex sampling of the $\Omega$-distributed random circuit.
For example, to benchmark a two-qubit gate set $\mathbb{G}_{2} = \{ \cs , \csd \}$ we could generate circuits containing $\cs$, $\csd$ and $
\cz$ gates and implement the $\cz$ gate by two consecutive $\cs$ or $\csd$ gates. This MRB experiment would be sensitive to the $H_{Z,Z}$ errors on the $\cs$ and $\csd$ gates that our MRB experiment is insensitive to (see above).

Our MRB protocol performs inversion layer-by-layer, and an alternative method to guarantee sensitivity to all errors is to use more complex inversion strategies that reduce the correlation in the gate layers in the two halves of a mirror circuit. One option is to invert multiple circuit layers at a time, through computing the inverse of the layers and compiling an inverse circuit---and similar ideas to this have recently been used to implement RB of continuously parameterized gates \cite{Shaffer2022-ho}. However, compilation can be computationally-intensive with many qubits. Alternatively, we can modify the inversion layers by adding in additional gates, while maintaining a circuit that is logically equivalent to the inverse.

\section{Simulations of MRB}
\label{app:simulations}

In this appendix, we provide further details about our simulations of MRB, which are discussed in Section~\ref{sec:theory-nc-2Q-gates}.

\subsection{Error Models for MRB Simulations}
\label{app:sim_models}
We simulated MRB with three classes of error models---stochastic, Hamiltonian, and stochastic+Hamiltonian. Our models are defined based on the error generator formalism in \cite{blume2021taxonomy}. Error rates are specified as elementary error generators of a post-gate error map. We include qubit-dependent Hamiltonian errors and Pauli stochastic errors on the $\gate{x}_{\nicefrac{\pi}{2}}$ and single-qubit idle gates with Hamiltonian error rates sampled in the range $[0, \nicefrac{h}{10}]$, and stochastic Pauli error rates sampled in the range $[0, \nicefrac{s}{10}]$. The stochastic and Hamiltonian errors are each split randomly across the three Paulis. We also include qubit-dependent Hamiltonian errors and Pauli stochastic errors on the $\cs$ and $\csd$ gates with Hamiltonian error rates sampled in the range $[0, h]$, and Pauli stochastic error rates sampled in the range $[0, s]$, spread at random across the 15 two-qubit Pauli errors. 

To generate error models, we start with an overall error parameter $p$ and select $s,h$ such that $h^2+s=p$. We generate models with $p \in [0.001, 0.2475]$ for 150 evenly-spaced values for the 1-qubit models and $p \in [0.0001, 0.075]$ for 150 evenly-spaced values for the 2- and 4-qubit models. In the stochastic error models, we set $h=0$. In the Hamiltonian error models, we set $s=0$. In the stochastic+Hamiltonian error models, we generate $s \in [0, p]$ at random, and set $h = \sqrt{p-s}$. 

For each error model, we run a randomly-generated set of MRB circuits consisting of $K=300$ circuits at each benchmark depth $d \in \{2^j \mid 0 \leq j \leq 8\}$. We approximate the error rate in $\Omega$-distributed random circuits ($\epsilon_{\Omega}$) via sampling. For each depth  $d \in \{2^j \mid 0 \leq j \leq 8\}$, we ran $K$ randomly-generated depth-$\nicefrac{d}{2}$ $\Omega$-distributed random circuits, each followed by a perfect projective measurement onto the target state.

\subsection{Simulations of MRB with measurement error}
\label{app:spam_error}
\begin{figure}
    \centering  \includegraphics{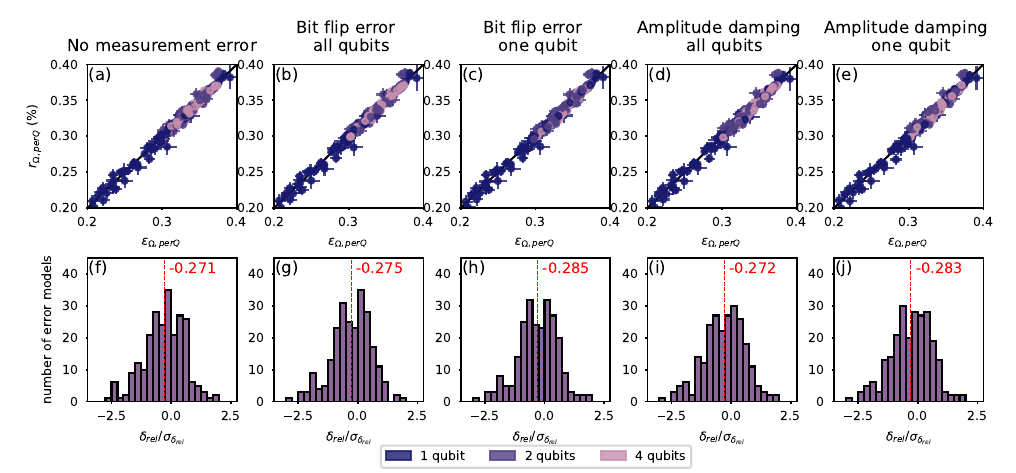}
    \caption{\textbf{Investigating the reliability of MRB in the presence of measurement error.} We simulated MRB with error models with no measurement error (a,f) and with four types of measurement error: (e,j) Amplitude-damping error on one qubit, (d,i) amplitude damping error distributed across all qubits, (c,h) bit flip error on one qubit, and (b,g) bit flip error distributed across all qubits.  For all simulations, we used the gate set $\mathbb{G} = (\{\cs, \csd \}, \mathbb{SU}(2))$ and randomly-generated Hamiltonian and stochastic gate errors. (a)-(e): We compare the MRB error rate per qubit $r_{\Omega, \textrm{perQ}}$ to the actual per-qubit error rate $\epsilon_{\Omega, \textrm{perQ}}$, estimated via sampling. We observe close agreement between the MRB error rate and $\epsilon_{\Omega}$ with each type of measurement error. (f)-(j): The distribution of the relative error $\delta_{\textrm{rel}} = \nicefrac{(r_{\Omega, \,\textrm{perQ}} - \epsilon_{\Omega, \,\textrm{perQ}})}{\epsilon_{\Omega, \,\textrm{perQ}}}$ divided by its uncertainty $\sigma_{\delta_{\textrm{rel}}}$, for models with each type of measurement error. The mean $\delta_{\textrm{rel}}$ (red dashed line) is shown for each distribution.}
    \label{fig:spam_error_1}
\end{figure}

\begin{figure}
    \centering
    \includegraphics{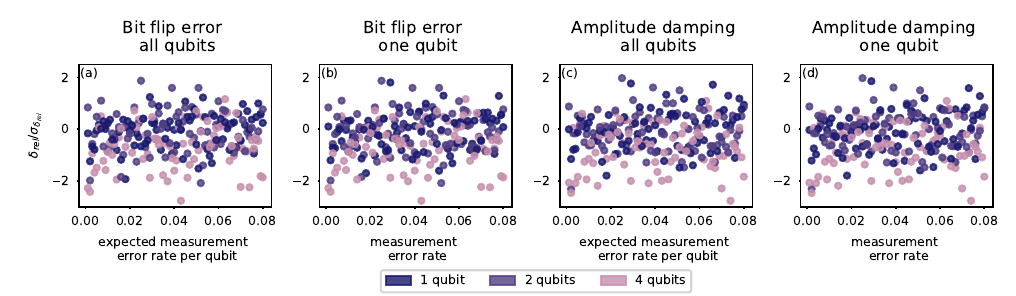}
    \caption{\textbf{MRB with varied-strength measurement error}  We simulated MRB with the gate set $\mathbb{G} = (\{\cs, \csd \}, \mathbb{SU}(2))$ with randomly-generated gate errors and four types of measurement error: (a) bit flip error distributed across all qubits, (b) bit flip error on one qubit, (c) amplitude damping error distributed across all qubits, (d) Amplitude-damping error on one qubit. We plot the relative error in the MRB error rate $\delta_{\textrm{rel}} = \nicefrac{(r_{\Omega, \,\textrm{perQ}} - \epsilon_{\Omega, \,\textrm{perQ}})}{\epsilon_{\Omega, \,\textrm{perQ}}}$ divided by its uncertainty $\sigma_{\delta_{\textrm{rel}}}$, versus the strength of the measurement error. We observe no systematic change in the error in the MRB error rate with as the strength of measurement error changes.}
    \label{fig:spam_error_2}
\end{figure}

We simulated MRB with two types of measurement error---amplitude damping error and bit flip error. For each type of measurement error, we performed simulations where only a single qubit had measurement error, and where each qubit had measurement error. We compare these results to simulations of MRB with no measurement error. 

To generate each error models, we sampled a stochastic+Hamiltonian gate error models via the method in Appendix~\ref{app:gate_errors} with overall error parameter $p$, with p=0.1 for single-qubit error models and p=0.02 for 2- and 4-qubit error models. We define our measurement error using the  single-qubit elementary error generators $S_X$, $S_Y$, and $A_{X,Y}$ defined in Ref.~\cite{blume2021taxonomy}. To add bit flip error of strength $p_m$ to a qubit, we apply the error $\exp(p_m S_x)$ immediately before measurement.  To add amplitude damping error to a qubit, we apply the error $\exp(p_m(S_X+S_Y-A_{X,Y}))$ immediately before measurement. We used 80 evenly-spaced values of $p_m \in [0.0001, 0.0801]$.  We ran simulations with measurement error on a single qubit and on all qubits. In models with measurement error on all qubits, we sampled a uniform random value $p \in [0, 2p_m]$ for each qubit, and apply error of strength $p$ to that qubit. In models with measurement error on one qubit, we apply a fixed measurement error of strength $p_m$ to one qubit.

Fig.~\ref{fig:spam_error_1} shows the results of these simulations. We observe that $r_{\Omega} \approx \epsilon_{\Omega}$ across all types of measurement error models we sampled, providing evidence that measurement error does not significantly impact the performance of MRB.  Furthermore, the distribution of $\delta_{\textrm{rel}}/\sigma_{\delta_{\textrm{rel}}}$, the relative error in MRB error rates divided by its uncertainty, is similar for all types of error model we tested. The mean $\delta_{\textrm{rel}}/\sigma_{\delta_{\textrm{rel}}}$ differs by under 0.015 between error models with and without measurement error. Fig.~\ref{fig:spam_error_2} shows $\delta_{\textrm{rel}}/\sigma_{\delta_{\textrm{rel}}}$ as a function of the measurement error parameter $p$. We observe no systematic change in the error in MRB with the rate of measurement error. 

\section{AQT experiments}\label{app:aqt}

In this appendix we provide further details about our experiments on AQT, which are discussed in Section~\ref{sec:aqt}.

\subsection{Experiment design}
A specific set of MRB circuits for a given $(\mathbb{G}_2,\mathbb{G}_1,\mathbb{Q}, \xi)$ is obtained by sampling $K$ circuits at a set of benchmark depths. We used $K=30$ and a set of exponentially spaced benchmarking depths ($d = 0,2,4,8\dots$). For the MRB designs in which $\xi=\nicefrac{1}{2}$ we did not independently sample the circuits for the three different gate sets. Instead, to sample a depth-$d$ circuit on qubits $\mathbb{Q}$ for each of our three gate sets:
\begin{enumerate}
    \item We sampled a depth $\nicefrac{d}{2}$ circuit $C$ for the $(\{\cz\}, \mathbb{SU}(2))$ gate set. 
    \item We created a correlated sample for the $( \{\cz\}, \mathbb{C}_1)$ gate set by replacing the $\mathbb{SU}(2)$ gates in $C$ with gates sampled from $\mathbb{C}_1$. \item We created a correlated sample for the $(\{\cs, \csd \}, \mathbb{SU}(2))$ gate set by replacing each $\cz$ gate in $C$ with either $\cs$ or $\csd$ at random.
    \item We independently converted each of the three circuits in (1)-(3) into a randomized mirror circuit.
\end{enumerate}
 Because the marginal distribution for the sampling of each circuit set is unaffected by this procedure, it does not impact the RB error rates we estimate (except by correlating their uncertainties), but it allows us to perform an interesting per-circuit comparison (see Appendix \ref{app:per_c}). 

To enable comparison to an established technique, we also ran \emph{direct} RB circuits, which are described in Section~\ref{sec:exp_drb}. This resulted in a total of 16,194 circuits. In the experiment, we randomized the order of this circuit list, and ran each circuit 1000 times in turn. We repeated this three times in succession, to enable us to look for substantial changes in a circuit's success probability that signify drift. We used standard statistical testing methods \cite{rudinger2018probing} to identify circuits in which the success probability changed between the three runs, and discarded that data. We performed qutrit classification in the readout, i.e., the readout was calibrated to resolve the `2' leakage state from the two computational basis states. Whenever a circuit output `2', the result was discarded.

\subsection{Comparing MRB circuits with Clifford and Haar-random single-qubit gates}
\label{app:per_c}

\begin{figure}
    \centering
    \includegraphics{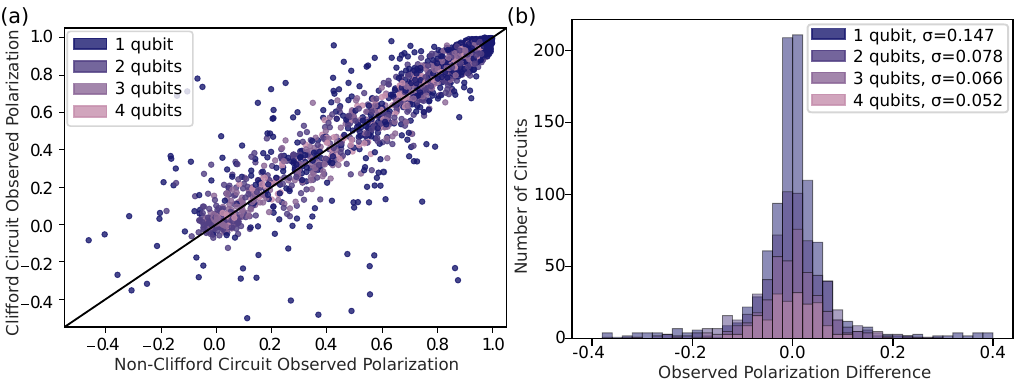}
    \caption{\textbf{Comparing the success rates of circuits that differ only by virtual phase gates.} Our experiments were designed so that each randomly sampled circuit containing $\cz$ and single-qubit Clifford gates [the $(\{\cz\},\mathbb{C}_1)$ gate set] differs from a randomly sampled circuit containing $\cz$ and Haar-random single-qubit unitaries [the $(\{\cz\},\mathbb{SU}(2))$ gate set] only by the angles in its $\z_{\theta}$ gates. Here we compare the observed polarization of each pair of circuits that differ only by the angles in these virtual $\z_{\theta}$ gates. For many of these circuit pairs $(C_1,C_2)$, $C_1$ and $C_2$ have very different observed polarizations, meaning that they have very different success rates. (a): The observed polarization for the $(\{\cz\},\mathbb{SU}(2))$ circuits and their corresponding Clifford circuits. (b): The difference in observed polarization between $(\{\cz\},\mathbb{SU}(2))$ circuits and their corresponding Clifford circuits.}
    \label{fig:per-circuit-comp}
\end{figure}

The observed similarity between the average success rates of circuits in which the single qubit-gate gates $\g{u}(\theta, \phi, \lambda)$ are sampled from two different distributions (see Section~\ref{sec:aqt_results}) does \emph{not} imply that the success rate of an individual circuit is independent of the values of $\theta$, $\phi$ and $\lambda$ in its $\g{u}(\theta, \phi, \lambda)$ gates. We designed our experiments so that each circuit in the $(\mathbb{SU}(2),\{\cz\},\mathbb{Q},\nicefrac{1}{2})$ experiment is identical to a circuit in the $(\mathbb{C}_1,\{\cz\},\mathbb{Q},\nicefrac{1}{2})$ experiment except for the values of each of the single-qubit gate's parameters. We can therefore use our data for each such circuit pair $(C_1,C_2)$ to investigate whether circuit success rates depend on the values of the phase shifts. Figure~\ref{fig:per-circuit-comp} shows the observed polarization $S$ for $C_1$ versus $S$ for $C_2$ for each pair of circuits $(C_1,C_2)$ that differ only by the values of the phases in $\z_{\theta}$ gates. There are many circuit pairs that have very different $S$, e.g., there is a circuit pair for which $S \approx 0.9$ for one circuit and $S \approx -0.3$ for the other (note that $-\nicefrac{1}{2} \leq S \leq 1$). Figure~\ref{fig:per-circuit-comp}(b) shows that the spread in the differences between the observed polarization of circuit pairs is largest for single-qubit circuits ($\sigma = 0.147$) and decreases as the number of qubits increases ($\sigma=0.052$ for $n=4$). The substantial variance in observed polarization differences implies strongly structured errors, e.g., coherent errors. Even for perfect $\z_{\theta}$ gates, the value of each phase shift impacts how errors in other gates propagate through a circuit \cite{Mavadia2018-ki, Ball2016-gk}.

\begin{table*}[b]
    \centering
    \begin{tabular}{|l|l|l|l|l|l|l|}
    \cline{1-6}
    Qubit subset $(\mathbb{Q})$         & $r (\{ \cs, \csd\},\mathbb{SU}(2),\mathbb{Q}, \nicefrac{1}{8})$& $r (\{ \cs, \csd\},\mathbb{SU}(2),\mathbb{Q}, \nicefrac{1}{2})$ &  $r_{DRB}(\{ \cs, \csd\},\mathbb{SU}(2),\mathbb{Q}, \nicefrac{1}{2})$ & $r(\cz\},\mathbb{SU}(2),\mathbb{Q}, \nicefrac{1}{2})$ & $r(\{ \cz\},\mathbb{C},\mathbb{Q}, \nicefrac{1}{2})$ \\ \cline{1-6}
    $\Q{4}$ &   & $0.25(3)\%   $            & $0.18(2)\%$  & & $0.27(2)\%$ \\ \cline{1-6}
    $\Q{5}$ &   & $0.12(1)\% $              & $0.12(1)\% $ & & $0.12(1)\%$ \\ \cline{1-6}
    $\Q{6}$& & $0.118(4)\%  $             & $0.108(5)\%$  & & $0.113(3)\%$ \\ \cline{1-6}
    $\Q{7}$& & $0.079(1)\%    $           & $0.080(2)\%$   & & $0.080(1)\%$ \\ \cline{1-6}
    $(\Q{4}, \Q{5})$&  $0.50(3)/$ & $0.77(3)\%  $             & $0.87(4)\% $  &  $0.73 (2)\%$ & $0.67(2)\%$\\ \cline{1-6}
    $(\Q{5}, \Q{6})$& $0.54(3)\%$ & $0.86(3)\%  $             & $0.81(3)\% $  &  $0.81(2)\%$ & $0.76(2)\%$  \\ \cline{1-6}
    $(\Q{6}, \Q{7})$&   $1.05(4)\%$ & $1.05(4)\%$        & $0.99(4)\%$  &  $1.04(3)\%$ & $0.98(2)\%$ \\ \cline{1-6}
    $(\Q{4}, \Q{5}, \Q{6})$&   $1.04(4)\%$ & $1.64(5)\%$        &   &  $1.48(4)\%$ & $1.51(4)\%$ \\\cline{1-6}
    $(\Q{5}, \Q{6}, \Q{7})$& $0.97(3)\%$  & $1.63(3)\%$        &   &  $1.61(4)\%$ & $1.61(3)\%$ \\\cline{1-6}
    $(\Q{4}, \Q{5}, \Q{6}, \Q{7})$&  $1.36(3)\%$ & $2.48(5)\%$        &   &  $2.45(5)\%$ & $2.34(5)\%$ \\\cline{1-6}
\end{tabular}
    \caption{\textbf{RB error rates on AQT.} The RB error rates for every RB experiment we ran on AQT. We benchmarked each connected subset of four linearly-connected qubits and used three different gate sets.}
    \label{tab:error_rates}
\end{table*}

\begin{figure}
    \centering
    \includegraphics{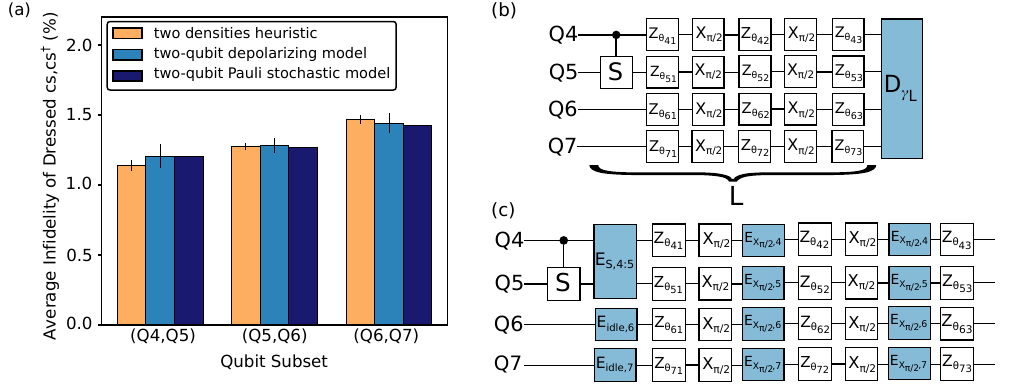}
    \caption{\textbf{Fitting error models to MRB data and estimating gate error rates.} We fit two types of error models to MRB data to estimate the infidelity of individual circuit layers. (a) By running two MRB experiments with two different two-qubit gate densities $\xi$, we can estimate the mean infidelity of a set of one or more two-qubit gates---here $\cs$ and $\csd$---using basic linear algebra (see Appendix \ref{app:two_densities}). We call this procedure the \emph{two densities heuristic}. The estimates of the average gate error obtained from the two densities heuristic (orange) are compared to independent estimates obtained from two more rigorous but more complex and computationally intensive procedures: fitting each set of two-qubit MRB data to (1) a depolarizing model (light blue), and (2) a stochastic Pauli errors model (dark blue). (b) To fit a depolarizing model, we assign an error rate to each dressed layer and an error rate to each qubit's readout. (c) To fit a Pauli stochastic model, we assign a Pauli stochastic channel to each possible gate except the virtual $\z_{\theta}$ gates. }
    \label{fig:error_models}\label{fig:2Q_layer_pairs}
\end{figure}

\subsection{Estimating the error rates of individual gates}\label{app:gate_errors}
A single MRB experiment is designed to estimate a single error rate $r_{\Omega}$ that quantifies the average rate at which an $n$-qubit layer causes an error in $\Omega$-random circuits. But we can also use MRB to extract information about the error rates of particular layers. In Section \ref{sec:gate_errors} we present one method for doing so---fitting to a depolarizing model. In this appendix we explain this method and present two alternative methods. These methods are complementary, as they trade off rigor for complexity. Note that one possible method for estimating the error rates of individual gates using MRB is to run an \emph{interleaved} \cite{magesan2012efficient} version of MRB (and interleaved standard RB has been previously used to measure the error rate of a $\cs$ gate \cite{Garion2020-gi}). We do not explore this here, although we note that interleaved MRB would inherit all of the known problems with interleaved standard RB \cite{dugas2016efficiently, kimmel2014robust}.

\subsubsection{Estimating gate error rates using a varied-densities heuristic}\label{app:two_densities}
 MRB uses flexible sampling of the circuit layers, as each composite layer is sampled from some distribution $\Omega$. By running MRB experiments with the same layer set $\mathbb{L}$ but different sampling distributions $\{\Omega_1,\Omega_2,\dots\}$ over $\mathbb{L}$, we can (approximately) ascertain the average error rates of different subsets of gates by applying basic linear algebra to $\{r_{\Omega_1}, r_{\Omega_2},\dots\}$ \cite{proctor2018direct}. In our experiments, we ran MRB for the gate set $(\{\cs, \csd\}, \mathbb{SU}(2))$ with two different $\Omega$ defined by two different two-qubit gate densities: $\xi=\nicefrac{1}{2}$ and $\xi=\nicefrac{1}{8}$. We focus on the three two-qubit sets of connected qubits. Using Eq.~\eqref{eq:r_omega_2Q}, for each two-qubit subset $\mathbb{Q}$, we have that 
\begin{equation}
\begin{pmatrix}
r_{\nicefrac{1}{2}}\\ 
r_{\nicefrac{1}{8}}
\end{pmatrix} = \begin{pmatrix} \nicefrac{1}{2} & \nicefrac{1}{2}\\ 
\nicefrac{7}{8} & \nicefrac{1}{8}
\end{pmatrix} 
\begin{pmatrix} \epsilon_1 \\ \epsilon_2\end{pmatrix},
\end{equation}
where $r_{\xi} = r(\mathbb{G}_1,\{\cs, \csd\},\textbf{}\mathbb{Q},\xi)$, $\epsilon_1$ is the infidelity of the dressed layer consisting of dressed idles on each qubit in $\mathbb{Q}$, and $\epsilon_2$ is the mean of the infidelities of $\mathbb{G}_1$-dressed $\cs$ and $\csd$ gates applied to the qubits $\mathbb{Q}$. We solve these linear equations to estimate $\epsilon_{2}$, for all three connected qubit pairs. These estimates are shown in Fig.~\ref{fig:2Q_layer_pairs}(a), and we call this method the \emph{two densities heuristic}, as it is based on the approximate relation of Eq.~\eqref{eqn:r_omega} \footnote{The two densities heuristic is not the only method for estimating $\epsilon_2$ from Eq.~\eqref{eq:r_omega_2Q}. An alternative is to first estimate $\epsilon_1$ from single-qubit MRB error rates, and to then put this $\epsilon_1$ into Eq.~\eqref{eq:r_omega_2Q} to find $\epsilon_2$ (this is what we did in Section~\ref{sec:aqt-crosstalk}). However, this will underestimate $\epsilon_1$ if there is crosstalk errors in the one-qubit gates (causing an overestimate of $\epsilon_2$), because the single-qubit MRB error rates are not impacted by these errors (as only one qubit is driven). This would not be the case if the single-qubit MRB experiments were run simultaneously \cite{gambetta2012characterization}. This is what we did in the experiments reported in Section~\ref{sec:ibm}.}.

\begin{figure*}

\centering
\includegraphics{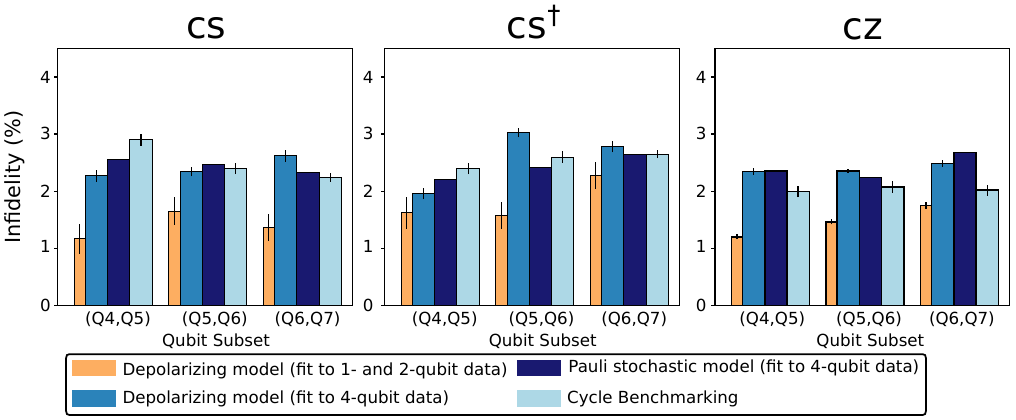}
\caption{\textbf{Estimating the infidelity of dressed 4-qubit layers.} By fitting error models to MRB data, we can estimate the infidelity of each $\mathbb{G}_1$-dressed layer used in the MRB circuits. Here we show four different estimates of the infidelities of 4-qubit layers containing a single $\cs$, $\csd$ or $\cz$ gate on one of the three connected pairs of qubits.
We fit a simple $n$-qubit depolarizing model to (1) the 4-qubit data, and (2) the 1- and 2-qubit data, and use both models to estimate the infidelity of 4-qubit $\mathbb{G}_1$-dressed layers. The estimates from fitting to the 1- and 2-qubit data do not account for any additional crosstalk errors that occur in 4-qubit layers, so the additional error estimated when fitting to the 4-qubit data is a quantification of crosstalk. We also fit a more sophisticated stochastic Pauli error model to the 4-qubit circuit data, resulting in comparable estimates to those obtained from the simple depolarizing model (which uses a scalable, less computationally intensive analysis). To validate our results against an established technique, we compare to infidelities independently estimated using cycle benchmarking \cite{erhard2019characterizing}. We observe qualitative agreement. The cycle benchmarking experiments measure the infidelities of layers dressed with one-qubit gates sampled from a different gate set (the Pauli group) to that used in our MRB experiments [$\mathbb{SU}(2)$ or $\mathbb{C}_1$, the single-qubit Clifford group], and these experiments were implemented on a different day than the MRB circuits, so exact agreement is not expected.}
\label{fig:cb}
\end{figure*}

\subsubsection{Estimating gate error rates by fitting depolarizing error models}
The two densities heuristic is built on the standard MRB data analysis, which extracts a single error rate ($r_{\Omega}$) from each MRB experiment design. But data from even a single MRB experiment contains a lot more information about each gate's errors than is contained in $r_{\Omega}$, e.g., RB data can contain sufficient information for complete tomography \cite{kimmel2014robust,Nielsen2021-cb}. In principle, this information can be extracted by fitting error models to MRB data---as has been demonstrated in simulations with 2-qubit standard RB \cite{Nielsen2021-cb}. However, fitting an error model to data typically requires simulating the circuits under that error model and this simulation is, in general, exponentially expensive in the number of qubits. Simplified, scalable approximations are therefore useful. One model that satisfies these criteria is a model in which each gate's error is modelled by a single error rate, and the error map for a layer of gates is a depolarizing channel \cite{proctor2020measuring}.

Our depolarizing model summarizes the errors in each dressed one- and two-qubit gate ($\gate{g}$) with a single, independent error rate $\epsilon_{\gate{g}}$. The error channel for each dressed $n$-qubit layer $\gate{L}$ is modelled by an $n$-qubit depolarizing channel with an infidelity $\epsilon_{\gate{L}}$ given by $\epsilon_{\gate{L}}=1-\prod_{\gate{g}\in\gate{L}}(1-\epsilon_{\gate{g}})$. This means modelling the error channel for each dressed $n$-qubit layer $\gate{L}$ by
\begin{equation}
\mathcal{D}_{\gamma_{\gate{L}}}[\rho] = \gamma_{\gate{L}}\, \rho + (1-\gamma_{\gate{L}}) \frac{\mathbb{I}}{2^n},
\end{equation}
with 
\begin{equation}
    \gamma_{\gate{L}} = \frac{1}{4^n-1}\left[4^n\prod_{\gate{g} \in \gate{L}}(1-\epsilon_{\gate{g}}) - 1\right].
\end{equation}
This error model is illustrated in Fig \ref{fig:error_models}(b). We also model the readout on each qubit with an independent error rate $\epsilon_{\Q{i}}$, where the readout error on an $n$-qubit circuit is an $n$-qubit depolarizing channel with infidelity $\epsilon_{\gate{R}}=1-\prod_{\Q{i}\in\mathbb{Q}}(1-\epsilon_{\Q{i}})$. Under this error model, the observed polarization of a circuit $\gate{C}=\gate{L}_1\gate{L}_2\dots \gate{L}_d$ is
\begin{equation}
S(\gate{C}) = \gamma(\gate{L}_1)\gamma(\gate{L}_2) \dots \gamma(\gate{L}_d)\gamma(\gate{R}).
\label{eq:S_depolarizing}
\end{equation}

The parameters of this depolarizing model are a set of error rates---an $\epsilon_{\gate{g}}$ for each $\mathbb{G}_1$-dressed one- and two-qubit gate $\gate{g}$ and an $\epsilon_{\Q{i}}$ for the readout on each qubit $\Q{i}$. To estimate these parameters we use a least-squares fit of Eq.~\eqref{eq:S_depolarizing} to the observed polarizations of the MRB circuits. We separately fit the parameters of the depolarizing model to the data from MRB circuits on different numbers of qubits ($n=1,2,3,4$), so that we can study how the error rates of the gates change with $n$, due to crosstalk errors.

Fitting Eq.~\eqref{eq:S_depolarizing} to the data from two-qubit MRB circuits results in estimates of the infidelity of each two-qubit dressed layer containing a two-qubit gate from $\mathbb{G}_2$ (and an estimate for $\epsilon_1$, the dressed two-qubit idle). We can therefore use these fits to compare to the two densities heuristic (above). Figure~\ref{fig:2Q_layer_pairs}(a) compares the mean of the entanglement infidelities of the dressed $\cs$ and $\csd$ gates, obtained from this fit, with the estimate from the two densities heuristic. The estimates of both methods are between 1.1\% and 1.5\%, with the estimates differing by 0.5\%--2.1\%, which cross-validates the two methods.

Fitting Eq.~\eqref{eq:S_depolarizing} to the data from 4-qubit MRB circuits provides estimates for the entanglement infidelities of all 15 dressed 4-qubit layers used in our experiments. Figure~\ref{fig:cb} shows the estimated infidelity for each of the nine $\mathbb{G}_1$-dressed layers that consist of a single dressed two-qubit gate applied to one of the three connected qubit pairs (in parallel with dressed idles on the other two qubits). These  infidelities are between $2\%$ and $3.1\%$, and they vary between qubit pairs and between gates ($\cs$, $\csd$ and $\cz$). We quantify the contribution of crosstalk errors to these infidelities by also predicting the infidelities of these 4-qubit layers from the dressed gate error rates obtained from fitting the depolarizing model to the one- and two-qubit data. Shown in Fig.~\ref{fig:cb}, these predicted infidelities are smaller than those estimated from the 4-qubit data by up to 60\%. Crosstalk errors are a large proportion of the total infidelity in these 4-qubit layers.

To validate our results, we compare the infidelities we estimate to independent estimates obtained from an established technique: \emph{cycle benchmarking} \cite{erhard2019characterizing}, a technique for estimating the infidelity of individual many-qubit gate layers. Fig.~\ref{fig:cb} shows that our estimates are broadly similar to those obtained from cycle benchmarking, differing by at most 23\%. Moreover, we only expect rough agreement with the cycle benchmarking estimates, for two reasons: (1) the cycle benchmarking experiments were implemented on a different day (they were run immediately after the gates were calibrated), and (2) cycle benchmarking estimates the error rate of layers that are dressed by random Pauli gates (whereas our layers are dressed by Haar-random gates or random single-qubit Clifford gates). 

\begin{figure*}[h!]
    \centering
    \includegraphics{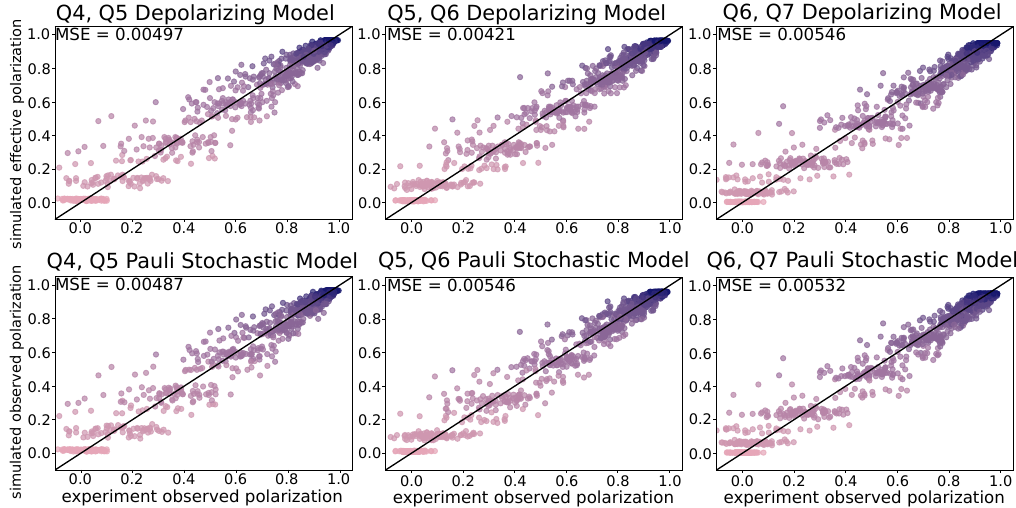}
    \caption{\textbf{Comparing Error Models for 2-Qubit MRB Data.} For each pair of qubits we benchmarked, we fit two error models, a depolarizing error model and a Pauli stochastic error model, to the data from the 2-qubit RB experiments.  Here, we compare the simulated observed polarization based on the fit models to the observed observed polarization for each circuit.}
\end{figure*}

\begin{figure*}[h!]
    \centering
    \includegraphics{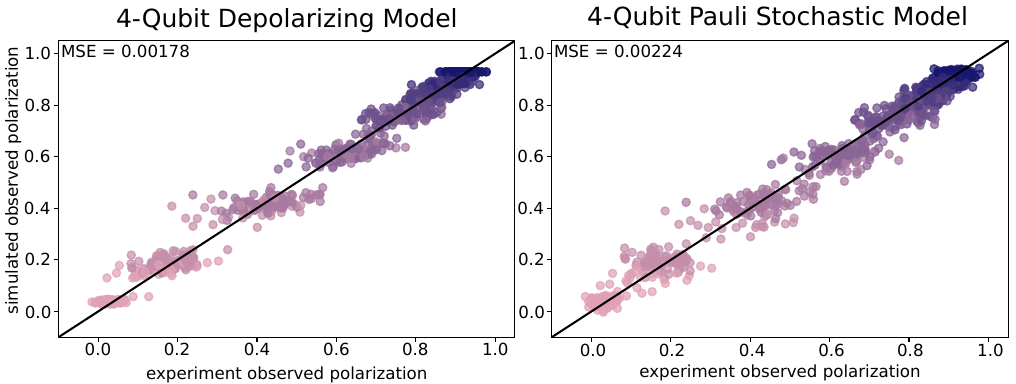}
    \caption{\textbf{Comparing Error Models for 4-Qubit MRB Data.} We fit two error models, a depolarizing error model and a Pauli stochastic error model, to the data from our 4-qubit MRB experiments. Here, we compare the simulated observed polarization based on the fit models to the observed observed polarization for each circuit. The mean-squared error in the observed polarization is approximately 40\% smaller for the Pauli stochastic model than the depolarizing model.}
\end{figure*}

\subsubsection{Estimating gate error rates by fitting Pauli error models}
Fitting data to an $n$-qubit depolarizing model is scalable, but the actual error map for each layer is unlikely to be global depolarization.  For example, a global depolarizing channel causes highly correlated errors, whereas physically we expect many errors to be local errors. We therefore fit a more physically well-motivated model against which to compare our estimates for each dressed layer's infidelity. Arbitrary Markovian errors on a set of $n$-qubit layers $\mathbb{L}$ can be modelled by an $n$-qubit process matrix for each $L \in \mathbb{L}$ \cite{Nielsen2021gate}. But each of these process matrices has $O(16^n)$ parameters, resulting in an infeasible number of parameters to estimate when $n=4$. Instead, we fit to a process matrix model of reduced complexity.
This error model is illustrated in Fig.~\ref{fig:error_models}(c). We model the error in each one- or two-qubit native gate (i.e., each $\g{x}_{\nicefrac{\pi}{2}}$ etc, not each dressed gate, or each element of $\mathbb{G}_1$) by a one- or two-qubit stochastic Pauli channel [Eq.~\eqref{eq:pauli_channel}], respectively. We allow the Pauli error rates  $\{\gamma_{P}\}$ to be gate- and qubit-dependent. We fix the error rates of the $\g{z_{\theta}}$ gate on each qubit to zero (because it is a virtual gate). We model state preparation error as a tensor product of single-qubit stochastic Pauli channels before the circuit, and we model measurement error as a tensor product of single-qubit stochastic Pauli channels immediately before readout. We estimate the error rates of all gates besides $\g{z_{\theta}}$ gates ($\cz$, $\cs$, $\csd$ on each of three qubit pairs, and $\x_{\nicefrac{\pi}{2}}$ and $\g{idle}$ on each of the four qubits) and on state preparation and readout. We fit this model (which has 159 gate error parameters and 24 SPAM error parameters) to the data from the 4-qubit MRB circuits, using maximum likelihood estimation.

The best-fit model contains a process matrix for each gate present in our circuits. These process matrices imply infidelities for each dressed layer. Figure \ref{fig:cb} shows the estimated infidelity for each of the 9 four-qubit $\mathbb{G}_1$-dressed layers containing a single two-qubit gate. The estimates we obtained from the stochastic Pauli errors model are comparable to those obtained from the depolarizing model. Maximum likelihood estimation of a Pauli stochastic model using data from general circuits is exponentially expensive in $n$ (due to the circuit simulation cost), whereas fitting a global depolarizing model is not. Note, however, that there are a variety of powerful techniques for efficient estimation of Pauli errors on Clifford gates, using data from Clifford circuits \cite{erhard2019characterizing, harper2019efficient, flammia2019efficient, Flammia2021-dn} (including a technique that uses data from Clifford randomized mirror circuits \cite{Flammia2021-dn}).

\section{IBM Q demonstrations}

The error rates from all of our MRB experiments on \texttt{ibmq\_montreal} are shown in Tables \ref{tab:ibm_many_q} and \ref{tab:1_2_q_ibm}. Calibration data from the time of our demonstration is shown in Table \ref{tab:ibm_calibration}

\begin{table*}[h!]
\centering
\begin{tabular}{|p{6cm}|l|}
\cline{1-2}
qubit subset & $r_{\Omega}$  \\\cline{1-2}
(Q0, Q1, Q2) & 0.88(2)  \\\cline{1-2} 
(Q0, Q1, Q2, Q3) & 1.39(3)  \\\cline{1-2} 
(Q0, Q1, Q2, Q3, Q4) & 1.96(5) \\\cline{1-2} 
(Q0, Q1, Q2, Q3, Q4, Q5) & 2.53(6)  \\\cline{1-2} 
(Q0, Q1, Q2, Q3, Q4, Q5, Q7) & 3.36(8)  \\\cline{1-2} 
(Q0, Q1, Q2, Q3, Q4, Q5, Q6 ,Q7) & 6.7(3)  \\\cline{1-2} 
(Q0, Q1, Q2, Q3, Q4, Q5, Q6, Q7, Q8, Q9) & 8.4(4) \\\cline{1-2} 
(Q0, Q1, Q2, Q3, Q4, Q5, Q6, Q7, Q8, Q9, Q10, Q11) & 11.6(4)  \\\cline{1-2} 
(Q0, Q1, Q2, Q3, Q4, Q5, Q6, Q7, Q8, Q9, Q10, Q11, Q12, Q13, Q14) & 14.4(7) \\\cline{1-2} 
(Q0, Q1, Q2, Q3, Q4, Q5, Q6, Q7, Q8, Q9, Q10, Q11, Q12, Q13, Q14, Q15, Q16, Q18) & 20.4(6) \\\cline{1-2} 
(Q0, Q1, Q2, Q3, Q4, Q5, Q6, Q7, Q8, Q9, Q10, Q11, Q12, Q13, Q14, Q15, Q16, Q18, Q19, Q20, Q21) & 23.6(9) \\\cline{1-2} 
(Q0, Q1, Q2, Q3, Q4, Q5, Q6, Q7, Q8, Q9, Q10, Q11, Q12, Q13, Q14, Q15, Q16, Q18, Q19, Q20, Q21, Q22, Q23, Q24, Q25) & 28(1) \\\cline{1-2} 
(Q0, Q1, Q2, Q3, Q4, Q5, Q6, Q7, Q8, Q9, Q10, Q11, Q12, Q13, Q14, Q15, Q16, Q18, Q19, Q20, Q21, Q22, Q23, Q24, Q25, Q26) &  27.9(9) \\\cline{1-2} 
\end{tabular}
    \caption{\textbf{Many-Qubit MRB on IBM Q.} The MRB error rates for every MRB experiment with $n>2$ qubits we ran on \texttt{ibmq\_montreal}. We benchmarked a single qubit subset $\mathbb{Q}$ containing $n$ qubits for 13 exponentially spaced $n$.}
    \label{tab:ibm_many_q}
\end{table*}

\begin{table*}[b]
\begin{tabular}{|l|l|l|}
\cline{1-3}
qubit subset & $r_{\Omega}$ (isolated MRB) & $r_{\Omega}$ (simultaneous MRB)   \\\cline{1-3}
\Q{0}   &           0.103(3)                            & 0.104(2)                                             \\\cline{1-3}
\Q{1}   &                                         & 0.113(3)        \\\cline{1-3} 
\Q{2}   &                                    & 0.106(2)        \\\cline{1-3} 
\Q{3}   &                                   & 0.105(1)        \\\cline{1-3} 
\Q{4}   &                                     & 0.087(2)       \\\cline{1-3} 
\Q{5}   &  & 0.113(3)    \\\cline{1-3} 
\Q{6}   & & 0.33(1)    \\\cline{1-3} 
\Q{7}   & & 0.38(1)    \\\cline{1-3} 
\Q{8}   & & 0.19(1)    \\\cline{1-3} 
\Q{9}   & & 0.149(2)    \\\cline{1-3} 
\Q{10}   & & 0.30(1)    \\\cline{1-3} 
\Q{11}   & & 0.135(3)    \\\cline{1-3} 
\Q{12}   & & 0.165(6)    \\\cline{1-3} 
\Q{13}   & & 0.118(4)    \\\cline{1-3} 
\Q{14}   & & 0.106(3)    \\\cline{1-3} 
\Q{15}   & & 0.208(7)    \\\cline{1-3} 
\Q{16}   & & 0.30(2)    \\\cline{1-3} 
\Q{17}   & & 0.127(3)    \\\cline{1-3} 
\Q{18}  & & 0.31(1)    \\\cline{1-3} 
\Q{19}   & & 0.148(4)    \\\cline{1-3} 
\Q{20}   & & 0.097(2)    \\\cline{1-3} 
\Q{21}   & & 0.205(7)    \\\cline{1-3} 
\Q{22}   & & 0.118(3)    \\\cline{1-3} 
\Q{23}   & & 0.140(5)    \\\cline{1-3} 
\Q{24}   & & 0.121(2)    \\\cline{1-3} 
\Q{25}   & & 0.391(9)    \\\cline{1-3} 
\Q{26}   & & 0.168(2)    \\\cline{1-3} 
\end{tabular}
\begin{tabular}{|l|l|l|}
\cline{1-3}
qubit subset & $r_{\Omega}$ (isolated MRB) & $r_{\Omega}$ (simultaneous MRB)   \\\cline{1-3}
(\Q{0}, \Q{1}) & 0.418(5) & 0.82(2)    \\\cline{1-3} 
(\Q{24}, \Q{25}) & & 2.7(1)    \\\cline{1-3} 
(\Q{14}, \Q{16}) & & 1.23(5)    \\\cline{1-3} 
(\Q{18}, \Q{21}) & & 1.61(7)    \\\cline{1-3} 
(\Q{3}, \Q{5}) & 0.389(5) & 0.75(2)    \\\cline{1-3} 
(\Q{4}, \Q{7}) & & 1.26(4)    \\\cline{1-3} 
(\Q{12}, \Q{15}) & & 0.99(3)    \\\cline{1-3} 
(\Q{19}, \Q{20}) & & 0.72(1)    \\\cline{1-3}
(\Q{1}, \Q{2}) & & 0.81(2)    \\\cline{1-3}  
(\Q{12}, \Q{13}) & & 0.92(3)    \\\cline{1-3} 
(\Q{22}, \Q{25}) & & 3.15(1)    \\\cline{1-3} 
(\Q{2}, \Q{3}) & & 0.64(1)    \\\cline{1-3} 
(\Q{8}, \Q{9}) & 0.499(8) & 0.85(2)    \\\cline{1-3} 
(\Q{25}, \Q{26}) & & 1.95(6)    \\\cline{1-3} 
(\Q{1}, \Q{4}) & & 1.34(3)    \\\cline{1-3} 
(\Q{6}, \Q{7}) & & 3.31(3)    \\\cline{1-3} 
(\Q{15}, \Q{18}) & & 1.98(7)    \\\cline{1-3} 
(\Q{23}, \Q{24}) & & 1.19(2)    \\\cline{1-3} 
(\Q{7}, \Q{10}) & & 1.30(4)    \\\cline{1-3} 
(\Q{11}, \Q{14}) & & 0.98(2)    \\\cline{1-3} 
(\Q{16}, \Q{19}) & 0.77(2)& 1.74(6)    \\\cline{1-3} 
(\Q{21}, \Q{23}) & & 1.45(6)    \\\cline{1-3} 
(\Q{8}, \Q{11}) & & 1.10(4)    \\\cline{1-3} 
(\Q{10}, \Q{12}) & & 0.66(1)    \\\cline{1-3} 
(\Q{19}, \Q{22}) & & 0.78(2)    \\\cline{1-3} 
(\Q{5}, \Q{8}) & & 1.00(4)    \\\cline{1-3} 
(\Q{17}, \Q{18}) & 0.530(8) & 1.29(5)    \\\cline{1-3} 
(\Q{13}, \Q{14}) & & 0.79(2)    \\\cline{1-3} 
\end{tabular}
    \caption{\textbf{1- and 2-qubit isolated and simultaneous MRB on IBMQ.} We performed simultaneous one-qubit MRB on all 27 individual qubits of \texttt{ibmq\_montreal}. We also performed simultaneous two-qubit MRB on each connected qubit pair of IBMQ Montreal, in eight groups. We ran isolated MRB experiments on five qubit pairs to compare the error rates from simultaneous and isolated two-qubit mirror RB. Isolated MRB experiments had an error rate approximately 50\% smaller than simultaneous MRB experiments.}
    \label{tab:1_2_q_ibm}
\end{table*}

\begin{table*}
\begin{tabular}{|l|l|l|l|l|l|l|l|l|}
\cline{1-9}
qubit & $T_1$ (us) & $T_2$ (us) & frequency (GHz)& anharmonicity  (GHz) & readout error & Pr(prep 1, measure 0)& Pr(prep 0, measure 1)& readout length  (ns) \\\cline{1-9}
\Q0 & 99.80 & 29.03 & 4.91 & -0.34 & 0.010 & 0.015 & 0.005 & 5201.78 \\\cline{1-9}
\Q1 & 159.43 & 24.69 & 4.83 & -0.32 & 0.015 & 0.024 & 0.005 & 5201.78 \\\cline{1-9}
\Q2 & 88.31 & 107.64 & 4.98 & -0.34 & 0.016 & 0.019 & 0.013 & 5201.78 \\\cline{1-9}
\Q3 & 65.10 & 68.28 & 5.10 & -0.34 & 0.010 & 0.011 & 0.009 & 5201.78 \\\cline{1-9}
\Q4 & 128.29 & 147.35 & 5.00 & -0.34 & 0.013 & 0.017 & 0.008 & 5201.78 \\\cline{1-9}
\Q5 & 68.79 & 100.11 & 5.03 & -0.34 & 0.013 & 0.015 & 0.011 & 5201.78 \\\cline{1-9}
\Q6 & 161.18 & 23.97 & 4.95 & -0.39 & 0.063 & 0.060 & 0.066 & 5201.78 \\\cline{1-9}
\Q7 & 141.25 & 101.48 & 4.91 & -0.32 & 0.063 & 0.061 & 0.065 & 5201.78 \\\cline{1-9}
\Q8 & 78.10 & 117.90 & 4.91 & -0.32 & 0.017 & 0.020 & 0.014 & 5201.78 \\\cline{1-9}
\Q9 & 96.97 & 101.53 & 5.04 & -0.34 & 0.009 & 0.013 & 0.006 & 5201.78 \\\cline{1-9}
\Q10 & 120.94 & 86.65 & 5.08 & -0.34 & 0.009 & 0.014 & 0.004 & 5201.78 \\\cline{1-9}
\Q11 & 162.14 & 46.60 & 5.03 & -0.34 & 0.015 & 0.013 & 0.018 & 5201.78 \\\cline{1-9}
\Q12 & 127.39 & 172.19 & 4.97 & -0.32 & 0.015 & 0.025 & 0.005 & 5201.78 \\\cline{1-9}
\Q13 & 119.46 & 68.34 & 4.87 & -0.34 & 0.008 & 0.012 & 0.004 & 5201.78 \\\cline{1-9}
\Q14 & 130.38 & 105.47 & 4.96 & -0.32 & 0.009 & 0.013 & 0.005 & 5201.78 \\\cline{1-9}
\Q15 & 108.30 & 78.65 & 5.03 & -0.34 & 0.016 & 0.024 & 0.008 & 5201.78 \\\cline{1-9}
\Q16 & 98.09 & 44.07 & 5.09 & -0.34 & 0.014 & 0.023 & 0.004 & 5201.78 \\\cline{1-9}
\Q17 & 119.27 & 116.98 & 5.07 & -0.34 & 0.013 & 0.015 & 0.010 & 5201.78 \\\cline{1-9}
\Q18 & 87.46 & 31.94 & 4.98 & -0.33 & 0.029 & 0.038 & 0.020 & 5201.78 \\\cline{1-9}
\Q19 & 116.18 & 164.59 & 4.98 & -0.32 & 0.010 & 0.014 & 0.007 & 5201.78 \\\cline{1-9}
\Q20 & 131.79 & 148.40 & 5.08 & -0.34 & 0.006 & 0.008 & 0.005 & 5201.78 \\\cline{1-9}
\Q21 & 131.59 & 51.46 & 5.07 & -0.31 & 0.018 & 0.023 & 0.014 & 5201.78 \\\cline{1-9}
\Q22 & 77.81 & 145.47 & 5.06 & -0.34 & 0.019 & 0.028 & 0.010 & 5201.78 \\\cline{1-9}
\Q23 & 146.96 & 63.95 & 4.97 & -0.34 & 0.009 & 0.015 & 0.003 & 5201.78 \\\cline{1-9}
\Q24 & 121.03 & 70.31 & 5.05 & -0.34 & 0.018 & 0.024 & 0.012 & 5201.78 \\\cline{1-9}
\Q25 & 30.19 & 38.93 & 4.93 & -0.32 & 0.026 & 0.043 & 0.009 & 5201.78 \\\cline{1-9}
\Q26 & 66.93 & 127.66 & 5.00 & -0.34 & 0.013 & 0.022 & 0.005 & 5201.78 \\\cline{1-9}
\end{tabular}
\caption{\textbf{IBMQ Montreal calibration data.} Calibration data from \texttt{ibmq\_montreal} from the time of our MRB demonstrations (September 7, 2021).}
    \label{tab:ibm_calibration}
\end{table*}
\end{document}